\documentclass[journal=jacsat,manuscript=article]{achemso}
\usepackage{chemformula} 
\usepackage[T1]{fontenc} 
\pdfoutput=1 
\usepackage{multirow}
\usepackage[version=3]{mhchem}
\usepackage{graphicx}
\usepackage{subcaption}
\usepackage{color,ulem}

\usepackage{dblfloatfix}
\usepackage{amssymb}
\usepackage{amsmath}
\usepackage[compact]{titlesec}
\usepackage{comment}

\author{Kamil M. Ciesielski}
\affiliation[CSM]
{Department of Physics, Colorado School of Mines, Golden 80401, Colorado, United States}

\email{kciesielski@mines.edu}
\alsoaffiliation[INTiBS]
{Institute of Low Temperature and Structure Research, Polish Academy of Sciences, 50-422 Wroclaw, Poland}

\author{Brenden R. Ortiz}
\affiliation[CSM]
{Department of Physics, Colorado School of Mines, Golden 80401, Colorado, United States}
\alsoaffiliation[UCSB]
{Materials Department, University of California, Santa Barbara 93106, California, United States}

\author{Lidia C. Gomes}
\affiliation[SP]
{Instituto de F\'isica Te\'orica, S\~ao Paulo State University (UNESP), 01049-010 S\~ao Paulo, Brazil}

\author{Vanessa Meschke}
\affiliation[CSM]
{Department of Physics, Colorado School of Mines, Golden 80401, Colorado, United States}

\author{Jesse M. Adamczyk}
\affiliation[CSM]
{Department of Physics, Colorado School of Mines, Golden 80401, Colorado, United States}

\author{Tara L. Braden}
\affiliation[CSM]
{Department of Physics, Colorado School of Mines, Golden 80401, Colorado, United States}

\author{Dariusz Kaczorowski}
\affiliation[INTiBS]
{Institute of Low Temperature and Structure Research, Polish Academy of Sciences, 50-422 Wroclaw, Poland}

\author{Elif Ertekin}
\affiliation[IUC]
{Department of Mechanical Science and Engineering, University of Illinois Urbana Champaign, Champaign 61820, Illinois, United States}

\author{Eric S. Toberer}
\affiliation[CSM]
{Department of Physics, Colorado School of Mines, Golden 80401, Colorado, United States}
\email{etoberer@mines.edu}

\title[An \textsf{achemso} demo]
  {Origin of ultra-low thermal conductivity in unconventional clathrates: Strong scattering from extremely low-frequency rattling modes}
  
\begin{document}

\begin{abstract}
Recent discoveries of materials with ultra-low thermal conductivity open a pathway to significant developments in the field of thermoelectricity. Here, we conduct a comparative study of three chemically similar antimonides to establish the root causes of their extraordinarily low thermal conductivity ($0.4-0.6$ Wm$^{-1}$K$^{-1}$ at 525\,K). The materials of interest are:  the  unconventional type-XI clathrate K$_{58}$Zn$_{122}$Sb$_{207}$,  the tunnel compound K$_{6.9}$Zn$_{21}$Sb$_{16}$,  and the type-I clathrate  K$_8$Zn$_{15.5}$Cu$_{2.5}$Sb$_{28}$ discovered herein. Calculations of the phonon dispersions show that the type-XI compound exhibits localized (\textit{i.e.}, rattling) phonon modes with unusually low frequencies that span the entire acoustic regime. In contrast, rattling in the type-I clathrate is observed only at higher frequencies, and no rattling modes are present in the tunnel structure. Modeling reveals that low-frequency rattling modes profoundly limit the acoustic scattering time; the scattering time of the type-XI clathrate is half that of the type-I clathrate and a quarter of the tunnel compound. For all three materials, the thermal conductivities are additionally suppressed by soft framework bonding that lowers the acoustic group velocities, and structural complexity that leads to diffusonic character of the optical modes. Understanding details of thermal transport in structurally complex materials  will be crucial for developing the next generation of thermoelectrics.
\end{abstract}

\section{Introduction}

The discoveries of materials with ultra-low thermal conductivity open a pathway towards breakthroughs in thermoelectric energy conversion efficiency and  thermal barriers coatings \cite{koley2021ultralow, mukhopadhyay2018two, lin2016concerted, zhao2014ultralow, schmitt2012probing, fulmer2013clathrate, cox2021clathrate}.
For example, Yb$_{14}$MnSb$_{11}$ \cite{brown2006yb14mnsb11} is a canonical example of a complex Zintl compound with lattice thermal conductivity that is half of silica glass (104 atoms in the primitive cell; 0.4-0.6 W m$^{-1}$K$^{-1}$); this discovery has enabled new, high efficiency radioisotope thermoelectric generators.  More recently, the discovery of the clathrate compound K$_{58}$Zn$_{122}$Sb$_{207}$ \cite{cox2021clathrate}, with lattice thermal conductivity below 0.4 W m$^{-1}$K$^{-1}$, inspires the current work.  However, structurally complex semiconductors, with dozens to hundreds of atoms in the unit cell, remain a challenging domain for predicting  thermal transport due to the mixture of localized, diffusive, and propagating vibrational modes \cite{agne2018minimum}. Developing an understanding of how crystal structure affects transport will require robust models of how these distinct transport channels contribute.

In general, structurally complex crystalline materials are not expected to exhibit high  thermal conductivity \cite{slack1979thermal}. This primarily origins from the massive number of optical branches that arise in large unit cells; the optical modes typically exhibit both low group velocity and strong scattering. However, to achieve thermal conductivity below 0.4 Wm$^{-1}$K$^{-1}$ in a dense crystalline material requires additional mechanisms beyond structural complexity.  

In particular, our focus turns to suppressing heat transfer from low-frequency vibrational modes in or near the acoustic branches. Such modes retain high group velocity, are less sensitive to point defects, and have fewer scattering channels. While nanostructures have historically been the focus of curtailing transport in these modes \cite{kim2015strategies}, we instead seek to explicitly design the phonon dispersion to lower the group velocities and enhance scattering. The emergence of semiconductors based on rigid cages containing  under-constrained, `rattling' atoms, \textit{e.g.}, clathrates \cite{dolyniuk2016clathrate} or skutterudites \cite{nolas1999skutterudites}, provided one of the earliest ways to rationally alter the dispersion. For example, changing the mass or the rattler-cage interaction strength allows the rattling frequencies to be adjusted.
Beyond classic rattlers, other moieties known for low frequency modes that interrupt acoustic transport include  antimony dumbbells in Zn$_4$Sb$_3$~\cite{schweika2007dumbbell} and ionic (Ag$^-$) sublattice in CsAg$_5$Te$_3$~\cite{lin2016concerted}. 

Focusing on rattling, the phenomenon is defined as enhanced atomic displacement parameters of the rattling atom with respect to the surrounding cage. Rattling can lead to formation of low-lying optical phonon branches, which interact with acoustic modes via avoided crossing~\cite{christensen2008avoided}. Originally, the effect of rattling on thermal conductivity was considered primarily in the view of resonant scattering \cite{tse2001phonon, euchner2012phononic}. However, meticulous studies of phononic properties in clathrates revealed, that optical modes from filler atoms can at least partially hybridize with the host lattice, which leads to enhanced scattering in rather wide range of frequencies \cite{tadano2015impact}. Furthermore, the interaction between guest and host modes results in softening of the host phonon modes \cite{tadano2015impact, euchner2012phononic, norouzzadeh2017phonon}. 

\begin{figure}[t]
	\centering
	\includegraphics[width=8.6cm]{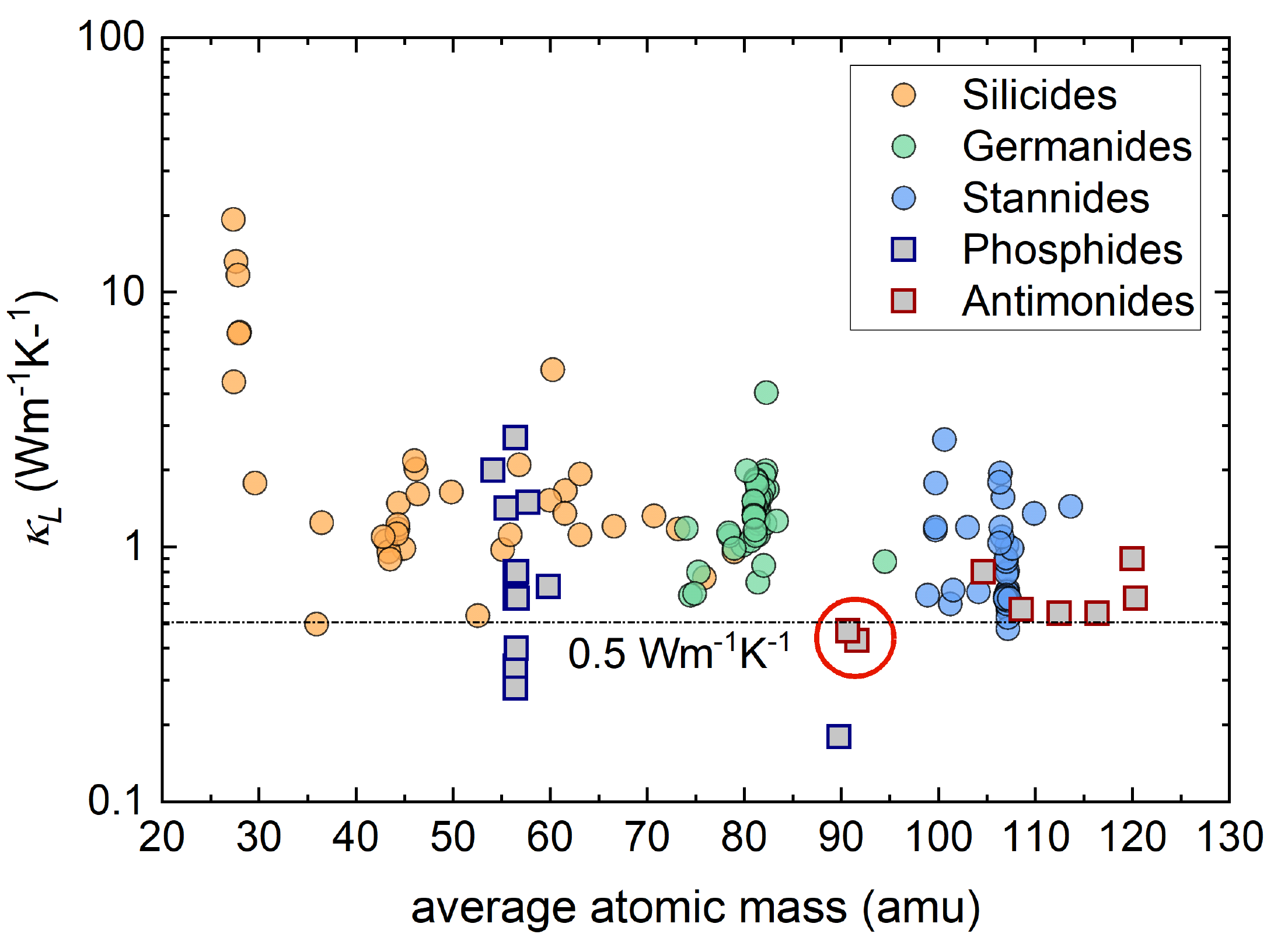}
	\caption{
	The experimental lattice thermal conductivity of clathrates at room temperature exhibits curiously low values for P- and Sb-based compounds. The circles and squares indicate conventional and unconventional clathrates, respectively. The points within the red open oval  at the bottom of the graph denote two clathrates studied in this article, K$_{55}$Zn$_{122}$Sb$_{207}$ and  K$_{8}$Zn$_{15.5}$Cu$_{2.5}$Sb$_{28}$. The
	majority of the data was adapted from Ref. \cite{dolyniuk2016clathrate}, while the data for phosphides and antimonides was updated from recent works \cite{fulmer2013clathrate, dolyniuk2017breaking, wang2017high, wang2018smaller, dolyniuk2018chemical, huo2003thermoelectric, dolyniuk2015twisted, owens2020chemically, owens2020iii}. 
	}
	\label{Fig1} 
\end{figure}

Clathrate compounds are  ideal for in-depth studies of how structural complexity impacts thermal conductivity. They comprise a wide range of available chemical compositions and a plethora of interesting transport phenomena, as well as potential for thermoelectric applications \cite{dolyniuk2016clathrate, takabatake2014phonon}.   `Conventional' semiconducting clathrates typically form in the type-I structure ($A_8X_{46}$, with rattler $A$ and framework $X$) and have frameworks built from covalently bound tetrels (Si, Ge, Sn) and elements from group 13 of the periodic table (Al, Ga, In). The resulting framework is based on face-sharing cages of pentagonal decahedra and tetrakaidecahedra.  Charge balance is achieved with cationic guest atoms, namely alkali metals or alkali-earth elements (Na, K, Rb, Cs, Ba).   The excellent thermoelectric performance of materials from this family has been recognized for two decades \cite{takabatake2014phonon, dolyniuk2016clathrate}. In particular,  Ba$_8$Ga$_{16}$Ge$_{30}$, a type-I clathrate, is known for high energy conversion efficiency in both \textit{n}-type \cite{toberer2008high} and \textit{p}-type \cite{wang2017thermoelectric} regimes. Conventional clathrates exhibit wide range of lattice thermal conductivity 0.5-20 Wm$^{-1}$K$^{-1}$~\cite{dolyniuk2016clathrate}. 
As can be seen from Fig.~\ref{Fig1}, the lattice thermal conductivity ($\kappa_L$) correlates inversely with their average atomic mass ($m_{av}$). The highest values of $\kappa_L$ are observed for conventional clathrates marked by circles on Fig. \ref{Fig1}, decreasing from silicides, through germanides down to stannides.  This trend stems from changes in both host network and guest atoms behavior with $m_{av}$. Heavier conventional clathrates are known to comprise softer cage networks -- increase in $m_{av}$ leads to linear decrease in Debye temperature, which limits propagation of acoustic phonons \cite{dolyniuk2016clathrate}. From the  perspective of filler atoms, bigger cages characteristic for heavier clathrates leads to lower Einstein temperatures of rattling modes~\cite{suekuni2007cage}, which in turn results in stronger acoustic-optical phonon scattering  \cite{ikeda2019kondo}. As expected based on a simple mechanical oscillator model, somewhat similar effect can be achieved by increasing the mass of the filler atom \cite{norouzzadeh2017phonon}.

New and uncommon clathrates such as Ba(Cu/Zn)$_{24}$P$_{28-\delta}$ \cite{dolyniuk2017breaking} or Cs$_8$In$_{27}$Sb$_{19}$ \cite{owens2020iii} occasionally emerge, driven by either pseudo-tetrel ($sp^3$-like) behavior of the elements from other groups of the periodic table or new bonding arrangements. Their host network is usually built from transition metals or post-transition metals (Ni, Cu, Zn, Au) paired with pnictides (P, As, Sb). Greater flexibility in the local coordination of these elements, with respect to classic tetrel-based clathrates, leads to new structural arrangements and bonding patterns \cite{wang2018unconventional}.  
These include extraordinary cavities, \textit{eg.} truncated octahedra [$4^6 6^8$] with  a square-planar coordination for Ni and Cu, and P$_3$ and P$_4$ rings for phosphorus \cite{wang2016enclathration}, or a 22-vertex cage with rhomboid faces [$4^5 5^6 6^2$] accompanied by five- and six-fold coordination of host atoms \cite{dolyniuk2017breaking}. 
Thermal conductivity of phosphides can be lower than that of conventional clathrates, despite rather low $m_{av}$, see Fig.~\ref{Fig1}. The recently reported group of antimonide clathrates with high $m_{av}$ attains almost universally ultra-low $\kappa_L$. Fig. S1 shows also similar plot for the total thermal conductivity. Data in such representation merges electronic and lattice contributions, yet benefits from higher accuracy than $\kappa_L$, which might suffer from error of Lorenz number calculations. Fig. S1 clearly shows, that the antimonides are characterized overall by the most suppressed thermal transport within the clathrate family. The origin of this important observation has not yet been quantified.

Clathrates based on Sb are still rare, yet they attract a significant attention as antimony is one of the most common building blocks for thermoelectric materials \cite{shi2011multiple,  fu2015band, ohno2018phase, ohno2017achieving, kraemer2015high, brown2006yb14mnsb11}. Antimonide clathrates are known to attain extremely intricate structures. Cs$_8$In$_{27}$Sb$_{19}$, $A_8$Ga$_{27}$Sb$_{19}$ (\textit{A}~=~Cs, Rb) \cite{owens2020iii}, and Cs$_8$Cd$_{18}$Sb$_{28}$  \cite{owens2020chemically} crystallize in 8-times enlarged supercells of archetypal type-I structure, while Cs$_{8}$Zn$_{18}$Sb$_{28}$ attain 18-fold enlarged type-I clathrate supercell  \cite{owens2020chemically}. The last cell is the biggest known for any clathrate to date. Very recently J. Zaikina \textit{et al.} reported the discovery of K$_{58}$Zn$_{122}$Sb$_{207}$, a compound with a novel type of atomic arrangement structure, that was called type-XI clathrate \cite{cox2021clathrate}. The noncentrosymmetric unit cell comprises cages that were observed before: pentagonal decahedra and tetrakaidecahedra, as well as by 23- and 24-vertex polyhedra unique to this composition.  Lattice thermal conductivity of Sb-based clathrates can be as low as 0.35 Wm$^{-1}$K$^{-1}$, whose values were obtained for type-XI clathrate \cite{cox2021clathrate}. The known antimonides clathrates are narrow band-gap semiconductors $E_g$ = 25-340 meV  \cite{owens2020iii, owens2020chemically, cox2021clathrate}. Last, but not least, some of the representatives of this family of compounds are also known for  excellent hole mobility, reaching 880 cm$^2$V$^{-1}$s$^{-1}$ at room temperature \cite{owens2020iii}. The last finding is particularly notable, as mobility for clathrates reported so far was usually well below 50 cm$^2$V$^{-1}$s$^{-1}$, apart of one exceptional case of 170 cm$^2$V$^{-1}$s$^{-1}$ for (K, Ba)$_{24}$(Ga, Sn)$_{136}$ \cite{kishimoto2015thermoelectric}. Despite the large promise for applications,  detailed studies of  thermoelectric transport at elevated temperatures for Sb-clathrates are yet to be performed.

In this paper, our primarily goal is to understand origin of extraordinarily suppressed phonon transport in the type-XI clathrate K$_{58}$Zn$_{122}$Sb$_{207}$.
We begin by preparing bulk, polycrystalline ingots of the type-XI clathrate and the  tunnel compound K$_{6.9}$Zn$_{21}$Sb$_{16}$ \cite{cox2018rapid}; the preliminary doping efforts of type-XI compound  revealed also the existence of K$_8$Zn$_{15.5}$Cu$_{2.5}$Sb$_{28}$, which crystalizes in type-I clathrate structure. Existence of these chemically similar materials with different unit cells enable a comparative study of the properties underlying their thermal transport \cite{lo2017synthesis, qin2016diverse, wu2009comparative}.

We begin with measurements of the crystal structure and bonding  through diffraction and speed of sound measurements, respectively.  The thermal conductivity is then presented and compared to a diffuson model. To reconcile the differences between the materials, \textit{ab-initio} calculations of the phonon dispersions highlight the difference in the rattling modes.  These calculations are validated with measurements of low temperature heat capacity.  Equipped with a comprehensive understanding of the low-frequency phononic dispersions, the thermal conductivities are then modeled analytically.  We conclude with the high temperature electronic properties to consider the thermoelectric potential of these materials.

\section{Methods}

\subsection{Experimental techniques}
\label{sect:exp-methods}

The samples were synthesized by combination of high-energy mechanical milling and reactive hot pressing. Raw elements with high purity: K 99.95\%, Zn 99.999\%, Sb 99.999\%, were weighted to a total mass of 10g and milled in tungsten carbide vials in high energy Spex 8000D ball mill. Two tungsten carbide balls with a diameter of 12.7~mm were used as the grinding media. Milling was done in two intervals of 30 min with 15 min of break in between to avoid overheating of the powder from friction. Shorter grinding times (2$\times$10 min and 2$\times$20 min) were found insufficient for proper mixing of the elements. The  powders were sieved through a 106 $\mu$m sieve and loaded into a graphite die. All procedures were performed in an inert nitrogen atmosphere in a dry glove box (O$_2$ $<$ 1 ppm, H$_2$O $<$ 1 ppm).

The die with the powder was put into a hot press and pre-baked for 30 min at 150$^\circ$C under dynamic vacuum (\textit{p} $<$ 1\,mTorr) in order to get rid of possible organic contaminants and traces of oxygen. The optimal sintering conditions were chosen based on the plunger displacement measured along the densification process. We found that the displacement reaching \textit{ca.} 1 mm for 2~g of loaded powder into 12.7 mm diameter die was sufficient for the samples to be synthesized in dense ($>$94\%, see below), virtually phase-pure form. The displacement was observed as a function of temperature at fixed pressure, starting from 30 MPa. If the proper shift was not observed until 550$^{\circ}$C, \textit{i.e.} the technical limitation of our equipment, the pressure was increased and the process was repeated on a separate powder. The exemplary schematic graph of displacement curve is shown in Fig. S2. The resultant optimal sintering conditions are  400$^\circ$C, 38 MPa for K$_8$Zn$_{15.5}$Cu$_{2.5}$Sb$_{28}$, 500$^\circ$C, 50 MPa for K$_{58}$Zn$_{122}$Sb$_{207}$ and    300$^\circ$C and 50 MPa for K$_{6.9}$Zn$_{21}$Sb$_{16}$. The densification time was 2 hours. Total time of synthesis did not exceed 5 hours. Bulk samples obtained by this method were found to be  stable in air for several months in Golden, Colorado. 

X-ray diffraction for both clathrates was performed at room temperature in Advanced Photon Source (APS), Argonne National Laboratory, 11-BM, using wavelength of 0.457917~\AA. For K$_{6.9}$Zn$_{21}$Sb$_{16}$ we carried out laboratory X-ray diffraction with a Bruker D2 Phaser diffractometer. All powders for XRD measurements were prepared from ground pellets resultant from hot pressing. This approach introduces more strain to the powder, which makes the Rietveld refinement more difficult, yet ensures, that structural data represents the very same material, that was measured from the perspective of thermoelectricity.  Rietveld refinements were performed with FullProf software \cite{rodriguez1993recent}. Heat capacity measurements were performed on a PPMS-9 Quantum Design system with 2$\tau$-relaxation method. Temperature dependent electrical resistivity and Hall effect measurements were carried out using the Van der Pauw geometry on a custom built device \cite{Borup2012HallApparatus, Vanderpauw1958ResistivityApparatus}. The current supplied for the Hall effect measurements was 100 mA in a magnetic field of 1 T. Measurement of the temperature dependent Seebeck coefficient was performed on a custom built apparatus \cite{Iwanaga2011SeebeckApparatus}. Multiple high temperature cycles of both the Hall effect and Seebeck coefficient measurements were performed in order to determine if any chemical evolution or hysteresis was present. Sound velocity measurements were carried out using a pulse-echo transducer setup with both longitudinal and transverse transducers (Olympus 5072PR Pulser/Receiver).  Thermal diffusivity was measured with a Netzsch LFA 467 Flash Diffusivity system. The specimens were coated with a graphite spray before the diffusivity measurements in order to suppress parasitic emissivity. The thermal conductivity was obtained with relation $\kappa$ =  $D d_{exp} C_p$, where \textit{D} denotes the  diffusivity, $d_{exp}$ stands for the density obtained with geometrical method, and $C_p$ corresponds to the heat capacity. 

\subsection{Theoretical Methods}

Phonon dispersions of the clathrates were obtained using density functional theory (DFT), as implemented in the Vienna Ab initio Simulation Package (VASP)~\cite{kressePRB1996}. First, we generated the structures of the three different materials (type-XI and type-I clathrates, as well as the tunnel compound). Given the disorder in these compounds, site occupations approximated to the majority elements, which led to elimination of negative phonon modes with respect to preliminary calculations. The reader is referred to Supporting Information, Section 3 for the details. This finding is in line with intuition that crystallographic disorder is causing structural instability in the materials.

\begin{table*}[t]
  \caption{Information about the unit cells of K$_8$Zn$_{15.5}$Cu$_{2.5}$Sb$_{28}$, K$_{58}$Zn$_{122}$Sb$_{207}$, and K$_{6.9}$Zn$_{21}$Sb$_{16}$ compounds for the DFT phonon calculations.}
  \label{tab:cells-info}
  \begin{tabular}{cccc}
  \hline
     & Number of atoms &Unit-cell volume (\AA$^{3}$) & Number of images \\
     \hline 
     K$_8$Zn$_{15.5}$Cu$_{2.5}$Sb$_{28}$ (type-I) & 54 & 1661.72 & 324 \\ 
     K$_{58}$Zn$_{122}$Sb$_{207}$ (type-XI) & 389 & 11374.14 & 2316 \\ 
     K$_{6.9}$Zn$_{21}$Sb$_{16}$ (Tunnel) & 44 & 1198.62 & 264 \\ 
  \hline
  \end{tabular}
  \end{table*}

Structural optimization was then performed using the generalized gradient approximation (GGA) of Perdew-Burke-Ernzerhof (PBE) \cite{Perdew1996generalized} and the projector augmented wave formalism~\cite{blochl1994projector}. The Kohn-Sham orbitals were expanded using a plane-wave basis with a cutoff energy of 400~eV. Lattice vectors and atomic positions were relaxed until forces were lower than 0.1~meV/\AA~on each atom. For type-I clathrate and the tunnel compound, the Brillouin zone was sampled using a $\Gamma$-centered $2\times2\times2$ Monkhorst-Pack k-point grids, respectively~\cite{monkhorst1976special}. For the larger unit-cell of type-XI clathrate we used the $\Gamma$-point only~\cite{monkhorst1976special}. The number of atoms and volume of the cells, after optimization, are summarized in Table~\ref{tab:cells-info}. The total energy convergence criteria was 10$^{-6}$ eV. The POSCAR and CONTCAR files used for our calculations are uploaded in the Mendeley database with DOI: 10.17632/t3s52hxm6h.1.

To compute phonon dispersions, and projected density of states (PDOS), the supercell approach with the finite displacement method~\cite{chaput2011} was applied using Phonopy~\cite{togo2015first}. The large size of the unit cells allows us to use 1$\times$1$\times$1 cell to generate the finite displacement for type-XI clathrate. For the tunnel compound and type-I clathrate 2$\times$2$\times$2 supercells were used. The number of images generated for each compound (corresponding to a symmetry reduced set of atomic displacements) are also presented in Table~\ref{tab:cells-info}. Force constants are calculated from the sets of atomic forces computed from DFT and are used to build the dynamical matrix. Phonon frequencies and eigenvectors are then calculated from the dynamical matrices, for specified q-points~\cite{parlinski1997first}.

We also performed tests using the DFT + U approach, as proposed by Dudarev et al.~\cite{dudarev1998electron}. Such method seeks to treat the on-site Coulomb interactions of localized orbitals. Here, we applied an additional Hubbard-like term of $U$~=~+6eV (a typical value used in the literature~\cite{lutfalla2011calibration, weber2012vanadium}) to the $d$-orbitals of Zn atoms. This value was chosen using our previous work for a zinc antimonide ZnVSb \cite{bensen2021anomalous}. There, we checked the typical range of the parameter $U$ for Zn in the +5, +6, +7 eV \cite{goh2017effects, kanoun2012magnetism}. The eventual values of $U$ were chosen based on comparison of theoretical density of states with experimental X-ray photoemission spectroscopy results \cite{bensen2021anomalous}. No significant impact of exchange-correlation potential on the phonon dispersion was observed for the type-I and type-XI clathrates. For the tunnel compound, however, such correction significantly reduced the number of negative modes.

\section{Results and Discussion}

In the following comparative analysis, we begin by describing the synthetic efforts and the associated structural chemistry of the three K-Zn-Sb compounds considered herein. Transport measurements are then conducted to understand how the structural differences manifest in terms of phononic dispersion and scattering as well as electrical transport. To understand differences in the thermal transport between these materials, theoretical calculations and low temperature heat capacity measurements are then combined with analytic transport models.

\subsection{Synthesis} The known type-XI clathrate K$_{58}$Zn$_{122}$Sb$_{207}$ \cite{cox2021clathrate} and the tunnel compound K$_{6.9}$Zn$_{21}$Sb$_{16}$ \cite{cox2018rapid} were successfully prepared by high energy ball milling and reactive hot pressing. 
Our lattice parameters for type-XI clathrate (\textit{a}~=~22.4985(6) \AA, \textit{c}~=~22.471(1)~\AA) and the tunnel compound (\textit{a}~=~12.340(4)~\AA, \textit{c}~=~7.345(2)~\AA) are close to those reported in previous literature \cite{cox2021clathrate, cox2018rapid}.
The comparative structural discussion herein is indebted to the prior single crystal diffraction studies \cite{cox2018rapid, cox2021clathrate}.

\begin{figure}[t]
	\centering
	\includegraphics[width=8.6cm]{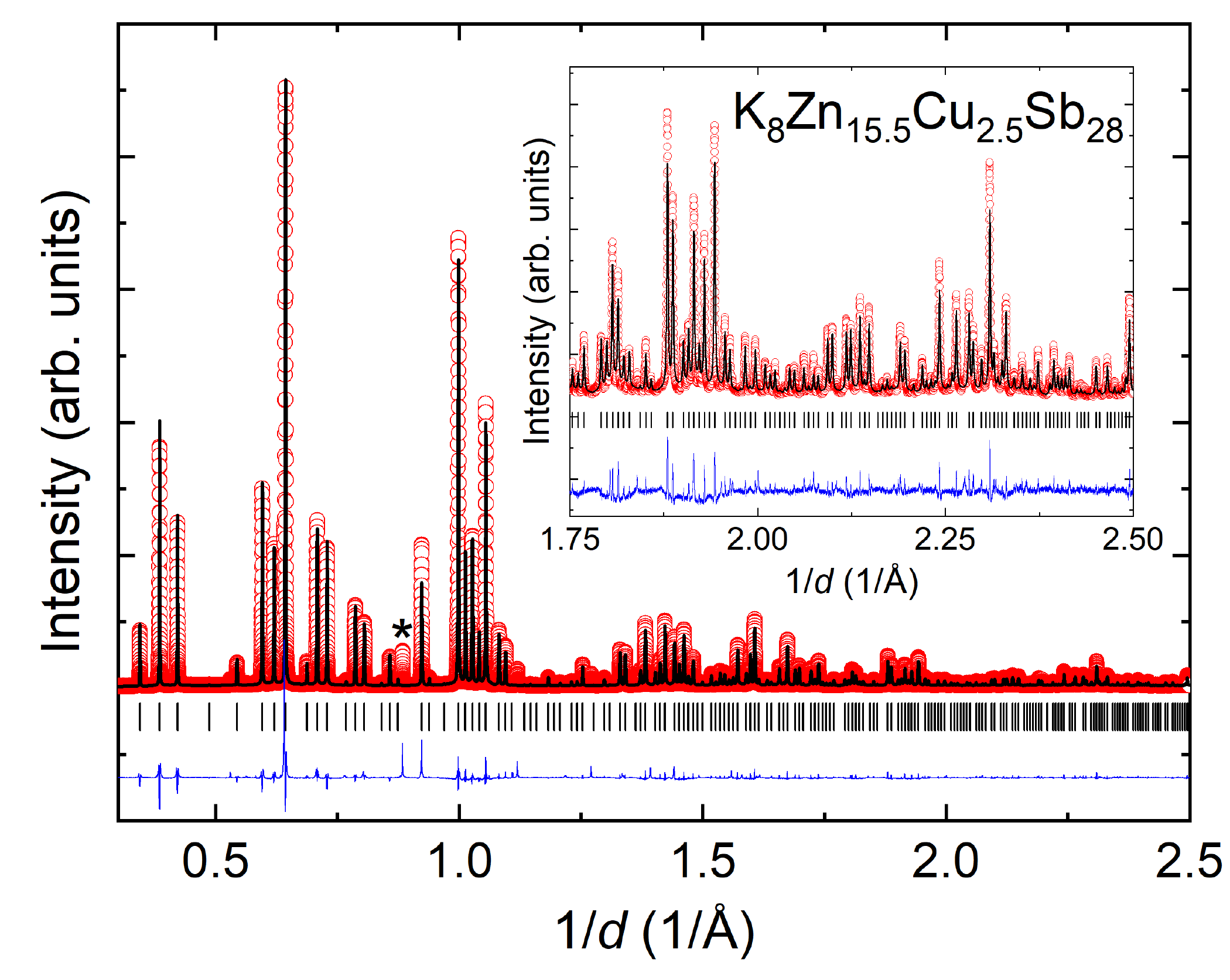}
	\caption{Synchrotron powder X-ray diffraction for K$_{8}$Zn$_{15.5}$Cu$_{1.5}$Sb$_{28}$ indicates that this quaternary forms in the type-I clathrate structure. The sample is virtually phase-pure; the only minuscule impurity is denoted by the asterisk. The inset presents the data from the low lattice spacing region. Data was obtained from BM-11 of Advanced Photon Source in Argonne National Laboratory.}
	\label{Fig2} 
	\end{figure}

Preliminary efforts to dope the type-XI clathrate with Cu led to the discovery of third compound of this investigation: K$_{8}$Zn$_{15.5}$Cu$_{2.5}$Sb$_{28}$.  The material crystallizes in a cubic type-I clathrate structure, \textit{a} = 11.59424(2) \AA. The synchrotron XRD pattern of K$_{8}$Zn$_{15.5}$Cu$_{2.5}$Sb$_{28}$ does not show any impurities from elements or known binary or ternary compounds from K-Zn-Sb-Cu space. The plot includes only a few minuscule unidentified maxima, and the most significant one is denoted by asterisk in Fig.~\ref{Fig2}. The reader is also referred to Fig. S3, which due to logarithmic Y axis underlines the low intensity unidentified peaks.  These maxima can be either signatures of superstructure that we could not solve based on powder XRD data, or traces of impurity phases. The latter issue is referred to during discussion of EDS chemical mapping (see Supporting Information, Sect. 1). Atomic positions, occupancies, and the displacements parameters ($U_{iso}$) resultant from the refinement for type-I clathrate are gathered in Tab. \ref{tab:RietveldDetails}. The other two structures, namely the type-XI and tunnel compounds, had diffraction patterns, lattice parameters and overall stoichiometries in close agreement with prior studies by Cox \textit{et al.} \cite{cox2021clathrate, cox2018rapid}, see Fig. S4 and S5, respectively and Tab. S1, S2 for the structural data. Assuming highest quality of structural data is obtainable from single-crystal refinement, the atomic coordinates, occupancies and atomic displacement coefficients in the final model were used from the works of Cox et al. \cite{cox2021clathrate, cox2018rapid}. Regarding phase purity, the only impurity in our type-XI clathrate sample  was a minuscule amount of K$_{6.9}$Zn$_{21}$Sb$_{16}$, while the tunnel compound is virtually a phase-pure specimen.

The other two structures, namely the type-XI and tunnel compounds, had diffraction patterns and lattice parameters resultant from our refinements in close agreement with prior studies by Cox \textit{et al.} \cite{cox2021clathrate, cox2018rapid}, see Fig. S4 and S5, respectively and Tab. S1, S2 for the structural data. The atomic coordinates, positions and displacement parameters were used from single-crystal Rietveld refinements from previous literature \cite{cox2021clathrate, cox2018rapid}.  The only impurity in the type-XI compound specimen was a minuscule amount of K$_{6.9}$Zn$_{21}$Sb$_{16}$, while the tunnel compound is virtually a phase-pure specimen. 

The resulting hot pressed ingots exhibit densities of $\sim95\%$ compared to the theoretical density obtained via XRD, see Tab. \ref{tab:dens_vs}. EDS mapping of the resulting ingots confirms the XRD results and identifies  minuscule Cu-rich precipitates as the main impurity of the type-I clathrate. The secondary phase might speculatively be an unidentified ternary compound from the space K-Zn-Cu-Sb. Furthermore, the chemical mapping confirms high phase purity of the tunnel compound and indicates  that for type-XI clathrate sample K$_{6.9}$Zn$_{21}$Sb$_{16}$ is the only impurity phase. The reader is referred to Supporting Information, Sect. 1 for EDS maps and their detailed description.

The synthetic approach for the type-XI and tunnel compounds differs from the prior literature, which focused on milling of hydride precursors and spark plasma sintering \cite{cox2021clathrate, cox2018rapid}. Instead, the synthesis was inspired by other reactive hot pressing efforts. For example, challenging carbides and borides have been synthesized through high energy milling and reactive hot pressing \cite{zhang2004reactive, chamberlain2009reactive} and as well as thermoelectric materials. Among thermoelectrics, La$_3$Te$_4$ \cite{may2008thermoelectric} and Mg$_2$Si \cite{alinejad2020activated} have previously been prepared by mechanical alloying followed by reactive hot pressing; in these cases, the milling alone takes the reaction nearly to completion. Given the incredible speed and simplicity of this synthesis route (see Sect. \ref{sect:exp-methods}, Experimental Techniques), we suggest that the presented approach might be used for exploring new chemical spaces, especially those containing alkali or alkali-earth metals that pose difficulties for traditional techniques and are likely to yield a bounty of novel, structurally intricate materials.

\begin{figure*}[t]
	\centering
	\includegraphics[width=0.75\textwidth]{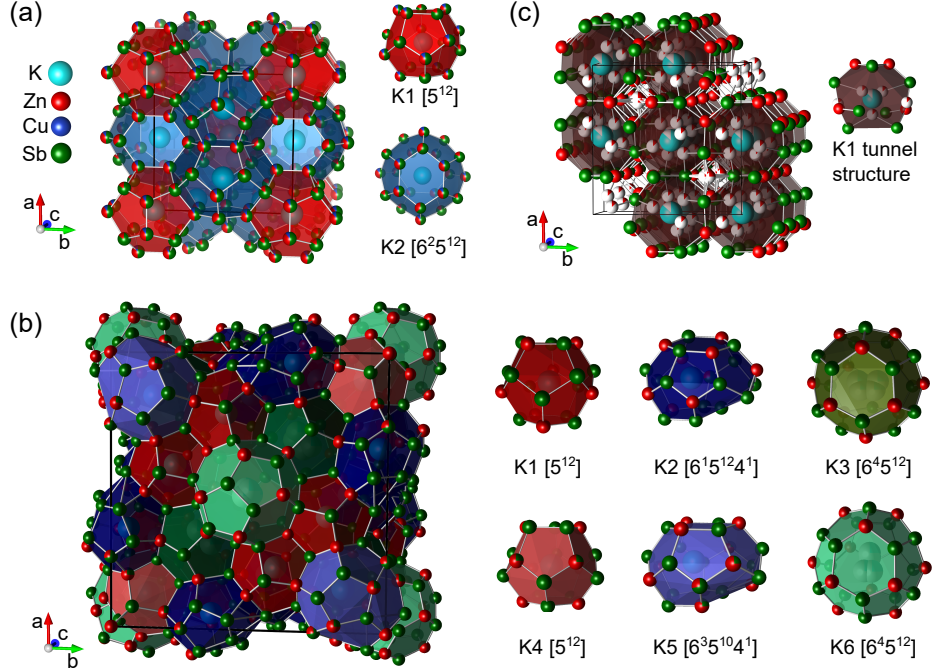}
	\caption{The focus of this work concerns the (a) K$_8$Zn$_{15.5}$Cu$_{2.5}$Sb$_{28}$, (b) K$_{58}$Zn$_{122}$Sb$_{207}$ and (c) K$_{6.9}$Zn$_{21}$Sb$_{16}$.  The red cages are small (122-157 \AA$^3$) while the blue (208-216 \AA$^3$) and green ($\sim$292 \AA$^3$) colors represent oversized cavities with respect to the filler - K atom. 
 See Tab. \ref{tab:RietveldDetails}, and Tab S1, S2 for atomic coordinates.}
	\label{Fig3} 
\end{figure*}

\subsection{Crystal Structures} 
\label{sect:struct}

The three crystal structures considered in this investigation are shown in Fig. \ref{Fig3}. All three involve a polyanionic `framework' constructed from Zn-Sb (and Cu, in the case of the type-I) surrounding K cations.  The two clathrate structures (I and XI) have isolated K whereas the tunnel structure forms chains of K. Given the rather subtle changes in stoichiometry between the two clathrates (I: K$_{0.15}$Zn$_{0.29}$Cu$_{0.05}$Sb$_{0.52}$ vs XI: K$_{0.15}$Zn$_{0.31}$Sb$_{0.53}$), the emergence of two distinct crystal structures is intriguing. In previous literature, changes from extraordinary geometries towards high symmetry of type-I clathrate structure with doping were observed \textit{e.g.} for Ba$_8$Cu$_{1-x}$Zn$_{x}$P$_{30-y}$ \cite{dolyniuk2017controlling} or in solid solutions  Cs$_{8}$Zn$_{18-x}$Cd$_x$Sb$_{28}$ \cite{owens2020chemically}.  

The type-I, type-XI, and the tunnel crystal structures have been discussed by Refs. \cite{dolyniuk2016clathrate} \cite{cox2021clathrate}, \cite{cox2018rapid}, respectively. 
Nevertheless, understanding the structural differences are crucial to this comparative investigation in terms of both electronic and phononic transport.  Beginning with the type-I clathrate (Fig. \ref{Fig3}a), the framework forms a 4-coordinated polyanionic  $sp^3$-hybridized network. Two different cages are found in the type-I: pentagonal decahedra [$5^{12}$] -- 20 vertices, and tetrakaidecahedra [$5^{12} 6^2$] -- 24 vertices.  The  type-XI clathrate  retains the 4-fold coordination for the  Zn, however, the  Sb atoms have 3, 4, or 5 nearest neighbours. The 3-coordinated Sb atoms are located exclusively on Sb3-16\textit{j} Wyckoff positions (see Tab. S1), with angles of 99.9$^{\circ}$ being close to those predicted for the $sp^3$ arrangement (109.5$^{\circ}$). Such bond angles suggest a lone pair extends from the 3-coordinate Sb atoms.  The 5-coordinated atoms are located only on Sb16-8\textit{j} site. One of the neighbors for Sb16-8\textit{j} species are Zn10 atoms with significant amount of vacancies (occupation of 21.9 \%, see Tab. S2), which along with angles between the remaining bonds in range 110.9-112.9$^{\circ}$ suggests that $sp^3$ hybridization  for Sb16-8\textit{j} is achieved when Zn10 site is vacant. The type-XI framework  forms four classes of cages: [$5^{12}$] -- 20 vertices, [$4^1 5^{12} 6^1$] -- 23 vertices, [$4^1 5^{10} 6^3$] -- 24 vertices, and [$5^{12} 6^4$] -- 28 vertices. In Fig. \ref{Fig3}, the cages associated with different K Wyckoff sites are colored uniquely.  

Finally, the tunnel structure contains a polyanionic network of Zn and Sb with strictly Zn-Sb bonds. Here, the Zn atoms have $sp^3$-like coordination. Likewise for the Sb1 atoms $sp^3$-like coordination is observed, while the remaining antimony on Sb2 position have 5 nearest neighbors; see Tab. S2 for atomic positions. This unusual coordination for the Sb2 is likely more complex locally, as one of the Zn sits in a 4-fold split position (Zn3). The potassium atoms are surrounded by ``open cages'', that can be also viewed as tunnels made from Zn and Sb, see Fig. \ref{Fig3}c. The  positions of K are 86\% occupied -- some of the potassium atoms are replaced with Zn triangular units (Zn6, Zn7 positions), similarly to the case of Sr$_2$Au$_6$Zn$_3$, where cations are replaced by Zn$_3$ triangles~\cite{gerke2013zn3}.

\subsubsection{Formal charge counting}
\label{sect:charge}

Charge counting in conventional clathrates typically views the polyanionic network as covalent with four electrons needed per atom. In the type-I clathrate (K$_8$Zn$_{15.5}$Cu$_{2.5}$Sb$_{28}$), the 46 atoms per formula unit in the framework are all 4-coordinate and  would thus satisfy their octets with 184 $e^-$. The framework nominally has 173.5 electrons from its constituent atoms and the potassium cations provide an additional 8 electrons. Thus, the polyanionic framework is electron deficient by only 2.5 $e^-$ per formula unit (\textit{i.e.}, 0.05 $e^-$ per framework atom).  In practice, subtle changes from nominal stoichiometry due to point defects will thus determine the actual free carrier concentration.  
Previous calculations of electron count for the type-XI clathrate \cite{cox2021clathrate} and the tunnel compound \cite{cox2018rapid} have found similar agreement.  For the type-XI clathrate, 0.3 electron excess per formula unit was obtained \cite{cox2021clathrate}, while the the tunnel compound is expected to have 1.1 electron deficiency per unit cell \cite{cox2018rapid}.

\subsubsection{Host network and bond distances to framework}
\label{sect:bonding}

To gain further insight into the framework bonding, we consider the Zn-Sb bond lengths of the compounds herein as well as other known Zn-Sb binary and ternary compounds.
The type-I clathrate has Zn-Sb bonds ranging from  2.66--2.73~\AA. Likewise, the bond distances of the tunnel compound were shown to range across  2.63--2.79 \AA~\cite{cox2018rapid}.
In contrast, the type-XI clathrate has bond lengths spanning in 2.61--2.96 \AA~\cite{cox2021clathrate}, which is a wider range than two other considered phases. This feature might suggest bigger variety of chemical bonding in K$_{58}$Zn$_{122}$Sb$_{207}$. 

\begin{table*}[t]
\caption{Atomic positions, occupancies and displacement parameters from Rietveld refinement for K$_8$Zn$_{15.5}$Cu$_{2.5}$Sb$_{28}$.}
  \label{tab:RietveldDetails}
  \begin{tabular}{ccccccc}
  	 \hline 
atom & Wyckoff site	&	\textit{x/a}	&	\textit{y/b}	&	\textit{z/c}		&	occupancy 	&	$U_{eq}$ [\AA$^2$]	\\ \hline 
	K1 & 2\textit{a}	    &	0	        &	0	            &	0	            	    &	1	        &	0.015(2)   	\\ 
	K2	    &6\textit{d}	    &	0.25	        &	0.5	            &	0	            &	1	        &	0.082(2)   	\\ 
Zn1	&\multirow{2}{*}{6\textit{c}}	&	\multirow{2}{*}{0.25}	&	\multirow{2}{*}{0}	    &	\multirow{2}{*}{0.5}	&		0.360(4)   &	\multirow{2}{*}{0.0116(4)}	\\ 
Sb1	&                            	&	                    	&	                    	&	                    	&		0.640(6)   &	                           	\\ 

	Zn2	&\multirow{2}{*}{16\textit{i}}	&	\multirow{2}{*}{0.1820(5)}	&	\multirow{2}{*}{0.1820(5)}	&	\multirow{2}{*}{0.1820(5)}	&	0.156(1) &\multirow{2}{*}{0.0137(2)}	\\ 
Sb2	&                            	&	                        	&	                        	&	                         	&		0.844(5) &	        \\ 

Zn3	&\multirow{3}{*}{24\textit{k}}	&	\multirow{3}{*}{0}      	&	\multirow{3}{*}{0.3138(8)}	&	\multirow{3}{*}{0.1149(8)}	&		0.400(2) &\multirow{3}{*}{0.0112(2)}\\ 
Cu1	&                            	&	                        	&	                        	&	                         	&		0.104(1) &	        \\ 
Sb3	&                            	&	                        	&	                        	&	                         	&		0.494(4) &	        \\ \hline 
  \end{tabular}
\end{table*}

In all three cases, these results are consistent with the prior literature showing Zn-Sb bonding range of ZnSb: 2.64-2.90 \AA; Zn$_4$Sb$_3$: 2.55-2.98~\AA; KZnSb: 2.62 \AA; NaZnSb: 2.77 \AA; Rb$_2$Zn$_5$Sb$_4$: 2.64-2.90~\AA~\cite{cox2018rapid}, Cs$_{8}$Zn$_{18}$Sb$_{28}$: 2.68-2.75~\AA~\cite{owens2020chemically}. This survey highlights that the longer bonds of the type-XI clathrate are not particularly anomalous. 

\subsubsection{Site Occupation}
\label{sect:atomic_occ}

Understanding of the breakdown in periodicity due to mixed and partially occupied sites is vital for proper description of the thermoelectric transport properties. For type-I clathrate, all framework sites exhibited significant alloying between the transition metals and antimony (see Tab. \ref{tab:RietveldDetails}). Due to similar atomic masses and sizes of Cu and Zn, it was not possible to clearly identify the Wyckoff site preferred by copper atoms; we assumed, that Cu is located on 24\textit{k} site, which had the greatest Zn content from the preliminary refinements. The composition resultant from Rietveld analysis is K$_8$Zn$_{14.01}$Cu$_{2.50}$Sb$_{29.5}$, which is close to the nominal one. This slight change in stoichiometry would yield an electron-rich framework by 2 $e^-$ per formula unit. 
Here, it is necessary to underline, that our treatment of atomic occupation stems from assumption allowing for antisite defects on the all three Zn/Cu/Sb Wyckoff sites and it represents only one of possibly many approaches to refinement of this crystal structure.  

In contrast to the strong site disorder of the type-I clathrate, the type-XI clathrate's site occupation was previously shown to be well-ordered \cite{cox2021clathrate}. Specifically, only 2 of 28 Wyckoff sites exhibit mixed Zn-Sb occupancy, and one Zn position incorporates some amount of vacancies  (Table S1). In the case of the tunnel compound, single crystal studies \cite{cox2018rapid} revealed that the potassium position suffers from vacancies (86\% occupancy), and open spots were replaced by Zn$_3$ triangular units (Zn6 and Zn7 positions in Tab. S2).

\subsubsection{Atomic Displacement Parameters}
\label{sect:ADPs}

Studying the atomic displacement parameters ($U_{iso}$) of the guest site in clathrates has a rich history of providing lattice dynamics insights \cite{dolyniuk2016clathrate, falmbigl2014mechanical, takabatake2014phonon}.  For the type-I clathrate, the framework $U_{iso}$ values are rather similar and small, not exceeding 0.0137(2)~\AA$^2$, see Tab. \ref{tab:RietveldDetails}. The potassium in small decahedron cage, with a volume of 156~\AA$^3$ has only  slight enlarged $U_{iso}$, 0.015(2)~\AA$^2$.  In contrast, the guest atom in large tetrakaidecahedra (vol. 216~\AA$^3$)  has a $U_{iso}$ of  0.082(2) \AA$^2$. These results are consistent with previous observations of rattling in type-I clathrates  \cite{qiu2004structure}.  For the type-XI clathrate, authors of Ref. \cite{cox2021clathrate} obtained similar results (Table  S1), with K in the smaller  pentagonal decahedra (vol. $\approx$ 157~\AA$^3$) exhibiting $U_{iso}$ similar to the Zn-Sb lattice. The bigger 23- and 24-vertex polyhedra (vol. 209~\AA$^3$ and  217~\AA$^3$) enabled K2 and K5 atoms to attain atomic displacement parameters of 0.0227 \AA$^2$ and 0.027~\AA$^2$, respectively, which is roughly doubled average $U_{iso}$ in the covalent framework. The biggest hexakaidecahedra  (vol. 291-292 \AA$^3$) utilized 4-fold and 2-fold split positions for the  K3 and K6 atoms, respectively, within the cavity and still exhibiting significant rattling: $U_{iso}$ = 0.083, 0.044 \AA$^2$, respectively. Here, it is necessary to underline that our $U_{iso}$ values for type-I clathrate were obtained at 300 K, while data for type-XI compounds was gathered at 100 K \cite{cox2018rapid}. The above implies, that type-XI should exbibit even bigger $U_{iso}$ at 300 K, showing stronger rattling that its type-I counterpart.

In contrast to the clathrates, the previous work on the tunnel compound found no significant enhancement in potassium atomic displacement parameters compared to the Zn-Sb lattice \cite{cox2018rapid}.  As the volume of the tunnel is somewhat ill-defined, we instead consider K-framework distances.  For the tunnel, the distances from K atom to Zn-Sb framework is in range 3.50--3.66~\AA, which is smaller than even for the smallest cage in the type-I or type-XI clathrates (3.80~\AA).  Further contrast is provided by the largest cages (K3, K6) in the type-XI clathrate, where distances of 4.56--4.59~\AA~are found. Thus, it is not surprising that the tunnel compound exhibits no enhancement in ADP, the environment surrounding the potassium is far more constrained. 
 
\begin{table*}[t]
  \caption{ Sound velocity (longitudinal $v_L$, transverse $v_T$, average $v_s$), and density for  K$_8$Zn$_{15.5}$Cu$_{2.5}$Sb$_{28}$, K$_{58}$Zn$_{122}$Sb$_{207}$, and K$_{6.9}$Zn$_{21}$Sb$_{16}$}
  \label{tab:dens_vs}
  \begin{tabular}{cccccc}
  \hline
& $v_L$ (m/s) & $v_T$  (m/s) &  $v_{s}$ (m/s)  & $d$ (g cm$^{-3}$) & $d_{rel}$ (\%) \\
 \hline   
K$_8$Zn$_{15.5}$Cu$_{2.5}$Sb$_{28}$      &  3418       &    1943     &    2159       & 5.00 & 94.2 \\
K$_{58}$Zn$_{122}$Sb$_{207}$             &  3505       &    2028     &    2251       & 4.93 & 95.2 \\
K$_{6.9}$Zn$_{21}$Sb$_{16}$              &  3402       &    1963     &    2179       & 5.08 & 95.3 \\
      \hline
  \end{tabular}
  \end{table*}
 
In summary, diffraction analysis indicates these three K-Zn-Sb compounds all exhibit significant structural and bonding complexity. Considering the framework, all contain predominantly sp$^3$ polyanionic Zn-Sb networks. The type-I and tunnel compounds exhibit the most framework occupancy disorder. However, the type-XI framework still has some Zn-Sb alloying as well as  3-coordinate Sb. Considering the potassium, the cation resides in quite different sites as judged by both cage sizes and $U_{iso}$ value, ranging from fully constrained by the lattice (tunnel) to extremely under-constrained (type-XI). As such, comparing transport in these three compounds requires consideration of both the framework and cation contributions.

\subsection{Sound velocity}

The low frequency speed of sound provides an experimental avenue to interrogate bond strengths.  While the low frequency modes are collective oscillations that involve all atoms, we might expect the sub-components of the lattice that are weakly coupled to the framework to contribute less to the velocity (akin to soft and stiff springs in parallel).  As such, the behavior of sound velocity is expected to primarily reflect the Zn-Sb bonding.  Results of the measurements are shown in Tab.~\ref{tab:dens_vs}. Mean values were calculated from separate longitudinal ($v_L$) and transverse ($v_T$) measurements using $v_s = 3^{1/3}(v_L^{-3} + 2v_T^{-3})^{-1/3}$. The longitudinal and shear  elastic moduli are defined as, $M = v_L^2 d$, and $G = v_T^2 d$, respectively. The mean bulk moduli is then obtained as $B = M - \frac{4G}{3}$. Values of \textit{G, M,} and \textit{B} are gathered in Tab. S6. Performed calculations shown, that the moduli do not vary by more than 7\% between these compounds. This is close to the error of the speed of sound measurement ($\sim5\%$) and  confirms similarity of Zn-Sb bonding patterns on Zn-Sb sublattice in the there studied compounds. 

To compare these results to other thermoelectric compounds, Fig. S9 considers the  squared mean speed of sound  as a function of $\frac{1}{d}$. With the approximation  that $v_s^2\propto \frac{B}{d}$, the studied K-Zn-Sb clathrates and tunnel compound exhibit smaller elastic modulus ($B$) than most other clathrates and indeed most other thermoelectric materials. This softness likely highlights the more ionic character of the Zn-Sb framework compared to clathrates based on group III-IV framework atoms. These results also highlight how much softer the clathrates generally are compared to the half-Heusler and skutterudite intermetallic compounds. The reader is referred to the Supporting Information, Sect. 2 for further insight into elastic properties and Gruneisen parameter that are  approximated from longitudinal and transverse components of the speed of sound.

 \begin{figure}[t]
	\centering
	\includegraphics[width=8.6cm]{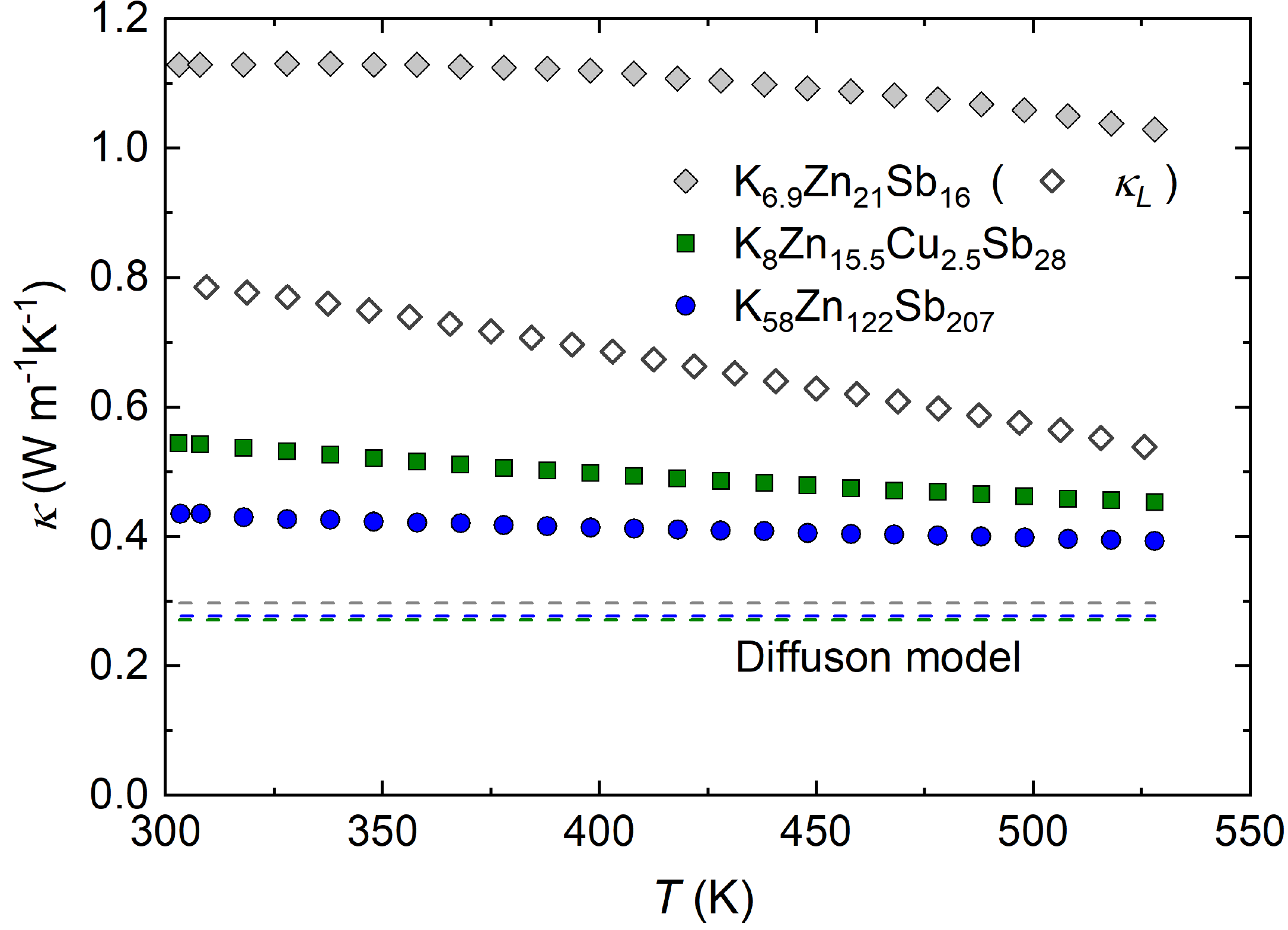}
	\caption{
	The total thermal conductivity of K$_{58}$Zn$_{122}$Sb$_{207}$ and  K$_{8}$Zn$_{15.5}$Cu$_{2.5}$Sb$_{28}$ is extraordinarily low; as these compounds are resistive, $\kappa\sim\kappa_L$.   For the tunnel compound, K$_{6.9}$Zn$_{21}$Sb$_{16}$, subtracting the electronic contribution yields the  lattice thermal conductivity (empty diamonds).  Dashed lines denote the bottom limit of thermal conductivity obtained with diffuson model.}
	\label{Fig4} 
\end{figure}

\subsection{Thermal conductivity}

The temperature dependence of the thermal conductivity is shown in Fig. \ref{Fig4}. 
The two previously known compounds were studied before only at low temperatures; at 300\,K they  exhibited thermal conductivity of 1.3 Wm$^{-1}$K$^{-1}$ for the tunnel compound \cite{cox2018rapid} and 0.36 Wm$^{-1}$K$^{-1}$ for the type-XI clathrate \cite{cox2021clathrate}, similar to the results herein.  The electronic contribution to thermal conductivity for both clathrates is negligible ($<$0.1 \%) due to their small electrical conductivity, so for transparency, on Fig. \ref{Fig4} we show only the original data for these materials.  For the tunnel compound, we calculated the lattice thermal conductivity ($\kappa_L$) by subtracting the electronic contribution ($\kappa_e$) from total thermal conductivity. The values of $\kappa_e$ were obtained using the Wiedemann-Franz law, where the Lorenz number was acquired according to Ref. \cite{kim2015characterization} (electronic transport is discussed below).  All three compounds show extraordinarily suppressed thermal transport at high temperature (0.4 - 0.6 Wm$^{-1}$K$^{-1}$ at 500\,K). These are among the lowest thermal conductivities observed among clathrates, and  among any inorganic crystalline solids \cite{qian2021phonon}. The temperature-independence of thermal conductivity for the studied here compounds is consistent with other clathrate materials, see \textit{e.g.} Refs. \cite{toberer2008high, candolfi2012high}. In contrast, the tunnel structure shows a decay indicative of phonon-phonon scattering. 

  \begin{figure}
\centering
	\includegraphics[width=8.6cm]{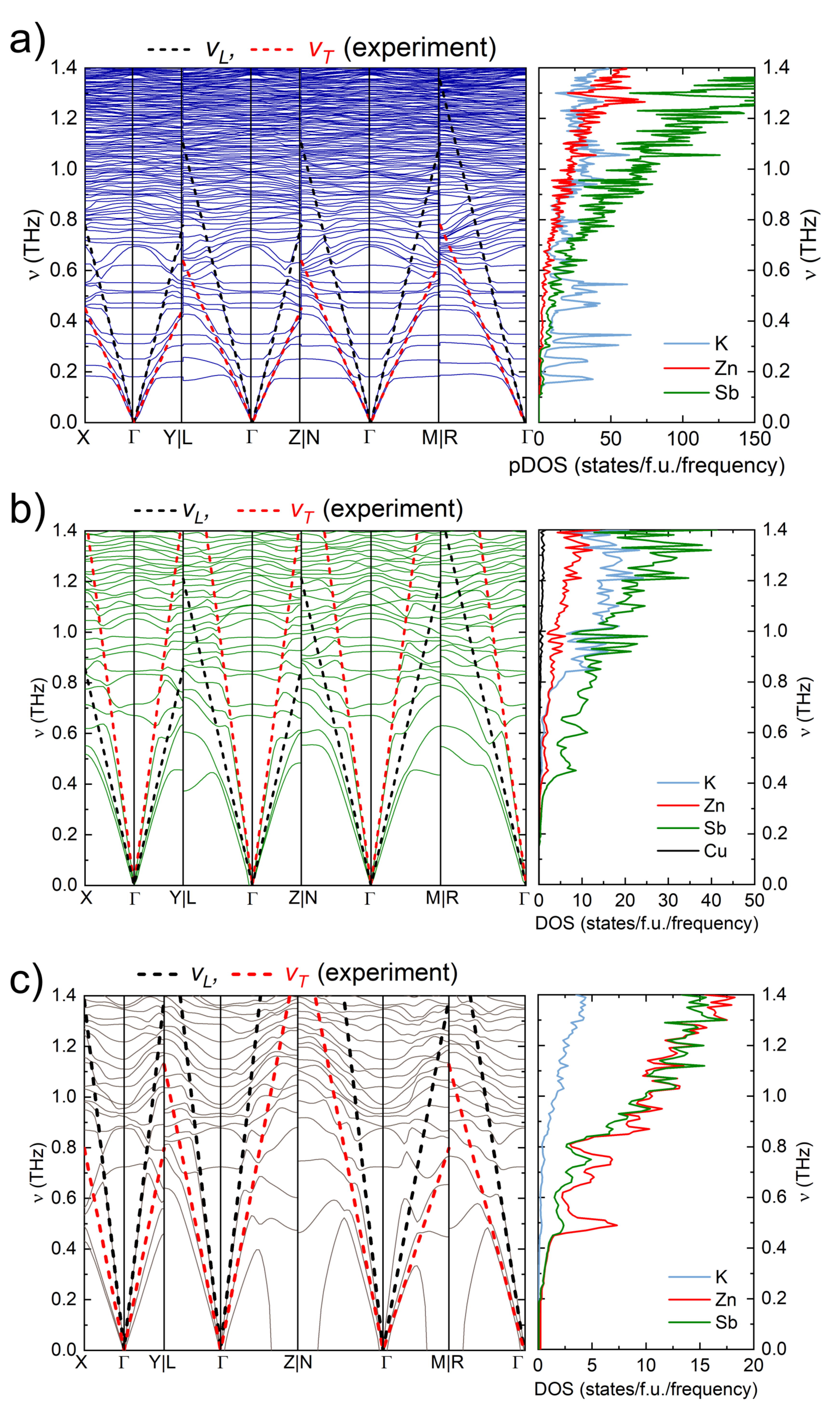}
	\caption{Phonon dispersion and atom-projected phonon density of states at low frequencies for (a) K$_{58}$Zn$_{122}$Sb$_{207}$, (b) K$_{8}$Zn$_{15.5}$Cu$_{1.5}$Sb$_{28}$, (c) K$_{6.9}$Zn$_{21}$Sb$_{16}$. K$_{58}$Zn$_{122}$Sb$_{207}$ displays a unique array of low lying optical modes in frequencies 0.17-0.70 THz. Dashed lines denote approximate acoustic modes calculated with experimental sound velocity and structural data from Rietveld refinement. The reader is referred to Fig. S10 for the graphs across the full frequency spectrum. }
	\label{Fig5}
\end{figure}
 
\subsection{Phonon dispersion}

To understand the sources of unusually low thermal conductivity in these materials, we can complement the above structural comparison with calculations of the vibrational properties. The combination of structural complexity and disorder renders accurate calculations of the phonon band structure challenging. To partially reduce the complexity, the calculations consider unit cells where the occupational disorder regarding Wyckoff positions with point defects was largely eliminated. Exemplary modification is reduction of antisite disorder in type-XI clathrate, where position 16j with nominal content Sba1 = 36.4\%, Zna1 = 63.6\% (see Tab. S1) was ascribed as fully occupied by zinc, while site 16j with nominal occupation Sba2 = 62.7\%, Zna2 = 37.3\% was assumed to be fully populated with Sb. The modifications did not alter significantly overall stoichiometry of any compound. The reader is referred to see Supporting Information, Sect. 3 for details of the procedure.

\begin{figure}[t]
	\centering
	\includegraphics[width=8.6cm]{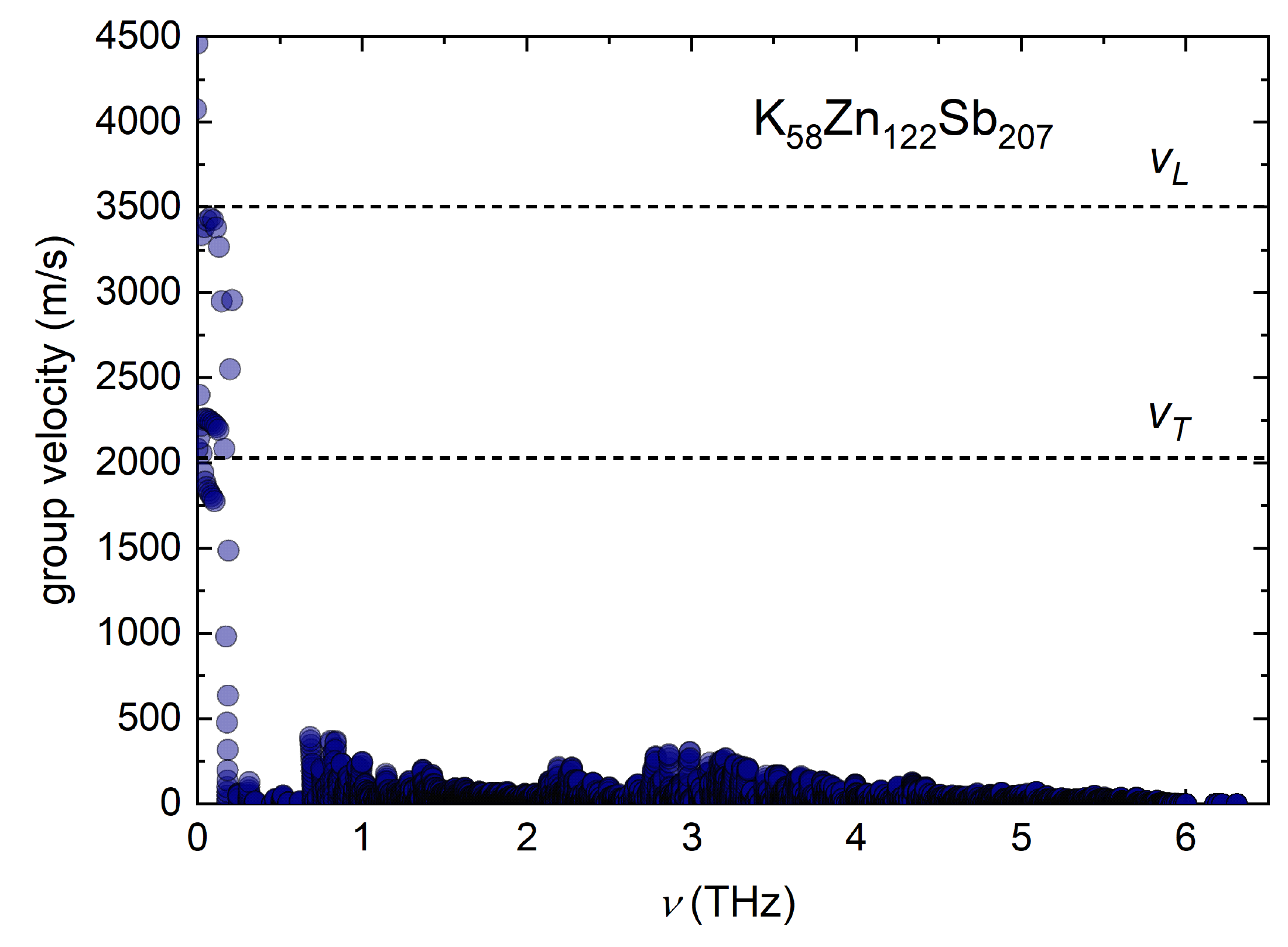}
	\caption{The group velocity obtained from phonon dispersion calculations along $\Gamma -$L for the type-XI clathrate shows good convergence of acoustic modes with the experimental values (dashed lines). For all optical phonons  above \textit{c.a.} 0.7 THz the group velocity is extremely low.}
	\label{Fig6} 
\end{figure}

Fig. \ref{Fig5} shows the lower frequency bands for the three compounds and the atom-projected phonon density of states (pDOS) up to 1.4 THz. Complete dispersions are shown in the Supporting Information, Fig. S10;  in all three cases, the significant structural complexity leads to a plethora of optical branches.  Within a phonon quasi-particle picture, these branches have low velocity and ample opportunities for scattering between states.   Fig. \ref{Fig6} shows the average group velocity of the type-XI clathrate with increasing frequency. Values obtained from theoretical calculations are in good agreement with experiment, considering both longitudinal and transverse acoustic phonons, marked as dashed lines on the figure. The group velocity significantly decreases at frequencies larger than \textit{ca.} 0.7 THz, which confirms diffusive character of optical modes in type-XI compound. For the group velocity for type-I clathrate and the tunnel compound, the reader is referred to Fig. S11.

Overlaid on the phonon dispersions, Fig. \ref{Fig5}, are the experimental speed of sound results using a Debye (\textit{i.e.}, linear) approximation.  The agreement with the slope of the acoustic branch and subsequent avoided crossings therein is excellent in all three cases.  For the tunnel structure, even after simplification of the unit cell, one phonon mode starting at $\Gamma$ point spans towards non-negligible imaginary frequencies, which makes comparison of this particular mode with experimental sound velocity difficult. The main contribution to these vibrations stems from Zn1 atoms located nominally on quarterly occupied position 8$g$ (see Tab. S2). Zn1 can be viewed as occupying split site, \textit{ie}. it is gently displaced from the high-symmetry location. In the phonon calculations we assumed for this atom fully occupied Wyckoff site 2\textit{a} to maintain stoichiometry and, afterwards, we relaxed its coordination. Correlation of this negative mode with atom occupying extraordinary position in the crystal structure allows for tentative interpretation of the negative phonon mode as a virtually unavoidable technical difficulty, rather than intrinsic instability of the crystal structure.

In materials with complex crystal structures, the acoustic branches typically carry a significant fraction of the heat due to their high group velocities and fewer scattering channels. As such, altering the acoustic branches remains a crucial focus for achieving ultra-low thermal conductivity. Next, we will discuss the most important optical branches. The flat bands that cross the acoustic branches can broadly be categorized into two cases. In the first case, the flat bands that intersect the upper frequencies of the acoustic branches (approx. 1-2 THz) are simply a product of framework vibrations that involve the heaviest atoms (\textit{i.e.}, Sb). These framework modes can be viewed as extensions of the acoustic branches and as such typically do not extend below the maximum of the transverse acoustic modes.  In the second case, we have low-lying optical branches arising from soft springs rather than heavy masses. These modes correspond to under-constrained K atoms in large cages  (\textit{i.e.}, rattling) and are not limited by the acoustic branches in their frequency.  

Considering Fig. \ref{Fig5}, all three compounds exhibit optical modes above $\sim$0.5 THz associated with  Zn-Sb framework motion. The pDOS shows a strong contribution from Sb in these modes, consistent with the heavy mass of Sb. However, the impact of K on the dispersion varies significantly.  In brief, only the type-XI clathrate exhibits rattling modes below the framework modes. Further, these rattling modes form a surprisingly  dense `ladder' starting at frequency of 0.17 THz.  
These modes are among the lowest-lying known for either classical \cite{lee2007neutron, takabatake2014phonon, koza2010vibrational, stefanoski2014inorganic}, or unconventional clathrates \cite{wang2020clathrate}, whose usually show the first flat phonon modes at \textit{ca}. 0.5-0.8 THz.  The observation is comparable results of calculations for type-II clathrate with the heaviest network available from clathrate-forming tetrel elements (Sn) and the heaviest filler atom among alkaline metals (Cs) Cs$_{24}$Sn$_{136}$. The authors suggest strongly anharmonic nature of low-lying vibrations calculated for this material \cite{myles2003rattling}. The experimentally known type-I counterpart of similar composition, Cs$_8$Sn$_{44}$, shows the lowest optical modes in its phonon dispersion at \textit{ca.} 0.5 THz \cite{tse2001phonon}.

To understand the source of these ultra-low frequency rattling modes in the type-XI clathrate, we consider the energy of the eigenvectors in detail. The time averaged total energy, $\left< E_{tot} \right>$, is the sum of the time averaged kinetic and potential energies, which are equally split ($\left<E_{tot} \right> = \left< KE \right> + \left< PE \right> = 2 \left< KE \right>$). As such, the time averaged kinetic energy of some atom, $i$, can thus be written as
$E_i\propto m_i |v_i|^2$ for some atom $i$ with mass $m_i$ and velocity $v_i$. The square of magnitude of the velocity can be written simply as $|v_i|^2 \propto |\vec{e}_i|^2$, where $|\vec{e}_i|$ is the magnitude of the eigenvector. This leaves the total energy of the system as $E_{tot} \propto \sum_i m_i |\vec{e}_i|^2$. Finally, the fraction of energy ($\xi$) contributed to $E_{tot}$ by all atoms with type $t$ can be computed as:

\begin{equation}
    \xi = \frac{\frac{1}{N_t}\sum_{i=1}^{N_t}m_i |\vec{e}_i|^2}{\sum_t \sum_{i=1}^{N_t} m_i|\vec{e}_i|^2}
\end{equation}

\noindent where and $N_t$ is the number of atoms with type $t$ in the system. For transparency, the type of atoms is distinguishing between each Wyckoff position of the potassium atoms (K1-K6, see Sect. \ref{sect:struct}), while for the Zn-Sb sublattice we gather Wyckoff positions of zinc and antimony atoms under the summary categories ``Zn'' and ``Sb'', respectively. The reader is referred to Fig. S12 for graphical presentation of the calculations. 

The procedure revealed that the lowest-lying optical modes correspond to almost exclusively atoms K3 (band numbers 1-15), while the second lowest (band numbers 15-30) are from K6. K3 and K6 atoms are located in the biggest cages (hexakaidecahedra, vol. 291-292~\AA$^{3}$) of the type-XI structure. Atoms K6 are characterized by the largest atomic displacement parameters ($U_{iso}$ = 0.083~\AA$^2$) in the type-XI structure, while K3 have the second largest $U_{iso}$ (0.044~\AA$^2$), see ADPs, Sect. \ref{sect:ADPs}. This near-perfect isolation of the flat optical modes for particular atoms in K$_{58}$Zn$_{122}$Sb$_{207}$ is in contrast with conventional clathrates, for which the rattling modes tend to be more strongly hybridized with the cage network \cite{euchner2012phononic, norouzzadeh2017phonon}. Such hybridization was shown to result in wide-frequency scattering of rattlers in conventional clathrates \cite{tadano2015impact}, while for type-XI compound studied here, we might expect that the resonant scattering scenario can be more relevant, which will be addressed during thermal conductivity modeling.

Going higher up in band index in the type-XI clathrate, one finds contribution of K2 and K5 atoms, which are located in middle-volume cages of type-XI clathrate, while the highest in frequency from all potassium species are K1 and K4 atoms -- fillers of the smallest [$5^{12}$] cavities. The positions of potassium modes from theoretical calculations nicely correlates with atomic displacement parameters from Rietveld refinement. K3 and K6 are characterized by the highest $U_{iso}$ values, while K2 and K5 exhibit moderate displacements. Atoms K1 and K4, in turn, exhibit the smallest ADPs among fillers of type-XI structure (see Tab. S1).  For the optical branches, the atomic contributions generally skew towards Zn and Sb, with the Zn being more prominent at high frequencies due to its smaller mass with respect to Sb. The atomic contribution to optical band  remains almost invariant through the inverse space path in the unit cell, see animations describing a path through inverse space in Fig. S12.

\subsection{Heat Capacity}

\begin{figure}
	\centering
	\includegraphics[width=8.6cm]{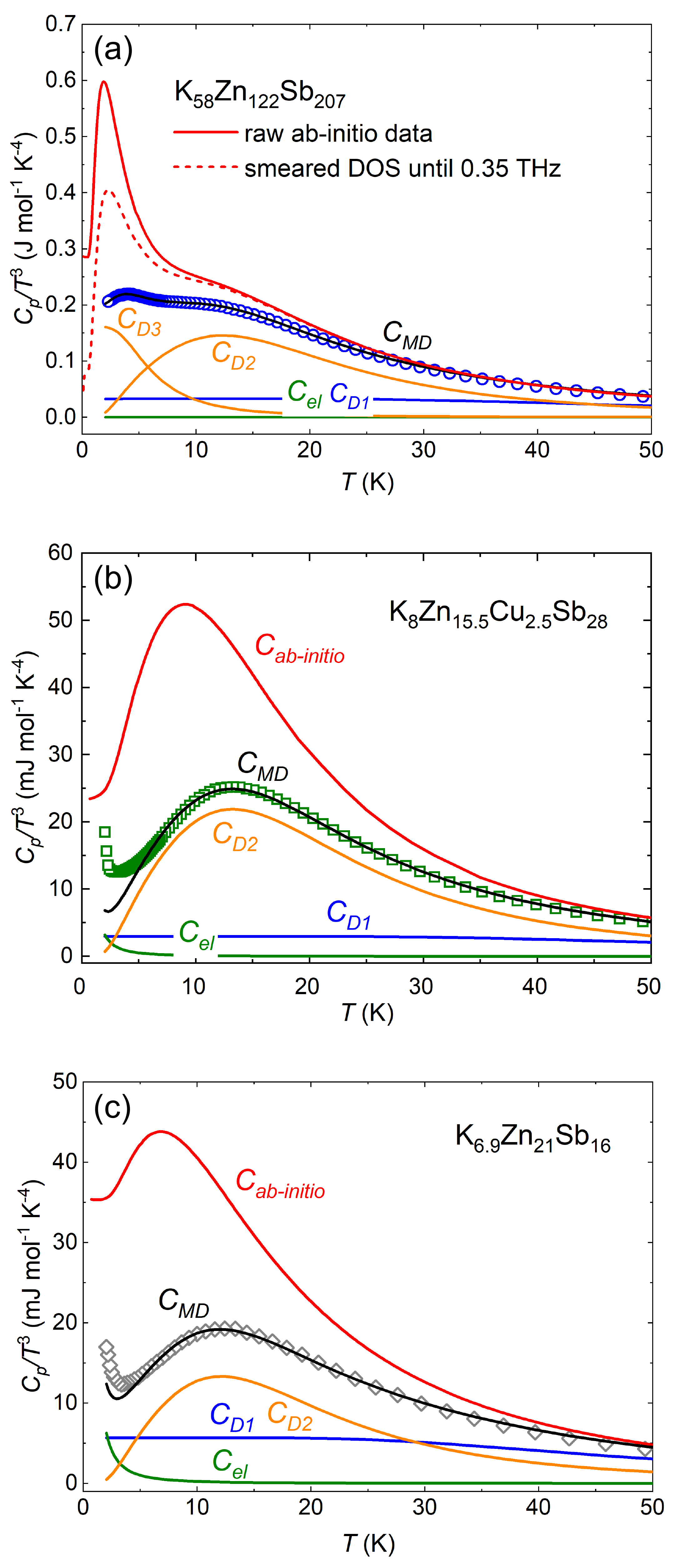}
	\caption{Low temperature heat capacity as a function of temperature for (a) type-XI clathrate (b) type-I clathrate, (c) tunnel compound. The experimental data is displayed with symbols while the heat capacity predicted from full \textit{ab-initio} calculation ($C_{ab-initio}$) is shown in red.  The analytic model $C_{MD}$ is obtained by summing the individual contributions. 
	 }
	\label{Fig7} 
\end{figure}

Low temperature heat capacity ($C_p$)  measurements of the three compounds (Fig. S13) are consistent with the soft lattice dynamics discussed above.  Focusing on the 2-50\,K regime, the experimental heat capacity can be used to validate the low-frequency theoretical phonon dispersion calculations.  Fig. \ref{Fig7} shows $C_p/T^3~vs.~T$ for the three materials; in a Debye model this would yield a flat region followed at higher temperatures by a gradual decay. Instead, we see that the experimental data for these materials produces broad humps at temperatures ranging from 4-15\,K.  

Maxima in this representation of heat capacity  denote deviations from the Debye model, specifically with respect to the density of states ($g_{D}(E)\propto\nu^2$).  In the simplest form, an Einstein model can be used to add a delta function to the Debye density of states.  While such a simple model does not quantitatively agree with the measured magnitudes, it can be used to estimate the primary vibrational frequency causing the observed humps. The reader is referred to Supporting Information, Sect. 4 and Fig. S13 for the basic Debye-Einstein analysis.  For the type-XI clathrate, we find two Einstein frequencies: 0.42 and 1.52 THz.  These frequencies are obtained by solving for the Einstein temperatures $\Theta_E$, with the conversion $\Theta_E \ \approx 5T_{max}$ \cite{bensen2021anomalous} and $\nu_{E1} = k_B \Theta_{E1}/ h$. The lower frequency corresponds  to the `ladder' in phonon dispersion created by K rattling modes in frequency range 0.17-0.70 THz (\textit{cf.}, Fig. \ref{Fig5}), while the upper frequency corresponds to a hump in phonon density of states due to Zn-Sb vibrations. The bottom Einstein temperature for type-XI compound ($\Theta_{E1} \approx$ 20 K) is among the lowest observed for any clathrate  \cite{dolyniuk2016clathrate, falmbigl2014mechanical, takabatake2014phonon}.

In contrast, type-I clathrate and tunnel compound exhibit only a single hump in the $C_p/T^3~vs.~T$ curve, with characteristic frequencies of \textit{ca.} 1.4 and 1.3 THz, respectively.  This simple model is qualitatively consistent with the calculated phonon dispersions:  all three compounds show a high concentration of optical branches near 1.5 THz corresponding to Zn-Sb modes, while only the type-XI exhibits K rattling modes below 0.7 THz.  

To develop a more quantitative analytic model, we begin by determining the electronic contribution to the heat capacity using a Sommerfeld fit of  the 2-4\,K regime (see Supporting Information, Sect. 4 for the details).  The impact of the electronic heat capacity is shown as the green curve in Fig. \ref{Fig7}.
After subtraction of the electronic the Debye parts to $C_p$, the remaining phononic part, is approximated by the perturbation as additional Debye-like branches with an  offset in the frequency space. 
Such a density of states is achieved by tilting an Einstein branch such that it has a non-zero group velocity:  $\nu(\vec{k}) = s|\vec{k}|+\nu_o$, where $s$ denotes the slope of the branch and $\nu_o$ corresponds to the frequency offset (see Fig. S14).   
 
Then, the density of states $g(\nu)$ for each polarization will be obtained according to classical formula as

\[   
g(\nu) = 
     \begin{cases}
      \frac{d}{d\nu} \left(\frac{k(\nu)^3}{6\pi^2 }  \right) =  \frac{(\nu-\nu_o)^2}{2\pi^2s^3} &  \nu_0\leq\nu\leq\nu_f  \\
      \text{0} &\mathrm{otherwise}\\
  
     \end{cases}
\]

\noindent Eventually, the heat capacity of new model with added Debye-like modes ($C_{D2}$ and $C_{D3}$) is computed using the  harmonic oscillator approach and integrating  between the limits of the optical mode:

\begin{equation}
    C_{D2,D3} = m \frac{d}{dT}\int_{\nu_o}^{\nu_f} g(\nu) h \nu f_{BE}  d\nu,
\end{equation}

\noindent with $f_{BE}$ being the Bose-Einstein distribution, and $m$ standing for the unitless weight coefficient of the studied optical modes with respect to total heat capacity. As such, the free parameters of these Debye-like contributions are $m$, $\nu_o$ and $\nu_f$ for each additional `branch'. 

\begin{figure}[t]
	\centering
	\includegraphics[width=8.6cm]{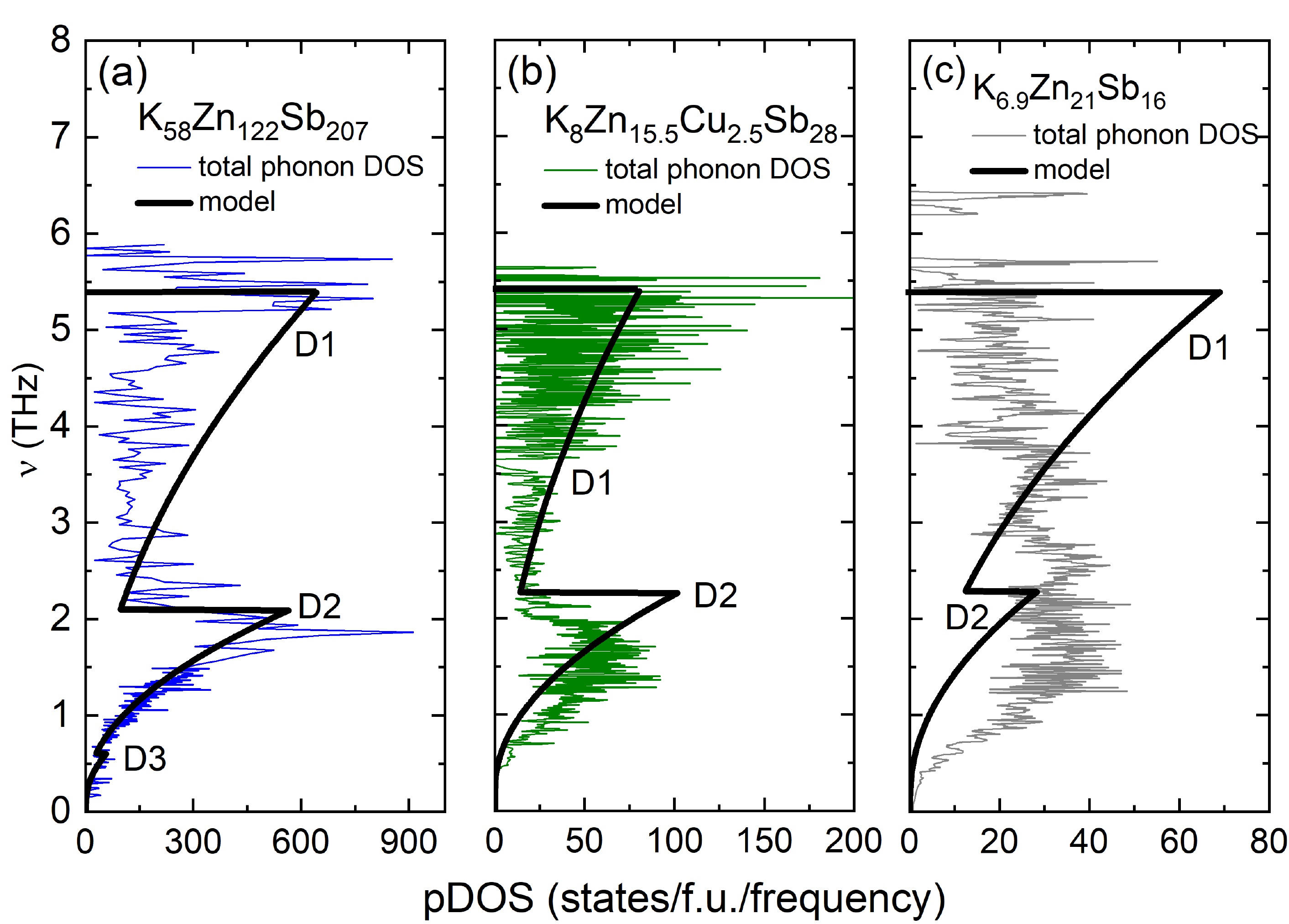}
	\caption{Overlay of total phonon density of states from \textit{ab-initio} calculations and its approximation obtained from analysis of experimental heat capacity. The overall agreement of the two types of data justifies the  model used for analysis of the experimental heat capacity.}
	\label{Fig8} 
\end{figure}

The phonon density of states used in our modeling is shown in Fig. \ref{Fig8} overlayed on the total \textit{ab-initio} phonon DOS. The main Debye model is denoted at D1, while additional Debye-like contributions are marked as D2 and D3. The simple analytical approach to the experiment is surprisingly well able to reproduce the complex phonon dispersion obtained from theoretical calculations.
As shown in Fig. \ref{Fig7}, the new analytic formalism is also able to reproduce the temperature variation of $C_p/T^3~vs.~T$ of all the studied compounds (Fig. \ref{Fig7}) significantly better than the original Einstein model (see Fig. S13c, f, i). The sum of electronic contribution $C_{el}$, main Debye model $C_{D1}$ and additional Debye-like modes $C_{D2,D3}$ is denoted as $C_{MD}$ = $C_{el}$ + $\sum\limits_{i} C_{Di}$. The parameters resultant from the analysis are gathered in Tab. S7. The above analytic model agrees well both with the experimental results and the calculated phonon dispersion.

As a final analysis, we seek to directly compare the phonon dispersion calculations with the experimental heat capacity. However, the negative frequency modes in the calculations must be addressed if physically reasonable heat capacity values are to be obtained. At the lowest frequencies (up to 0.15 THz), we substituted the numerical density of states with a Debye approximation ($g$ $\propto \nu^2$), so that the values of parabola match the calculated DOS at 0.15 THz. In doing so, we avoid interferring with the lowest-lying rattling modes. 

The theoretical $C_p$ was calculated accordingly to the procedure described in Ref. \cite{bensen2021anomalous}. In 2-300 K range the so-obtained theoretical and experimental $C_p$'s are in good agreement for all three compounds (Fig. S13a, d, g). When the data is replotted to $C_p/T^3~vs.~T$ for type-XI clathrate (Fig. \ref{Fig7}a), one can notice, that the theoretical heat capacity shows qualitatively similar humps to those observed in the experiment. 
Temperatures of the maxima are in decent agreement, yet the magnitude of theoretical $C_p$ is overestimated. Such feature is not particularly rare in theoretical calculations, see \textit{eg.} Refs. \cite{bedoya2016influence, paudel2009calculated}. Its most likely origin is the fact, that for \textit{ab-initio} calculations we use perfectly ordered structure. In real material with 0D, 1D and 2D defects, the density of states might be less sharply defined.  In type-XI clathrate we removed the atoms from partially occupied site Zn10 (occ. 21.9\%), as well as we simplified the positions suffering from Sb-Zn vacancies (see Supporting Information, Sect. 3). Differing cage environment for potassium atoms in real material (\textit{i.e.} broken periodicity) can easily influence frequency of the rattling, which might in turn is likely to lead to smearing of the rattling peaks in phonon density of states.

To explore the impact of structural disorder  on the heat capacity, we simulated smearing of the DOS peaks of type-XI clathrate below an arbitrary limit of 0.35 THz. The artificial DOS was produced as a sum of parabolic dispersion and Gaussian maxima located at frequencies predicted by the real phonon dispersion (see Fig. S15a). To correspond to the \textit{ab-initio} calculations, we maintained the integral below artificial DOS similar to the \textit{ab-initio} one (\textit{cf.} Fig. S15b). With the smeared DOS change in theoretical $C_p$ is as expected - the theoretical $C_p/T^3~vs.~T$ curve is lower, closer to the experimental one (see red dashed curve in Fig. \ref{Fig7}a). Overestimation of the theoretical heat capacity with respect to the experiment was also observed for the other studied K-Zn-Sb materials (see Fig. \ref{Fig7}). We hypothesize the physical source of smearing of the density of states at low frequency arises due to  chemical variation (\textit{i.e.}, vacancies and Zn/Sb disorder) in the framework surrounding each rattling K. 

In summary, the experimental heat capacity results are broadly consistent with the \textit{ab-initio} phonon calculations. The humps in the heat capacity at 10-20 K in all the three materials results from vibrations of Zn-Sb sublattice. Only type-XI clathrate shows in additional maximum at 4 K its heat capacity, which is correlated with its unique array of potassium rattling modes.   Evidence of smearing of the density of states is consistent with a greater degree of disorder than was utilized in the phonon calculations.

\subsection{Thermal transport model}

Equipped with both experimental  and \textit{ab-initio} results, we can now attempt to disentangle the sources of the low thermal conductivity in these K-Zn-Sb materials.  Our general approach is to subtract the contribution from the optical modes and then focus on modeling the acoustic contribution.

The optical branches in complex materials are often modeled as either phononic with a minimum phonon lifetime \cite{erhart2015microscopic} or as diffusionic \cite{agne2018minimum}; in both cases the intention is to highlight that the combination of low group velocity and strong scattering severely limits the ability of any single optical branch to conduct heat. Diffusons are non-propagating, non-localized quasi-particles, with movement that can be described by the theory of random walk \cite{agne2018minimum}. In materials studied here, the energy difference between bands is approximately 10\,$\mu$eV, which leads significant band overlap and suggests a breakdown of the phonon view of heat transport.  The thermal conductivity predicted for the studied compounds by the diffuson model is given by:

\begin{equation}
    \kappa_{diff} = 0.76n_{at}^{2/3}k_B\frac{1}{3} \left(2v_T+v_L\right)
\end{equation}

\noindent where $n_{at}$ denotes the number density of atoms.
As we seek to model just the diffusionic contribution of the optical branches, the magnitude is reduced by  1-3\% ($\frac{n-1-r}{n}$, where $n$ is the number of atoms in the primitive unit cell and $r$ is the number of rattling atoms).  
The result of this calculation is shown for each compound  at the bottom of Fig. \ref{Fig4}. 
We find that the diffusonic contribution to the lattice thermal conductivity ranges from 30-70\%. 
Here, the sheer complexity of the unit cells drives this large diffusonic contribution rather than any specific structural features.  

\begin{figure}
	\centering
	\includegraphics[width=8.6cm]{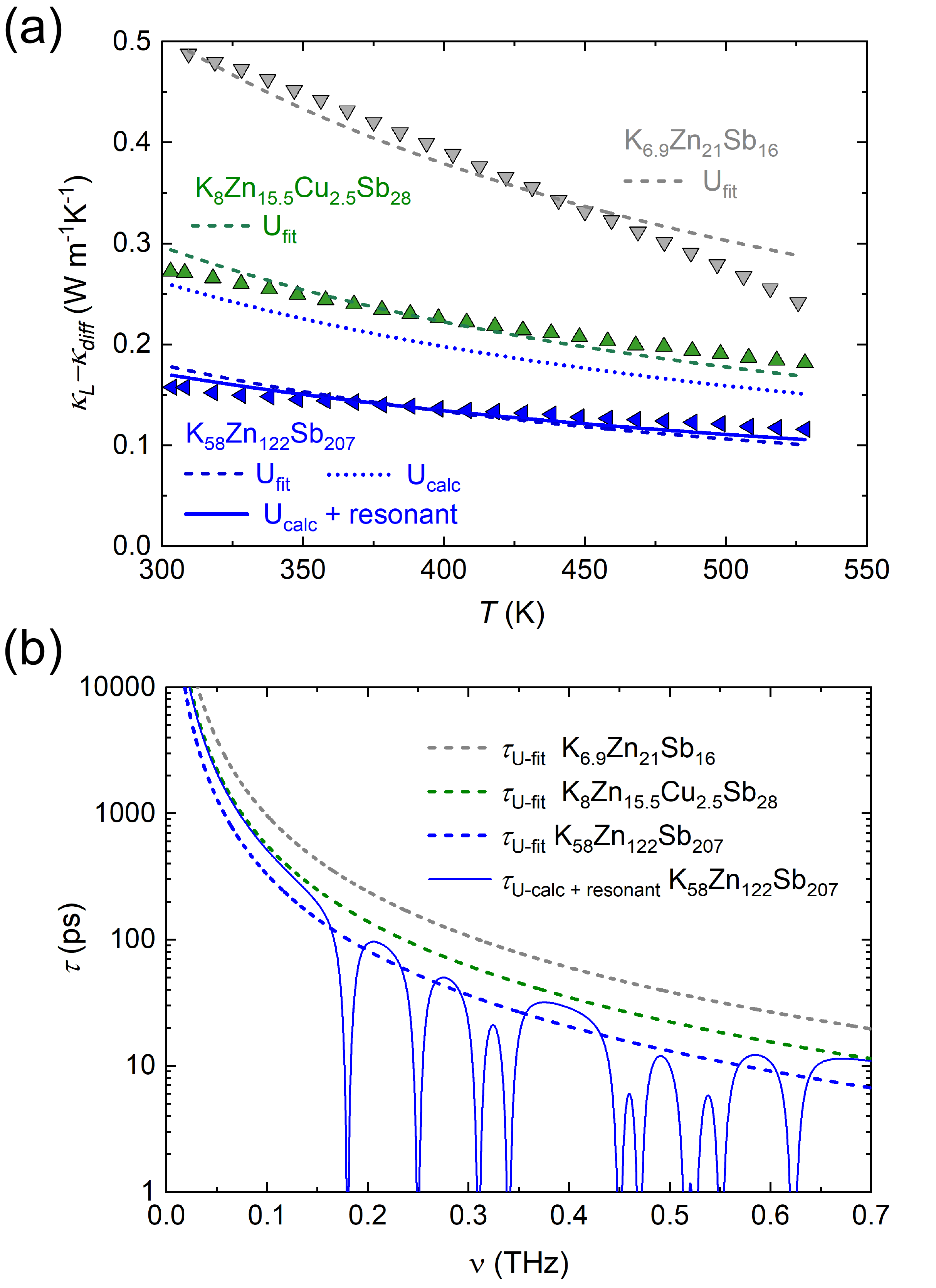}
	\caption{(a) 
	Callaway analysis closely fits the experimental thermal conductivity of the low frequency modes in the K-Zn-Sb materials. The tunnel and type-I compounds are well-fit just using Umklapp scattering ($\mathrm{U_{fit}}$).
	Equating the Umklapp scattering ($\mathrm{U_{calc}}$) of the  type-XI clathrate with that found for the type-I, an additional resonant scattering is important for a proper fit of the thermal transport ($\mathrm{U_{calc}}$+resonant). An alternative solution is extremely strong Umklapp scattering ($\mathrm{U_{fit}}$) for the type-XI clathrate. (b) The associated phonon relaxation times as a function of frequency at 300 K used in the fits above; the type-XI shows exceptionally short relaxation times regardless of the model used. The relaxation time $\tau_\mathrm{{U_{fit}}}$ for type-I clathrate is identical to that obtained  for type-XI clathrate in U$_{\mathrm{calc}}$ curve.}
	\label{Fig9} 
\end{figure}

The residual lattice thermal conductivity after subtracting out the diffusonic optical contribution should arise from the acoustic branches, see Fig. \ref{Fig9}. Here, we see acoustic contributions ranging from  0.1 W m$^{-1}$ K$^{-1}$ (type-XI) to 0.5 W m$^{-1}$ K$^{-1}$ (tunnel) across the temperature range considered.  
To gain insight into the phonon scattering mechanisms underlying the acoustic transport, we applied the Callaway model, which is simply a classical diffusion model of heat transport \cite{callaway1959model, ma2014examining}.
The general formula for thermal conductivity in the Callaway model is:
\begin{equation}
\kappa = \frac{1}{3} C_p v_s l
\end{equation}

\noindent where $C_p$ denotes the specific heat,  $v_s$ stands for the average speed of sound, while $l$ corresponds to the phonon mean free path. Hence, low thermal conductivity might be easily understood as a result of suppressed speed of sound due to softness of the bonds, or low $l$ due to intense scattering. 

Within an isotropic Debye model, substitutions can be made to yield:

\begin{equation}
\label{eq:Callaway-main}						\kappa_L=\frac{k_\mathrm{B}}{2\pi^2v_s}\left(\frac{k_b T}{\hbar}\right)^3 \int_{0}^{\Theta_{ac}/T} \frac{\tau_{tot} x^4e^x}{\left(e^x-1\right)^2}dx,
\end{equation}

\noindent where $x=h \nu/k_b T$ and $\tau_{tot}$ corresponds to the relaxation time. Here, it is necessary to underline that the upper limit of integral used is not $\Theta_D/T$, as in classical approach to such fitting, see \textit{eg}. Refs. \cite{petersen2015critical, ciesielski2020thermoelectric, bocher2017vacancy}, but its reduced version ($\Theta_{ac}/T$), which limits the contribution to just the dispersive component of the acoustic branches. The frequency above which the phonon group velocity is significantly reduced from the original speed of sound  is approximately $\nu_{ac}$ = 0.7 THz for all three compounds (\textit{cf.} Fig. \ref{Fig5} and Fig. \ref{Fig6}). This can be converted to   temperature units via $h\nu_{ac} = k_B \Theta_{ac}$.   Somewhat similar approach to Callaway-like model was proposed in Ref. \cite{schrade2018using}.

Considering Eq. \ref{eq:Callaway-main}, the only free parameter is $\tau$; as such we can now focus on extracting information about scattering within the acoustic branch.  
The phonon relaxation time $\tau_{tot}$ is calculated according to Matthiesen’s rule :
\begin{equation}
\label{eq:tau}
\tau_{tot} ^{-1}=\sum_i\tau_i
\end{equation}

\noindent where a variety of scattering mechanisms can be considered (\textit{e.g.}, Umklapp, grain boundary, point defect, or resonant scattering).  Below, we highlight three of these scattering sources that will be useful in our analysis. First, Umklapp scattering is considered, where \textit{U} denotes the Umklapp scattering parameter:

\vspace{-3mm}
\begin{equation}
\label{eq:tauU}
\tau_U ^{-1}=UT\left(2\pi \nu \right)^2 \textrm{exp}(-\Theta_{ac}/3T).
\end{equation}

Second, we consider grain boundary scattering with \textit{L} is the grain size; this term has neither frequency nor temperature dependence:
\begin{equation}
\label{eq:tauGB}
\tau_{GB} ^{-1}=\frac{v_s}{L}.
\end{equation}

 For rattling, we can also consider including resonant scattering.  The derivation of the simple resonant scattering model stems for formalism used for a mechanic oscillator. The model was successful in description of thermal conductivity in a variety of compounds with dynamical disorder \cite{pohl1962thermal, bagheri2020impact, pisoni2014ultra, michalski1995thermal} and was proposed for thermoelectric clathrates by Nolas \textit{et al.} \cite{nolas1998semiconducting}. In our treatment, we apply the commonly accepted assumption of negligible damping of the oscillator \cite{pohl1962thermal, bagheri2020impact, pisoni2014ultra, michalski1995thermal}, which results in the expression for relaxation time:
\begin{equation}
    \label{eq:Resonant}
 \tau_{res}^{-1} =  a_r \left(   \frac{\nu_0^2\nu^2}{(\nu_0^2-\nu^2)^2}    \right)  ,
\end{equation}
\noindent where parameter $a_r$ is a product of concentration of the scattering centers and strength of coupling between optical and acoustic modes, while $\nu_0$ denotes the  frequency of the flat mode.

The acoustic contribution to $\kappa_L$ for each K-Zn-Sb compound that  we seek to fit is shown in Fig. \ref{Fig9}a.  We begin with the tunnel compound, as this material shows no rattling modes and a strong temperature-dependence of thermal conductivity. This temperature dependence likely comes from phonon-phonon scattering; as such the Callaway model included both   Umklapp and grain boundary scattering. The grain size converged to its upper boundary set in the refinement (200 $\mu$m) and continued to grow if permitted. This behavior indicates, that low frequency scattering of phonons on grain boundaries is insignificant  for the tunnel compound. SEM images corroborate this conclusion, with large grains and no impurity phase precipitations at the grain boundaries, see Fig. S8.  Hence, further on we proceed to fitting with Umklapp process as the only scattering source. Considering the fit of Fig. \ref{Fig9}a for the tunnel compound, we find good agreement within the uncertainty introduced by subtracting the diffusonic contribution.  Following a similar procedure, we find fits for the two clathrate compounds (U$_{\mathrm{fit}}$ curves on Fig. \ref{Fig9}a).  The resultant values of parameter $U$ for the tunnel, type-I, and type-XI clathrates are: 9.6 $\times$ 10$^{-18}$ K$^{-1}$s,  and 1.66$\times$ 10$^{-17}$ K$^{-1}$s, and 2.64 $\times$ 10$^{-17}$ K$^{-1}$s respectively.

We highlight that the above analysis was simply a single parameter fit and does not incorporate point-defect scattering. The conventional approach to point-defect scattering incorporates Rayleigh treatment of spectral dependence ($\tau^{-1}_{PD} \propto \nu^{4}$) \cite{gurunathan2020analytical}, which makes is similar to Umklapp scattering ($\tau^{-1}_{U} \propto \nu^{2}$, see above), and might imply interdependence of the fitting parameters. Instead, we treat  $U$  as an \textit{effective scattering parameter} that describes the combination of both Umklapp and point defect scattering. In the Supporting Information, Sect. 5, the $U$ values are converted to an effective Gruinesen parameter ($\gamma_G^*$) that explicitly neglects point defect scattering. 

We also provide alternative approach, which allows to estimate possible influence of point-defect scattering. In order maintain the single-parameter approach to fitting,  we perform Callaway analysis including point-defect scattering with the mass-field fluctuations term. The reader is referred to Sect. 6 of Supporting Information for details of the procedure and Fig. S16a for the obtained curves of thermal conductivity. The Umklapp scattering coefficients  tunnel compound and type-I clathrate (9.4 $\times$ 10$^{-18}$ K$^{-1}$s and 1.62$\times$ 10$^{-17}$ K$^{-1}$s) are similar to the previously obtained \textit{effective scattering parameter} (see two paragraphs above), while for type-XI clathrate the point-defect has a negligible effect on final value of $U$, which converges due to previously obtained value of 2.64 $\times$ 10$^{-17}$ K$^{-1}$s, due to small amount of defects in type-XI structure. The relaxation times related to so-obtained point-defect scattering are at least order of magnitude higher for all studied compounds than relaxation times resultant from the Umklapp scattering, see Fig. S16b.

Fig. \ref{Fig9}b shows the resulting differences in relaxation times for the three compounds using a strictly Umklapp scattering view ($\tau_{U-fit}$) of the acoustic branches.  The graph spans up to cut-off frequency of 0.7 THz for the acoustic phonons considered in Eq.  \ref{eq:Callaway-main}, see above. 

The phonon relaxation time at $\nu_{ac}$ = 0.7 THz for the tunnel and type-I compound are 19 ps, and 11 ps. Reduction by almost a factor of 2 in $\tau$ is likely driven by the greater influence of structural defects in the tunnel compound (see Site Occupation, Sect. \ref{sect:atomic_occ}). 
From the type-I to the type-XI clathrate, we find a further average two-fold reduction in $\tau$.  This can be understood as a result of more intense phonon-phonon scattering due to the overlapping acoustic and optical modes at ultra-low frequencies. In contrast to the type-I clathrate, the point defect concentration is significantly smaller. Our order of magnitude in phonon relaxation time agrees with previous works on clathrates \cite{tadano2015impact, harkonen2016ab, euchner2012phononic, lory2017direct}.
 
The Umklapp-focused analysis does not explicitly incorporate the impact of the rattling modes; as such we consider a separate approach that includes resonant scattering, see Eq. \ref{eq:Resonant}.  To maintain single-parameter character of the analysis, for the type-XI clathrate, we fix the value of \textit{U} to that obtained above for the type-I clathrate.
Frequencies of the flat modes ($\nu_0$) below 0.7 THz were fixed based on our calculation of the phonon dispersion for the type-XI clathrate (Fig. \ref{Fig5}a). This limits the fitting to a single free parameter $a_r$, which was assumed to be uniform for all modes ($a_r$:  9.1 $\times$ $10^7$ Hz). The resulting $\tau(\nu)$ for the type-XI clathrate is shown in Fig.~\ref{Fig9}b with a series of resonant wells. 
The curve (U$_{\mathrm{calc}}$ + resonant) reproduces the experimental thermal conductivity approximately as well as the Umklapp-only model.  Fig.~\ref{Fig9}a also shows the thermal conductivity predicted using the same value of $U$ as the type-I clathrate (blue dotted line, U$_{\mathrm{calc}}$); the impact of incorporating resonant scattering (blue continuous line) is significant.  Presence of flat, low-lying optical modes suppress the phonon transport in this compounds by \textit{ca.} 35\% in the vicinity of room temperature within the approximation of similar $U$ values between the clathrate compounds considered.

To sum up, our analysis revealed, that ultra-low thermal conductivity of the K-Zn-Sb clathrates has three fundamental reasons:  (1)  soft bonding of covalent Zn-Sb sublattice,  (2) largely diffusionic overall character of thermal transport, and (3) intense phonon scattering. The soft bonding yields low group velocity of the acoustic phonons. The structural complexity, in turns, leads to abundance strongly overlapping optical modes, where the quasiparticle description no longer applies. In such situation, a more localized diffuson scenario emerges. Eventually, the strong scattering occurs for type XI clathrate due to extremely low frequency rattling modes affecting the acoustic phonon branches.  Further studies of phonon relaxation time, \textit{e.g.}, by means of inelastic neutron scattering, might provide more insight into phonon scattering behavior of these curious systems.

\subsection{Electronic transport}

\begin{figure}[t]
	\centering
	\includegraphics[width=8.6cm]{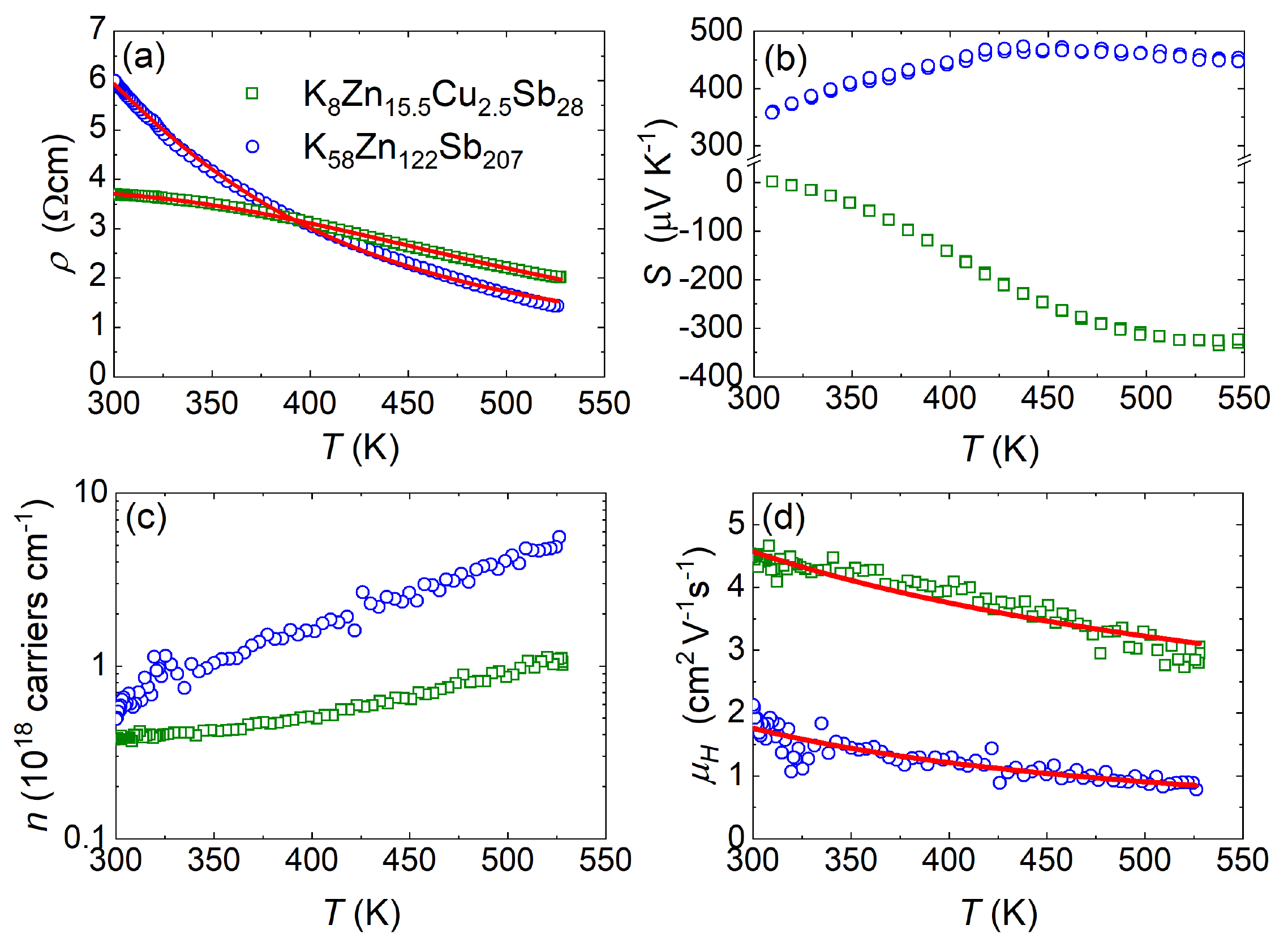}
	\caption{Temperature dependencies of (a) electrical resistivity, (b) thermopower, (c) Hall carrier mobility, and (d) carrier concentration for K$_{8}$Zn$_{15.5}$Cu$_{1.5}$Sb$_{28}$ and K$_{58}$Zn$_{122}$Sb$_{207}$. Both the materials show intrinsic electrical behavior. Solid line in panel (a) is least-squares fit with the Arrhenius equation; see text for details.}
	\label{Fig10} 
\end{figure}

Given the unusually low thermal conductivity at high temperature for all three K-Zn-Sb compounds, we consider the electronic properties and the potential for thermoelectric performance.  
The electronic transport properties of the undoped polycrystalline ingots of the type-I and XI clathrates reveal semiconducting properties. The tunnel compound exhibits semimetal behavior and will be discussed subsequently.  Electrical resistivity for the clathrates is displayed on Fig. \ref{Fig10}a. Both compounds show  a decrease of $\rho$ with increasing temperature, which suggests thermal activation of charge carriers.  To extract information about the band gap we applied the Arrhenius formula $\rho(T)^{-1} = \sigma_{0}+~\sigma_c\mathrm{exp}\left(\frac{-E_g}{2k_B T}\right)$. Fitting results in band gaps for the type-I and type-XI compounds of 0.40 and 0.25\,eV, respectively. The temperature  of the predicted maximum thermoelectric efficiency with such $E_g$'s should be located roughly at 500-700 K \cite{sofo1994optimum}. Remaining fit parameters are gathered in Tab. S8. In the previous study, the resistivity value at room temperature for K$_{58}$Zn$_{122}$Sb$_{207}$ was reported as 29 $\Omega$cm \cite{cox2018rapid}, which is factor of 5 higher than the current study. Discrepancies of this kind are typical for intrinsic, carrier-poor semiconductors as slight shift in native defect concentrations can change the number of free carriers. The former report on the band gap of type-XI clathrate shown very similar band gap value (0.22 eV) \cite{cox2021clathrate} to the $E_g$ reported here.

Fig. \ref{Fig10}b displays the temperature dependencies of thermopower (\textit{S}) for the K-Zn-Sb clathrates. The Seebeck coefficient attains sizable values for the type-XI compound and has positive sign, indicating hole-dominated transport. Maximum in  $S$ around 450 K reveals onset of bipolar phenomena. The previously reported Seebeck coefficient at 300 K for K$_{58}$Zn$_{122}$Sb$_{207}$ (380 $\mu$V K$^{-1}$) is very similar to \textit{S} obtained in the current study.  In contrast, the thermopower of the type-I clathrate shows a null value at room temperature, indicating strong bipolar effects in this semiconductor. 

Basic Goldsmid-Sharp formula  $E_g^* = 2e T_{max} |S|_{max}$ \cite{goldsmid1999estimation} can be used for calculations of the band gap. For type-I clathrate, the obtained value of $E_g^*$ is 0.36 eV, while for type-XI compound $E_g^*$ = 0.43 eV. For type-I clathrate $E_g$ from resistivity analysis and $E_g^*$ are similar, within the errors of the methods. In the case of the counterpart with type-XI structure, however, $E_g^*$ is almost twice the resistivity band gap. Differences between real band gap and value of $E_g^*$ can indicate on large differences between weighted mobilities between majority and minority charge carriers. The weighted mobility ratio is defined as $A = \frac{\left(m^*_{maj}\right)^{3/2}\mu_{0,maj}}{\left(m^*_{min}\right)^{3/2}\mu_{0,min}}$, where $\mu_{maj,min}$ corresponds to the mobility of majority and minority charge carriers and $m^*_{maj,min}$ denotes the effective mass of majority and minority carriers. The value of the weighted mobility defines maximum thermoelectric efficiency after optimization of the material \cite{snyder2020weighted}. Larger value of $E_g^*$ than the resistivity band gap in type-XI clathrate indicate on weighted mobility ratio in favor of majority charge carriers \cite{schmitt2015resolving, gibbs2015band}. This finding suggests that for thermoelectric optimization of type-XI clathrate doping should be performed in the $p$-type regime.

Temperature dependencies of Hall carrier concentration ($n_H$) for the type-I and XI clathrates are shown in Fig. \ref{Fig10}c. Low values of $n_H$ ($<10^{18}$ cm$^{-3}$) and exponential growth with temperature highlight the intrinsic character of the studied specimens. The sign of the Hall constant, from which we obtained values of $n_H$, was consistent with the sign of thermopower (+) for type-XI clathrate. Interestingly, this was not the case for the type-I compound. Below we discuss possible origin of this observation.

Within a two-band model, Seebeck coefficient is defined as:
\begin{equation}
S = \frac{S_h \sigma_h+ S_e\sigma_e}{\sigma_h+\sigma_e},
\end{equation}
\noindent where $S_{h,e}$ and $\sigma_{h,e}$ are thermopower and electrical conductivity contribution from holes and electrons, respectively. In the simplest view:
\begin{equation}
\sigma = (n,p)e\mu~~~~~~\mu \propto \frac{\tau}{m^*} ~~~~~~S \propto \frac{m^*}{(p,n)^{2/3}},
\end{equation}
\noindent where \textit{p,~n} are hole and electron concentrations and $m^*$ stands for the effective mass. So, one can observe negative Seebeck coefficient if 

\vspace{-3mm}
\begin{equation}
    S_h \sigma_h+ S_e\sigma_e~<~0,
\end{equation}
\vspace{-4mm}
\begin{equation}
    \frac{m^*_{h}}{p^{2/3}}~p~\frac{\tau_h}{m^*_h} -\frac{m^*_{e}}{n^{2/3}}~n~\frac{\tau_e}{m^*_e}~<~0,
\end{equation}
\vspace{-6mm}

\begin{equation}
    p^{1/3}\tau_h < n^{1/3}\tau_e.
\end{equation}

\noindent Assuming similar relaxation time of electrons and holes, we see that negative thermopower is observed in materials with larger concentration of electrons than holes.

Hall constant ($R_H$) in two-band model is defined as:
\vspace{-0.1mm}
\begin{equation}
R_H = \frac{p\mu_h^2-n\mu_e^2}{e(p\mu_h+n\mu_n)^2},
\end{equation}
\vspace{0.3mm}

\noindent where $\mu_{h,e}$ denote mobilities of holes and electrons, respectively.  Positive sign of Hall constant is observed when

\vspace{-5mm}
\begin{equation}
p\mu_h^2-n\mu_e^2>0,
\end{equation}
\vspace{-8mm}

\begin{equation}
\frac{\mu_h}{\mu_e} >\sqrt{\frac{n}{p}}.
\end{equation}

The above equation shows, that positive Hall constant (and positive $n_H$ in $R_H = 1/en_H$ approximation) might be observed even if concentration of electrons is bigger than holes, but holes significantly have higher mobility. Having said that, coexistence of negative $S$ and positive $n_H$ in type-I clathrate most likely results from prevalent concentration of electrons and somewhat higher mobility of holes. Due to intrinsic nature of the sample differences between carrier concentrations and mobilities between electrons and holes can be subtle.

The  repeatably low carrier densities for these materials are not a result of synthetic prowess at forming a charge-balanced composition. Instead, one can infer that both systems have compensating defects (\textit{e.g.}, Zn$_{\mathrm{Sb}}$ and Sb$_{\mathrm{Zn}}$) that place the Fermi level in the middle of the band gap.  As different synthetic conditions in the preliminary studies yielded similar results, we conclude that the chemical potential volume that these compounds occupy is small. If the elemental chemical potentials could be significantly varied within a single phase region, the energetic cost of these hypothetical defects would shift and carry the Fermi level along with them.  

Hall mobility ($\mu_H$) for both clathrates is shown on Fig.~\ref{Fig10}d. The overall values are rather low ($<5$ cm$^2$V$^{-1}$s$^{-1}$), most likely due to the combination of structural complexity of the electronic dispersion and significant crystallographic disorder. 
Temperature dependencies of mobility were fitted with power law $\mu_H \propto T^{\delta}$. The exponents we found are -0.68  and -1.29 for for type-I and type-XI, respectively; see red lines on  Fig. \ref{Fig10}d. For semiconductors, values between -1.5  to -1 are expected for electron-acoustic phonon scattering \cite{rowe2017materials}. Lower values of the exponent for type-I clathrate might be a result of point-defect scattering, which is characterized by a temperature exponent of  -0.5 \cite{xie2014intrinsic}. The finding is consistent with bigger amount of point defect revealed in the structure of this material with respect to type-XI clathrate (see Site Occupation, Sect. \ref{sect:atomic_occ}). The effective mass calculated for type-XI clathrate at 310 K was 0.38$m_e$, see Sect. 7 in Supporting Information. Parabolic band model was not applied for type-I compound due to its bipolar behavior in $S$ and $n_H$.

Electrical resistivity of the tunnel compound is shown in Fig. S17a. Its low values are consistent with the semimetallic band structure calculated for this compound \cite{cox2018rapid}. The weak overall temperature dependence of the resistivity suggest that  point-defect scattering plays an important role, which is known to be weakly temperature-dependent process. Seebeck coefficient is linear with temperature and attains rather moderate values (90 $\mu$VK$^{-1}$ at 550 K), consistently with semimetallic behavior of the resistivity (see Fig. S17b). Its positive sign corroborates the \textit{ab-initio} calculation indicating location of the Fermi level inside the valence band \cite{cox2018rapid}. The transport results of the tunnel compound is consistent with the room temperature values of resistivity and Seebeck coefficient previously reported below 300 K (3 m$\Omega$cm and 52 $\mu$VK$^{-1}$ \cite{cox2018rapid}).  The carrier concentration of the tunnel compound is two orders of magnitude higher that of K-Zn-Sb clathrates, see Fig.~S17c. The mobility is $\sim$7   cm$^2$V$^{-1}$s$^{-1}$ and temperature-independent at high temperatures (Fig.~S17d).

Thermoelectric properties were used for calculations of the figure of merit $zT = S^2T/\rho \kappa$. Both clathrates show rather low values of $zT$ (0.021 and 0.007 at 550 K for type-XI and type-I compounds, respectively), which indicated on necessity of doping to reduce their electrical resistivity. Their optimization seem feasible due to appropriate values of the band gap. The tunnel compound exhibits moderate performance with  $zT$ = 0.22 at 550 K. Its semimetallic-like band structure \cite{cox2018rapid}, indicate, that tuning charge carrier concentration might be difficult in this case.

\section{Summary}

The emergence of unconventional clathrates has led to the observation of extremely suppressed heat transport.  
However, a detailed understanding of how structural chemistry  affects thermal conductivity has not been provided up to date. 
Here, we address this challenge for K-Zn-Sb compounds, namely the type-XI clathrate K$_{58}$Zn$_{122}$Sb$_{207}$, the tunnel compound K$_{6.9}$Zn$_{21}$Sb$_{16}$, and the newly discovered type-I clathrate K$_8$Zn$_{15.5}$Cu$_{2.5}$Sb$_{28}$.
Both clathrates show ultra-low values of thermal conductivity even  at room temperature, 0.41 and 0.53 Wm$^{-1}$K$^{-1}$ for type-XI and type-I compounds, respectively. The tunnel structure exhibits higher thermal conductivity of 0.79 Wm$^{-1}$K$^{-1}$.

Focusing on the low frequency modes, all three antimonides exhibit very soft Zn-Sb bonding with a correspondingly small group velocity of acoustic phonons.
In the type-XI clathrate, further suppression of heat transfer arises due to  rattling modes. 
These localized optical modes are non-degenerate, \textit{ie.} they are arranged in a semi-regular ladder cutting through the acoustic branches  via avoided crossing in an unprecedentedly low frequency of range 0.17-0.70 THz.  
Calculations reveal that these modes are due to the rattling of K atoms encapsulated in oversized cages (hexakaidecahedra, vol. ~290 \AA$^3$). 
The presence of these rattling modes is confirmed by heat capacity measurements.

In contrast, the type-I clathrate and tunnel compound lack such huge cages and do not show optical phonon modes until 0.7 THz.  Potassium atoms from type-I clathrate enclosed in bigger of its two cavities (tetrakaidecahedra, 208 \AA$^3$) also exhibit rattling, however the associated modes are interwoven into the optical branches of the Zn-Sb network. No such rattling is observed via x-ray diffraction or phonon dispersion for the tunnel compound, which is consistent with the K-atoms being constrained in small sized cavities (\textit{ca.} 122 \AA$^3$). 
The above results are in line with low temperature heat capacity measurements. Type-I clathrate and the tunnel compound exhibit only one maximum in phonon density of states at \textit{ca}. 1.5 THz, which is primarily correlated with vibrations of Zn-Sb sublattice.

The combination of phonon calculations and experiments enables a cohesive picture of thermal transport in these K-Zn-Sb compounds to form.  The  ultra-low thermal conductivity in the type-XI clathrate arises from combination of ($i$)  structural complexity, ($ii$) soft bonding, and ($iii$)  rattling.
The structural complexity results in an abundance of low velocity, strongly overlapping optical modes. In such a scenario, the quasiparticle description no longer applies and a more localized diffuson scenario emerges. The soft bonding yields low acoustic mode group velocities. Finally, the extremely low frequency rattling modes lead to enhanced  scattering throughout the acoustic branches.  

The type-I clathrate retains the structural complexity and soft bonding, but the shift to higher frequencies for the rattling yields a doubling of the acoustic relaxation times compared to the type-XI clathrate. Similarly, the absence of any rattling modes in the tunnel compound leads to a further doubling  of the acoustic relaxation times compared to the type-I clathrate. Both type-I clathrate and the tunnel compound comprise point defects in their unit cells, which contribute to lowering the their thermal conductivity.

Having deeply probed the thermal behavior of these materials, significant opportunities remain for creating K-Zn-Sb thermoelectric clathrates with electronic performance as excellent as their thermal properties. 
The critical need is to significantly enhance the free carrier concentration from the non-degenerate levels found herein.
The discrepancy between resistivity and Goldsmid-Sharp band gaps for the type-XI clathrate indicates that \textit{p}-type doping will be the most fruitful for thermoelectric performance. Furthermore, unexplored opportunities of Zn/Cd/Hg and As/Sb/Bi  substitutions and alternative filler atoms open a pathway to discovery of new, structurally intricate clathrates.

\begin{acknowledgement}

This work was funded primarily with support from the
National Science Foundation (NSF) via grant DMR 1555340 sponsoring work of EST, VM, BRO, TB, JMA, and KC.  EE and LCG  acknowledge support from NSF grant DMR 1729149. This research is part of the Blue Waters sustained-petascale computing project, which is supported by the National Science Foundation (awards OCI-0725070 and ACI-1238993) the State of Illinois, and as of December, 2019, the National Geospatial-Intelligence Agency. Blue Waters is a joint effort of the University of Illinois at Urbana-Champaign and its National Center for Supercomputing Applications. This work also used the Extreme Science and Engineering Discovery Environment (XSEDE), which is supported by National Science Foundation grant number ACI-1548562 \cite{towns2014xsede}.

\end{acknowledgement}

\begin{suppinfo}

Additional data regarding structural, thermal, and electronic properties of the studied materials: atomic coordinates for type-XI clathrate and the tunnel compound; results of scanning electron microscopy imaging and energy dispersive X-ray spectroscopy for all the studied materials; (data for all compounds henceforth) analysis of elastic properties; details of unit cells used for phonon calculations; quantitative analysis of heat capacity; interpretation of Umklapp scattering coefficient from Callaway analysis; additional Callaway modeling incorporating presence of point-defect scattering; parabolic band modeling of electronic data; graphs presenting total thermal conductivity of clathrates known in the literature; visual presentation of sintering profile;  additional powder X-ray diffraction patterns, supplementary analysis of theoretical phonon dispersion by calculating group velocity (for type-I clathrate and the tunnel compound) and energy fraction per atom for type-XI clathrate; electrical properties of the tunnel compound.

\end{suppinfo}

\bibliography{references}

\providecommand{\latin}[1]{#1}
\makeatletter
\providecommand{\doi}
  {\begingroup\let\do\@makeother\dospecials
  \catcode`\{=1 \catcode`\}=2 \doi@aux}
\providecommand{\doi@aux}[1]{\endgroup\texttt{#1}}
\makeatother
\providecommand*\mcitethebibliography{\thebibliography}
\csname @ifundefined\endcsname{endmcitethebibliography}
  {\let\endmcitethebibliography\endthebibliography}{}
\begin{mcitethebibliography}{103}
\providecommand*\natexlab[1]{#1}
\providecommand*\mciteSetBstSublistMode[1]{}
\providecommand*\mciteSetBstMaxWidthForm[2]{}
\providecommand*\mciteBstWouldAddEndPuncttrue
  {\def\EndOfBibitem{\unskip.}}
\providecommand*\mciteBstWouldAddEndPunctfalse
  {\let\EndOfBibitem\relax}
\providecommand*\mciteSetBstMidEndSepPunct[3]{}
\providecommand*\mciteSetBstSublistLabelBeginEnd[3]{}
\providecommand*\EndOfBibitem{}
\mciteSetBstSublistMode{f}
\mciteSetBstMaxWidthForm{subitem}{(\alph{mcitesubitemcount})}
\mciteSetBstSublistLabelBeginEnd
  {\mcitemaxwidthsubitemform\space}
  {\relax}
  {\relax}

\bibitem[Koley \latin{et~al.}(2021)Koley, Lakshan, Raghuvanshi, Singh,
  Bhattacharya, and Jana]{koley2021ultralow}
Koley,~B.; Lakshan,~A.; Raghuvanshi,~P.~R.; Singh,~C.; Bhattacharya,~A.;
  Jana,~P.~P. Ultralow Lattice Thermal Conductivity at Room Temperature in
  \mbox{Cu$_4$TiSe$_4$}. \emph{Angew. Chem. Int. Ed.} \textbf{2021}, \emph{60},
  9106--9113\relax
\mciteBstWouldAddEndPuncttrue
\mciteSetBstMidEndSepPunct{\mcitedefaultmidpunct}
{\mcitedefaultendpunct}{\mcitedefaultseppunct}\relax
\EndOfBibitem
\bibitem[Mukhopadhyay \latin{et~al.}(2018)Mukhopadhyay, Parker, Sales,
  Puretzky, McGuire, and Lindsay]{mukhopadhyay2018two}
Mukhopadhyay,~S.; Parker,~D.~S.; Sales,~B.~C.; Puretzky,~A.~A.; McGuire,~M.~A.;
  Lindsay,~L. Two-channel Model for Ultralow Thermal Conductivity of
  Crystalline \mbox{Tl$_3$VSe$_4$}. \emph{Science} \textbf{2018}, \emph{360},
  1455--1458\relax
\mciteBstWouldAddEndPuncttrue
\mciteSetBstMidEndSepPunct{\mcitedefaultmidpunct}
{\mcitedefaultendpunct}{\mcitedefaultseppunct}\relax
\EndOfBibitem
\bibitem[Lin \latin{et~al.}(2016)Lin, Tan, Shen, Hao, Wu, Calta, Malliakas,
  Wang, Uher, Wolverton, \latin{et~al.} others]{lin2016concerted}
Lin,~H.; Tan,~G.; Shen,~J.-N.; Hao,~S.; Wu,~L.-M.; Calta,~N.; Malliakas,~C.;
  Wang,~S.; Uher,~C.; Wolverton,~C., \latin{et~al.}  Concerted Rattling in
  \mbox{CsAg$_5$Te$_3$} Leading to Ultralow Thermal Conductivity and High
  Thermoelectric Performance. \emph{Angew. Chem. Int. Ed.} \textbf{2016},
  \emph{128}, 11603--11608\relax
\mciteBstWouldAddEndPuncttrue
\mciteSetBstMidEndSepPunct{\mcitedefaultmidpunct}
{\mcitedefaultendpunct}{\mcitedefaultseppunct}\relax
\EndOfBibitem
\bibitem[Zhao \latin{et~al.}(2014)Zhao, Lo, Zhang, Sun, Tan, Uher, Wolverton,
  Dravid, and Kanatzidis]{zhao2014ultralow}
Zhao,~L.-D.; Lo,~S.-H.; Zhang,~Y.; Sun,~H.; Tan,~G.; Uher,~C.; Wolverton,~C.;
  Dravid,~V.~P.; Kanatzidis,~M.~G. Ultralow Thermal Conductivity and High
  Thermoelectric Figure of Merit in \mbox{SnSe} Crystals. \emph{Nature}
  \textbf{2014}, \emph{508}, 373--377\relax
\mciteBstWouldAddEndPuncttrue
\mciteSetBstMidEndSepPunct{\mcitedefaultmidpunct}
{\mcitedefaultendpunct}{\mcitedefaultseppunct}\relax
\EndOfBibitem
\bibitem[Schmitt \latin{et~al.}(2012)Schmitt, Haldolaarachchige, Xiong, Young,
  Jin, and Chan]{schmitt2012probing}
Schmitt,~D.~C.; Haldolaarachchige,~N.; Xiong,~Y.; Young,~D.~P.; Jin,~R.;
  Chan,~J.~Y. Probing the Lower Limit of Lattice Thermal Conductivity in an
  Ordered Extended Solid: \mbox{Gd$_{117}$Co$_{56}$Sn$_{112}$}, a Phonon
  Glass--Electron Crystal System. \emph{J. Am. Chem. Soc} \textbf{2012},
  \emph{134}, 5965--5973\relax
\mciteBstWouldAddEndPuncttrue
\mciteSetBstMidEndSepPunct{\mcitedefaultmidpunct}
{\mcitedefaultendpunct}{\mcitedefaultseppunct}\relax
\EndOfBibitem
\bibitem[Fulmer \latin{et~al.}(2013)Fulmer, Lebedev, Roddatis, Kaseman, Sen,
  Dolyniuk, Lee, Olenev, and Kovnir]{fulmer2013clathrate}
Fulmer,~J.; Lebedev,~O.~I.; Roddatis,~V.~V.; Kaseman,~D.~C.; Sen,~S.;
  Dolyniuk,~J.-A.; Lee,~K.; Olenev,~A.~V.; Kovnir,~K. Clathrate
  \mbox{Ba$_8$Au$_{16}$P$_{30}$}: The “Gold Standard” for Lattice Thermal
  Conductivity. \emph{J. Am. Chem. Soc.} \textbf{2013}, \emph{135},
  12313--12323\relax
\mciteBstWouldAddEndPuncttrue
\mciteSetBstMidEndSepPunct{\mcitedefaultmidpunct}
{\mcitedefaultendpunct}{\mcitedefaultseppunct}\relax
\EndOfBibitem
\bibitem[Cox \latin{et~al.}(2021)Cox, Gvozdetskyi, Bertolami, Lee, Shipley,
  Lebedev, and Zaikina]{cox2021clathrate}
Cox,~T.; Gvozdetskyi,~V.; Bertolami,~M.; Lee,~S.; Shipley,~K.; Lebedev,~O.~I.;
  Zaikina,~J.~V. Clathrate \mbox{XI K$_{58}$Zn$_{122}$Sb$_{207}$}: A New Branch
  on the Clathrate Family Tree. \emph{Angew. Chem. Int. Ed.} \textbf{2021},
  \emph{133}, 419--427\relax
\mciteBstWouldAddEndPuncttrue
\mciteSetBstMidEndSepPunct{\mcitedefaultmidpunct}
{\mcitedefaultendpunct}{\mcitedefaultseppunct}\relax
\EndOfBibitem
\bibitem[Brown \latin{et~al.}(2006)Brown, Kauzlarich, Gascoin, and
  Snyder]{brown2006yb14mnsb11}
Brown,~S.~R.; Kauzlarich,~S.~M.; Gascoin,~F.; Snyder,~G.~J.
  \mbox{Yb$_{14}$MnSb$_{11}$}: New High Efficiency Thermoelectric Material for
  Power Generation. \emph{Chem. Mater.} \textbf{2006}, \emph{18},
  1873--1877\relax
\mciteBstWouldAddEndPuncttrue
\mciteSetBstMidEndSepPunct{\mcitedefaultmidpunct}
{\mcitedefaultendpunct}{\mcitedefaultseppunct}\relax
\EndOfBibitem
\bibitem[Agne \latin{et~al.}(2018)Agne, Hanus, and Snyder]{agne2018minimum}
Agne,~M.~T.; Hanus,~R.; Snyder,~G.~J. Minimum Thermal Conductivity in the
  Context Of Diffuson-Mediated Thermal Transport. \emph{Energy Environ. Sci.}
  \textbf{2018}, \emph{11}, 609--616\relax
\mciteBstWouldAddEndPuncttrue
\mciteSetBstMidEndSepPunct{\mcitedefaultmidpunct}
{\mcitedefaultendpunct}{\mcitedefaultseppunct}\relax
\EndOfBibitem
\bibitem[Slack(1979)]{slack1979thermal}
Slack,~G.~A. The Thermal Conductivity of Nonmetallic Crystals. \emph{Solid
  State Physics} \textbf{1979}, \emph{34}, 1--71\relax
\mciteBstWouldAddEndPuncttrue
\mciteSetBstMidEndSepPunct{\mcitedefaultmidpunct}
{\mcitedefaultendpunct}{\mcitedefaultseppunct}\relax
\EndOfBibitem
\bibitem[Kim(2015)]{kim2015strategies}
Kim,~W. Strategies for Engineering Phonon Transport in Thermoelectrics.
  \emph{J. Mater. Chem. C} \textbf{2015}, \emph{3}, 10336--10348\relax
\mciteBstWouldAddEndPuncttrue
\mciteSetBstMidEndSepPunct{\mcitedefaultmidpunct}
{\mcitedefaultendpunct}{\mcitedefaultseppunct}\relax
\EndOfBibitem
\bibitem[Dolyniuk \latin{et~al.}(2016)Dolyniuk, Owens-Baird, Wang, Zaikina, and
  Kovnir]{dolyniuk2016clathrate}
Dolyniuk,~J.-A.; Owens-Baird,~B.; Wang,~J.; Zaikina,~J.~V.; Kovnir,~K.
  Clathrate Thermoelectrics. \emph{Mater. Sci. Eng. R Rep} \textbf{2016},
  \emph{108}, 1--46\relax
\mciteBstWouldAddEndPuncttrue
\mciteSetBstMidEndSepPunct{\mcitedefaultmidpunct}
{\mcitedefaultendpunct}{\mcitedefaultseppunct}\relax
\EndOfBibitem
\bibitem[Nolas \latin{et~al.}(1999)Nolas, Morelli, and
  Tritt]{nolas1999skutterudites}
Nolas,~G.; Morelli,~D.; Tritt,~T.~M. Skutterudites: A Phonon-Glass-Electron
  Crystal Approach to Advanced Thermoelectric Energy Conversion Applications.
  \emph{Annu. Rev. Mater. Sci.} \textbf{1999}, \emph{29}, 89--116\relax
\mciteBstWouldAddEndPuncttrue
\mciteSetBstMidEndSepPunct{\mcitedefaultmidpunct}
{\mcitedefaultendpunct}{\mcitedefaultseppunct}\relax
\EndOfBibitem
\bibitem[Schweika \latin{et~al.}(2007)Schweika, Hermann, Prager, Persson, and
  Keppens]{schweika2007dumbbell}
Schweika,~W.; Hermann,~R.; Prager,~M.; Persson,~J.; Keppens,~V. Dumbbell
  Rattling In Thermoelectric Zinc Antimony. \emph{Phys. Rev. Lett.}
  \textbf{2007}, \emph{99}, 125501\relax
\mciteBstWouldAddEndPuncttrue
\mciteSetBstMidEndSepPunct{\mcitedefaultmidpunct}
{\mcitedefaultendpunct}{\mcitedefaultseppunct}\relax
\EndOfBibitem
\bibitem[Christensen \latin{et~al.}(2008)Christensen, Abrahamsen, Christensen,
  Juranyi, Andersen, Lefmann, Andreasson, Bahl, and
  Iversen]{christensen2008avoided}
Christensen,~M.; Abrahamsen,~A.~B.; Christensen,~N.~B.; Juranyi,~F.;
  Andersen,~N.~H.; Lefmann,~K.; Andreasson,~J.; Bahl,~C.~R.; Iversen,~B.~B.
  Avoided Crossing of Rattler Modes in Thermoelectric Materials. \emph{Nat.
  Mater.} \textbf{2008}, \emph{7}, 811--815\relax
\mciteBstWouldAddEndPuncttrue
\mciteSetBstMidEndSepPunct{\mcitedefaultmidpunct}
{\mcitedefaultendpunct}{\mcitedefaultseppunct}\relax
\EndOfBibitem
\bibitem[Tse \latin{et~al.}(2001)Tse, Li, and Uehara]{tse2001phonon}
Tse,~J.; Li,~Z.; Uehara,~K. Phonon Band Structures and Resonant Scattering in
  \mbox{Na$_8$Si$_{46}$} and \mbox{Cs$_8$Sn$_{44}$} Clathrates. \emph{EPL
  (Europhysics Letters)} \textbf{2001}, \emph{56}, 261\relax
\mciteBstWouldAddEndPuncttrue
\mciteSetBstMidEndSepPunct{\mcitedefaultmidpunct}
{\mcitedefaultendpunct}{\mcitedefaultseppunct}\relax
\EndOfBibitem
\bibitem[Euchner \latin{et~al.}(2012)Euchner, Pailh{\`e}s, Nguyen, Assmus,
  Ritter, Haghighirad, Grin, Paschen, and de~Boissieu]{euchner2012phononic}
Euchner,~H.; Pailh{\`e}s,~S.; Nguyen,~L.; Assmus,~W.; Ritter,~F.;
  Haghighirad,~A.; Grin,~Y.; Paschen,~S.; de~Boissieu,~M. Phononic Filter
  Effect of Rattling Phonons In The Thermoelectric Clathrate
  \mbox{Ba$_{8}$Ge$_{40+x}$Ni$_{6-x}$}. \emph{Phys. Rev. B} \textbf{2012},
  \emph{86}, 224303\relax
\mciteBstWouldAddEndPuncttrue
\mciteSetBstMidEndSepPunct{\mcitedefaultmidpunct}
{\mcitedefaultendpunct}{\mcitedefaultseppunct}\relax
\EndOfBibitem
\bibitem[Tadano \latin{et~al.}(2015)Tadano, Gohda, and
  Tsuneyuki]{tadano2015impact}
Tadano,~T.; Gohda,~Y.; Tsuneyuki,~S. Impact of Rattlers on Thermal Conductivity
  of A Thermoelectric Clathrate a First-Principles Study. \emph{Phys. Rev.
  Lett.} \textbf{2015}, \emph{114}, 095501\relax
\mciteBstWouldAddEndPuncttrue
\mciteSetBstMidEndSepPunct{\mcitedefaultmidpunct}
{\mcitedefaultendpunct}{\mcitedefaultseppunct}\relax
\EndOfBibitem
\bibitem[Norouzzadeh \latin{et~al.}(2017)Norouzzadeh, Myles, and
  Vashaee]{norouzzadeh2017phonon}
Norouzzadeh,~P.; Myles,~C.~W.; Vashaee,~D. Phonon Dynamics in Type-\mbox{VIII}
  Silicon Clathrates: \mbox{B}eyond the Rattler Concept. \emph{Phys. Rev. B}
  \textbf{2017}, \emph{95}, 195206\relax
\mciteBstWouldAddEndPuncttrue
\mciteSetBstMidEndSepPunct{\mcitedefaultmidpunct}
{\mcitedefaultendpunct}{\mcitedefaultseppunct}\relax
\EndOfBibitem
\bibitem[Dolyniuk \latin{et~al.}(2017)Dolyniuk, Zaikina, Kaseman, Sen, and
  Kovnir]{dolyniuk2017breaking}
Dolyniuk,~J.-A.; Zaikina,~J.~V.; Kaseman,~D.~C.; Sen,~S.; Kovnir,~K. Breaking
  the Tetra-Coordinated Framework Rule: New Clathrate
  \mbox{Ba$_8$M$_{24}$P$_{28+\delta}$ ($M$ = Cu/Zn)}. \emph{Angew. Chem. Int.
  Ed.} \textbf{2017}, \emph{129}, 2458--2462\relax
\mciteBstWouldAddEndPuncttrue
\mciteSetBstMidEndSepPunct{\mcitedefaultmidpunct}
{\mcitedefaultendpunct}{\mcitedefaultseppunct}\relax
\EndOfBibitem
\bibitem[Wang \latin{et~al.}(2017)Wang, Lebedev, Lee, Dolyniuk, Klavins, Bux,
  and Kovnir]{wang2017high}
Wang,~J.; Lebedev,~O.~I.; Lee,~K.; Dolyniuk,~J.-A.; Klavins,~P.; Bux,~S.;
  Kovnir,~K. High-Efficiency Thermoelectric
  \mbox{Ba$_{8}$Cu$_{14}$Ge$_{6}$P$_{26}$}: Bridging the Gap Between
  Tetrel-Based and Tetrel-Free Clathrates. \emph{Chem. Sci.} \textbf{2017},
  \emph{8}, 8030--8038\relax
\mciteBstWouldAddEndPuncttrue
\mciteSetBstMidEndSepPunct{\mcitedefaultmidpunct}
{\mcitedefaultendpunct}{\mcitedefaultseppunct}\relax
\EndOfBibitem
\bibitem[Wang \latin{et~al.}(2018)Wang, He, Mordvinova, Lebedev, and
  Kovnir]{wang2018smaller}
Wang,~J.; He,~Y.; Mordvinova,~N.~E.; Lebedev,~O.~I.; Kovnir,~K. The Smaller the
  Better: Hosting Trivalent Rare-Earth Guests in \mbox{Cu--P} Clathrate Cages.
  \emph{Chem} \textbf{2018}, \emph{4}, 1465--1475\relax
\mciteBstWouldAddEndPuncttrue
\mciteSetBstMidEndSepPunct{\mcitedefaultmidpunct}
{\mcitedefaultendpunct}{\mcitedefaultseppunct}\relax
\EndOfBibitem
\bibitem[Dolyniuk \latin{et~al.}(2018)Dolyniuk, Wang, Marple, Sen, Cheng,
  Ramirez-Cuesta, and Kovnir]{dolyniuk2018chemical}
Dolyniuk,~J.-A.; Wang,~J.; Marple,~M.~A.; Sen,~S.; Cheng,~Y.;
  Ramirez-Cuesta,~A.~J.; Kovnir,~K. Chemical Bonding and Transport Properties
  in Clathrates-\mbox{I} with \mbox{Cu--Zn--P} Frameworks. \emph{Chem. Mater.}
  \textbf{2018}, \emph{30}, 3419--3428\relax
\mciteBstWouldAddEndPuncttrue
\mciteSetBstMidEndSepPunct{\mcitedefaultmidpunct}
{\mcitedefaultendpunct}{\mcitedefaultseppunct}\relax
\EndOfBibitem
\bibitem[Huo \latin{et~al.}(2003)Huo, Sasakawa, Muro, and
  Takabatake]{huo2003thermoelectric}
Huo,~D.; Sasakawa,~T.; Muro,~Y.; Takabatake,~T. Thermoelectric properties of a
  clathrate compound \mbox{Ba$_{8}$Cu$_{16}$P$_{30}$}. \emph{Appl. Phys. Lett.}
  \textbf{2003}, \emph{82}, 2640--2642\relax
\mciteBstWouldAddEndPuncttrue
\mciteSetBstMidEndSepPunct{\mcitedefaultmidpunct}
{\mcitedefaultendpunct}{\mcitedefaultseppunct}\relax
\EndOfBibitem
\bibitem[Dolyniuk \latin{et~al.}(2015)Dolyniuk, Wang, Lee, and
  Kovnir]{dolyniuk2015twisted}
Dolyniuk,~J.-A.; Wang,~J.; Lee,~K.; Kovnir,~K. Twisted Kelvin Cells and
  Truncated Octahedral Cages in the Crystal Structures of Unconventional
  Clathrates, \mbox{$AM_2$P$_4$ ($A$} = \mbox{Sr, Ba; $M$ = Cu, Ni)}.
  \emph{Chem. Mater.} \textbf{2015}, \emph{27}, 4476--4484\relax
\mciteBstWouldAddEndPuncttrue
\mciteSetBstMidEndSepPunct{\mcitedefaultmidpunct}
{\mcitedefaultendpunct}{\mcitedefaultseppunct}\relax
\EndOfBibitem
\bibitem[Owens-Baird \latin{et~al.}(2020)Owens-Baird, Yox, Lee, Carroll, Wang,
  Chen, Lebedev, and Kovnir]{owens2020chemically}
Owens-Baird,~B.; Yox,~P.; Lee,~S.; Carroll,~X.~B.; Wang,~S.~G.; Chen,~Y.-S.;
  Lebedev,~O.~I.; Kovnir,~K. Chemically Driven Superstructural Ordering Leading
  to Giant Unit Cells in Unconventional Clathrates
  \mbox{Cs$_{8}$Zn$_{18}$Sb$_{28}$} and \mbox{Cs$_{8}$Cd$_{18}$Sb$_{28}$}.
  \emph{Chem. Sci.} \textbf{2020}, \emph{11}, 10255--10264\relax
\mciteBstWouldAddEndPuncttrue
\mciteSetBstMidEndSepPunct{\mcitedefaultmidpunct}
{\mcitedefaultendpunct}{\mcitedefaultseppunct}\relax
\EndOfBibitem
\bibitem[Owens-Baird \latin{et~al.}(2020)Owens-Baird, Wang, Wang, Chen, Lee,
  Donadio, and Kovnir]{owens2020iii}
Owens-Baird,~B.; Wang,~J.; Wang,~S.~G.; Chen,~Y.-S.; Lee,~S.; Donadio,~D.;
  Kovnir,~K. \mbox{III--V} Clathrate Semiconductors With Outstanding Hole
  Mobility: \mbox{Cs$_8$In$_{27}$Sb$_{19}$} and \mbox{$A_8$Ga$_{27}$Sb$_{19}$
  (\textit{A} = Cs, Rb)}. \emph{J. Am. Chem. Soc.} \textbf{2020}, \emph{142},
  2031--2041\relax
\mciteBstWouldAddEndPuncttrue
\mciteSetBstMidEndSepPunct{\mcitedefaultmidpunct}
{\mcitedefaultendpunct}{\mcitedefaultseppunct}\relax
\EndOfBibitem
\bibitem[Takabatake \latin{et~al.}(2014)Takabatake, Suekuni, Nakayama, and
  Kaneshita]{takabatake2014phonon}
Takabatake,~T.; Suekuni,~K.; Nakayama,~T.; Kaneshita,~E. Phonon-Glass
  Electron-Crystal Thermoelectric Clathrates: Experiments and Theory.
  \emph{Rev. Mod. Phys} \textbf{2014}, \emph{86}, 669\relax
\mciteBstWouldAddEndPuncttrue
\mciteSetBstMidEndSepPunct{\mcitedefaultmidpunct}
{\mcitedefaultendpunct}{\mcitedefaultseppunct}\relax
\EndOfBibitem
\bibitem[Toberer \latin{et~al.}(2008)Toberer, Christensen, Iversen, and
  Snyder]{toberer2008high}
Toberer,~E.~S.; Christensen,~M.; Iversen,~B.; Snyder,~G.~J. High Temperature
  Thermoelectric Efficiency in \mbox{Ba$_8$Ga$_{16}$Ge$_{30}$}. \emph{Phys.
  Rev. B} \textbf{2008}, \emph{77}, 075203\relax
\mciteBstWouldAddEndPuncttrue
\mciteSetBstMidEndSepPunct{\mcitedefaultmidpunct}
{\mcitedefaultendpunct}{\mcitedefaultseppunct}\relax
\EndOfBibitem
\bibitem[Wang and Chang(2017)Wang, and Chang]{wang2017thermoelectric}
Wang,~L.-H.; Chang,~L.-S. Thermoelectric Properties of \textit{p}-type
  \mbox{Ba$_8$Ga$_{16}$Ge$_{30}$} Type-\mbox{I} Clathrate Compounds Prepared by
  the Vertical Bridgman Method. \emph{J. Alloys Compds.} \textbf{2017},
  \emph{722}, 644--650\relax
\mciteBstWouldAddEndPuncttrue
\mciteSetBstMidEndSepPunct{\mcitedefaultmidpunct}
{\mcitedefaultendpunct}{\mcitedefaultseppunct}\relax
\EndOfBibitem
\bibitem[Suekuni \latin{et~al.}(2007)Suekuni, Avila, Umeo, and
  Takabatake]{suekuni2007cage}
Suekuni,~K.; Avila,~M.; Umeo,~K.; Takabatake,~T. Cage-Size Control of Guest
  Vibration and Thermal Conductivity in
  \mbox{Sr$_{8}$Ga$_{16}$Si$_{30-x}$Ge$_{x}$}. \emph{Phys. Rev. B}
  \textbf{2007}, \emph{75}, 195210\relax
\mciteBstWouldAddEndPuncttrue
\mciteSetBstMidEndSepPunct{\mcitedefaultmidpunct}
{\mcitedefaultendpunct}{\mcitedefaultseppunct}\relax
\EndOfBibitem
\bibitem[Ikeda \latin{et~al.}(2019)Ikeda, Euchner, Yan, Tome{\v{s}}, Prokofiev,
  Prochaska, Lientschnig, Svagera, Hartmann, Gati, \latin{et~al.}
  others]{ikeda2019kondo}
Ikeda,~M.; Euchner,~H.; Yan,~X.; Tome{\v{s}},~P.; Prokofiev,~A.; Prochaska,~L.;
  Lientschnig,~G.; Svagera,~R.; Hartmann,~S.; Gati,~E., \latin{et~al.}
  Kondo-Like Phonon Scattering in Thermoelectric Clathrates. \emph{Nat.
  Commun.} \textbf{2019}, \emph{10}, 1--9\relax
\mciteBstWouldAddEndPuncttrue
\mciteSetBstMidEndSepPunct{\mcitedefaultmidpunct}
{\mcitedefaultendpunct}{\mcitedefaultseppunct}\relax
\EndOfBibitem
\bibitem[Wang \latin{et~al.}(2018)Wang, Dolyniuk, and
  Kovnir]{wang2018unconventional}
Wang,~J.; Dolyniuk,~J.-A.; Kovnir,~K. Unconventional Clathrates With Transition
  Metal--Phosphorus Frameworks. \emph{Acc. Chem. Res.} \textbf{2018},
  \emph{51}, 31--39\relax
\mciteBstWouldAddEndPuncttrue
\mciteSetBstMidEndSepPunct{\mcitedefaultmidpunct}
{\mcitedefaultendpunct}{\mcitedefaultseppunct}\relax
\EndOfBibitem
\bibitem[Wang \latin{et~al.}(2016)Wang, Kaseman, Lee, Sen, and
  Kovnir]{wang2016enclathration}
Wang,~J.; Kaseman,~D.; Lee,~K.; Sen,~S.; Kovnir,~K. Enclathration of \mbox{$X$@
  La4} Tetrahedra in Channels of \mbox{Zn--P} Crameworks in
  \mbox{La$_3$Zn$_4$P$_6 X$ ($X$ = Cl, Br)}. \emph{Chem. Mater.} \textbf{2016},
  \emph{28}, 4741--4750\relax
\mciteBstWouldAddEndPuncttrue
\mciteSetBstMidEndSepPunct{\mcitedefaultmidpunct}
{\mcitedefaultendpunct}{\mcitedefaultseppunct}\relax
\EndOfBibitem
\bibitem[Shi \latin{et~al.}(2011)Shi, Yang, Salvador, Chi, Cho, Wang, Bai,
  Yang, Zhang, and Chen]{shi2011multiple}
Shi,~X.; Yang,~J.; Salvador,~J.~R.; Chi,~M.; Cho,~J.~Y.; Wang,~H.; Bai,~S.;
  Yang,~J.; Zhang,~W.; Chen,~L. Multiple-Filled Skutterudites: High
  Thermoelectric Figure of Merit Through Separately Optimizing Electrical and
  Thermal Transports. \emph{J. Am. Chem. Soc} \textbf{2011}, \emph{133},
  7837--7846\relax
\mciteBstWouldAddEndPuncttrue
\mciteSetBstMidEndSepPunct{\mcitedefaultmidpunct}
{\mcitedefaultendpunct}{\mcitedefaultseppunct}\relax
\EndOfBibitem
\bibitem[Fu \latin{et~al.}(2015)Fu, Zhu, Liu, Xie, and Zhao]{fu2015band}
Fu,~C.; Zhu,~T.; Liu,~Y.; Xie,~H.; Zhao,~X. Band Engineering of High
  Performance p-type \mbox{FeNbSb} Based Half-\mbox{H}eusler Thermoelectric
  Materials for Figure of Merit \textit{zT} > 1. \emph{Energy Environ. Sci}
  \textbf{2015}, \emph{8}, 216--220\relax
\mciteBstWouldAddEndPuncttrue
\mciteSetBstMidEndSepPunct{\mcitedefaultmidpunct}
{\mcitedefaultendpunct}{\mcitedefaultseppunct}\relax
\EndOfBibitem
\bibitem[Ohno \latin{et~al.}(2018)Ohno, Imasato, Anand, Tamaki, Kang, Gorai,
  Sato, Toberer, Kanno, and Snyder]{ohno2018phase}
Ohno,~S.; Imasato,~K.; Anand,~S.; Tamaki,~H.; Kang,~S.~D.; Gorai,~P.;
  Sato,~H.~K.; Toberer,~E.~S.; Kanno,~T.; Snyder,~G.~J. Phase Boundary Mapping
  to Obtain n-type \mbox{Mg}$_3$\mbox{Sb}$_2$-Based Thermoelectrics.
  \emph{Joule} \textbf{2018}, \emph{2}, 141--154\relax
\mciteBstWouldAddEndPuncttrue
\mciteSetBstMidEndSepPunct{\mcitedefaultmidpunct}
{\mcitedefaultendpunct}{\mcitedefaultseppunct}\relax
\EndOfBibitem
\bibitem[Ohno \latin{et~al.}(2017)Ohno, Aydemir, Amsler, P{\"o}hls, Chanakian,
  Zevalkink, White, Bux, Wolverton, and Snyder]{ohno2017achieving}
Ohno,~S.; Aydemir,~U.; Amsler,~M.; P{\"o}hls,~J.-H.; Chanakian,~S.;
  Zevalkink,~A.; White,~M.~A.; Bux,~S.~K.; Wolverton,~C.; Snyder,~G.~J.
  Achieving \textit{zT} > 1 in inexpensive Zintl phase
  \mbox{Ca$_9$Zn$_{4+x}$Sb$_{9}$} by phase boundary mapping. \emph{Adv. Funct.
  Mater.} \textbf{2017}, \emph{27}, 1606361\relax
\mciteBstWouldAddEndPuncttrue
\mciteSetBstMidEndSepPunct{\mcitedefaultmidpunct}
{\mcitedefaultendpunct}{\mcitedefaultseppunct}\relax
\EndOfBibitem
\bibitem[Kraemer \latin{et~al.}(2015)Kraemer, Sui, McEnaney, Zhao, Jie, Ren,
  and Chen]{kraemer2015high}
Kraemer,~D.; Sui,~J.; McEnaney,~K.; Zhao,~H.; Jie,~Q.; Ren,~Z.; Chen,~G. High
  Thermoelectric Conversion Efficiency of \mbox{MgAgSb}-Based Material With
  Hot-Pressed Contacts. \emph{Energy Environ. Sci} \textbf{2015}, \emph{8},
  1299--1308\relax
\mciteBstWouldAddEndPuncttrue
\mciteSetBstMidEndSepPunct{\mcitedefaultmidpunct}
{\mcitedefaultendpunct}{\mcitedefaultseppunct}\relax
\EndOfBibitem
\bibitem[Kishimoto \latin{et~al.}(2015)Kishimoto, Koda, Akai, and
  Koyanagi]{kishimoto2015thermoelectric}
Kishimoto,~K.; Koda,~S.; Akai,~K.; Koyanagi,~T. Thermoelectric Properties of
  Sintered Type\mbox{-II} Clathrates \mbox{(K, Ba)$_{24}$(Ga, Sn)$_{136}$} with
  Various Carrier Concentrations. \emph{J. Appl. Phys.} \textbf{2015},
  \emph{118}, 125103\relax
\mciteBstWouldAddEndPuncttrue
\mciteSetBstMidEndSepPunct{\mcitedefaultmidpunct}
{\mcitedefaultendpunct}{\mcitedefaultseppunct}\relax
\EndOfBibitem
\bibitem[Cox \latin{et~al.}(2018)Cox, Gvozdetskyi, Owens-Baird, and
  Zaikina]{cox2018rapid}
Cox,~T.; Gvozdetskyi,~V.; Owens-Baird,~B.; Zaikina,~J.~V. Rapid Phase Screening
  via Hydride Route: A Discovery of \mbox{K$_{8-x}$Zn$_{18+3x}$Sb$_{16}$}.
  \emph{Chem. Mater.} \textbf{2018}, \emph{30}, 8707--8715\relax
\mciteBstWouldAddEndPuncttrue
\mciteSetBstMidEndSepPunct{\mcitedefaultmidpunct}
{\mcitedefaultendpunct}{\mcitedefaultseppunct}\relax
\EndOfBibitem
\bibitem[Lo \latin{et~al.}(2017)Lo, Ortiz, Toberer, He, Svitlyk, Chernyshov,
  Kolodiazhnyi, Lidin, and Mozharivskyj]{lo2017synthesis}
Lo,~C.-W.~T.; Ortiz,~B.~R.; Toberer,~E.~S.; He,~A.; Svitlyk,~V.;
  Chernyshov,~D.; Kolodiazhnyi,~T.; Lidin,~S.; Mozharivskyj,~Y. Synthesis,
  Structure, and Thermoelectric Properties of $\alpha$-\mbox{Zn$_3$Sb$_2$} and
  Comparison to $\beta$-\mbox{Zn$_{13}$Sb$_{10}$}. \emph{Chem. Mater.}
  \textbf{2017}, \emph{29}, 5249--5258\relax
\mciteBstWouldAddEndPuncttrue
\mciteSetBstMidEndSepPunct{\mcitedefaultmidpunct}
{\mcitedefaultendpunct}{\mcitedefaultseppunct}\relax
\EndOfBibitem
\bibitem[Qin \latin{et~al.}(2016)Qin, Qin, Fang, Zhang, Yue, Yan, Hu, and
  Su]{qin2016diverse}
Qin,~G.; Qin,~Z.; Fang,~W.-Z.; Zhang,~L.-C.; Yue,~S.-Y.; Yan,~Q.-B.; Hu,~M.;
  Su,~G. Diverse Anisotropy of Phonon Transport in Two-Dimensional Group
  \mbox{IV--VI} Compounds: A comparative Study. \emph{Nanoscale} \textbf{2016},
  \emph{8}, 11306--11319\relax
\mciteBstWouldAddEndPuncttrue
\mciteSetBstMidEndSepPunct{\mcitedefaultmidpunct}
{\mcitedefaultendpunct}{\mcitedefaultseppunct}\relax
\EndOfBibitem
\bibitem[Wu \latin{et~al.}(2009)Wu, Nyl{\'e}n, Naseyowma, Newman,
  Garcia-Garcia, and Haussermann]{wu2009comparative}
Wu,~Y.; Nyl{\'e}n,~J.; Naseyowma,~C.; Newman,~N.; Garcia-Garcia,~F.~J.;
  Haussermann,~U. Comparative Study of the Thermoelectric Properties of
  Amorphous \mbox{Zn$_{41}$Sb$_{59}$} and Crystalline \mbox{Zn$_4$Sb$_3$}.
  \emph{Chem. Mater.} \textbf{2009}, \emph{21}, 151--155\relax
\mciteBstWouldAddEndPuncttrue
\mciteSetBstMidEndSepPunct{\mcitedefaultmidpunct}
{\mcitedefaultendpunct}{\mcitedefaultseppunct}\relax
\EndOfBibitem
\bibitem[Rodr{\'\i}guez-Carvajal(1993)]{rodriguez1993recent}
Rodr{\'\i}guez-Carvajal,~J. Recent Advances in Magnetic Structure Determination
  by Neutron Powder Diffraction. \emph{Physica B} \textbf{1993}, \emph{192},
  55--69\relax
\mciteBstWouldAddEndPuncttrue
\mciteSetBstMidEndSepPunct{\mcitedefaultmidpunct}
{\mcitedefaultendpunct}{\mcitedefaultseppunct}\relax
\EndOfBibitem
\bibitem[Borup \latin{et~al.}(2012)Borup, Toberer, Zoltan, Nakatsukasa, Errico,
  Fleurial, Iversen, and Snyder]{Borup2012HallApparatus}
Borup,~K.~A.; Toberer,~E.~S.; Zoltan,~L.~D.; Nakatsukasa,~G.; Errico,~M.;
  Fleurial,~J.-P.; Iversen,~B.~B.; Snyder,~G.~J. Measurement of the Electrical
  Resistivity and \mbox{H}all Coefficient at High Temperatures. \emph{Rev. Sci.
  Instrum.} \textbf{2012}, \emph{83}, 123902\relax
\mciteBstWouldAddEndPuncttrue
\mciteSetBstMidEndSepPunct{\mcitedefaultmidpunct}
{\mcitedefaultendpunct}{\mcitedefaultseppunct}\relax
\EndOfBibitem
\bibitem[van~der Pauw(1958)]{Vanderpauw1958ResistivityApparatus}
van~der Pauw,~L.~J. A method of measuring the resistivity and \mbox{H}all
  coefficient on lamellae of arbitrary shape. \emph{Philips Tech. Rev.}
  \textbf{1958}, \emph{20}, 220--224\relax
\mciteBstWouldAddEndPuncttrue
\mciteSetBstMidEndSepPunct{\mcitedefaultmidpunct}
{\mcitedefaultendpunct}{\mcitedefaultseppunct}\relax
\EndOfBibitem
\bibitem[Iwanaga \latin{et~al.}(2011)Iwanaga, Toberer, LaLonde, and
  Snyder]{Iwanaga2011SeebeckApparatus}
Iwanaga,~S.; Toberer,~E.~S.; LaLonde,~A.; Snyder,~G.~J. A High Temperature
  Apparatus for Measurement of the \mbox{S}eebeck Coefficient. \emph{Rev. Sci.
  Instrum.} \textbf{2011}, \emph{82}, 063905\relax
\mciteBstWouldAddEndPuncttrue
\mciteSetBstMidEndSepPunct{\mcitedefaultmidpunct}
{\mcitedefaultendpunct}{\mcitedefaultseppunct}\relax
\EndOfBibitem
\bibitem[Kresse and Furthm\"uller(1996)Kresse, and
  Furthm\"uller]{kressePRB1996}
Kresse,~G.; Furthm\"uller,~J. Efficient Iterative Schemes for Ab Initio
  Total-Energy Calculations Using a Plane-Wave Basis Set. \emph{Phys. Rev. B}
  \textbf{1996}, \emph{54}, 11169--11186\relax
\mciteBstWouldAddEndPuncttrue
\mciteSetBstMidEndSepPunct{\mcitedefaultmidpunct}
{\mcitedefaultendpunct}{\mcitedefaultseppunct}\relax
\EndOfBibitem
\bibitem[Perdew \latin{et~al.}(1996)Perdew, Burke, and
  Ernzerhof]{Perdew1996generalized}
Perdew,~J.~P.; Burke,~K.; Ernzerhof,~M. Generalized Gradient Approximation Made
  Simple. \emph{Phys. Rev. Lett.} \textbf{1996}, \emph{77}, 3865--3868\relax
\mciteBstWouldAddEndPuncttrue
\mciteSetBstMidEndSepPunct{\mcitedefaultmidpunct}
{\mcitedefaultendpunct}{\mcitedefaultseppunct}\relax
\EndOfBibitem
\bibitem[Bl\"ochl(1994)]{blochl1994projector}
Bl\"ochl,~P.~E. Projector Augmented-Wave Method. \emph{Phys. Rev. B}
  \textbf{1994}, \emph{50}, 17953--17979\relax
\mciteBstWouldAddEndPuncttrue
\mciteSetBstMidEndSepPunct{\mcitedefaultmidpunct}
{\mcitedefaultendpunct}{\mcitedefaultseppunct}\relax
\EndOfBibitem
\bibitem[Monkhorst and Pack(1976)Monkhorst, and Pack]{monkhorst1976special}
Monkhorst,~H.~J.; Pack,~J.~D. Special Points for Brillouin-zone Integrations.
  \emph{Phys. Rev. B} \textbf{1976}, \emph{13}, 5188--5192\relax
\mciteBstWouldAddEndPuncttrue
\mciteSetBstMidEndSepPunct{\mcitedefaultmidpunct}
{\mcitedefaultendpunct}{\mcitedefaultseppunct}\relax
\EndOfBibitem
\bibitem[Chaput \latin{et~al.}(2011)Chaput, Togo, Tanaka, and Hug]{chaput2011}
Chaput,~L.; Togo,~A.; Tanaka,~I.; Hug,~G. Phonon-phonon interactions in
  transition metals. \emph{Phys. Rev. B} \textbf{2011}, \emph{84}, 094302\relax
\mciteBstWouldAddEndPuncttrue
\mciteSetBstMidEndSepPunct{\mcitedefaultmidpunct}
{\mcitedefaultendpunct}{\mcitedefaultseppunct}\relax
\EndOfBibitem
\bibitem[Togo and Tanaka(2015)Togo, and Tanaka]{togo2015first}
Togo,~A.; Tanaka,~I. First Principles Phonon Calculations in Materials Science.
  \emph{Scr. Mater.} \textbf{2015}, \emph{108}, 1--5\relax
\mciteBstWouldAddEndPuncttrue
\mciteSetBstMidEndSepPunct{\mcitedefaultmidpunct}
{\mcitedefaultendpunct}{\mcitedefaultseppunct}\relax
\EndOfBibitem
\bibitem[Parlinski \latin{et~al.}(1997)Parlinski, Li, and
  Kawazoe]{parlinski1997first}
Parlinski,~K.; Li,~Z.~Q.; Kawazoe,~Y. First-Principles Determination of the
  Soft Mode in Cubic ${\mathrm{ZrO}}_{2}$. \emph{Phys. Rev. Lett.}
  \textbf{1997}, \emph{78}, 4063--4066\relax
\mciteBstWouldAddEndPuncttrue
\mciteSetBstMidEndSepPunct{\mcitedefaultmidpunct}
{\mcitedefaultendpunct}{\mcitedefaultseppunct}\relax
\EndOfBibitem
\bibitem[Dudarev \latin{et~al.}(1998)Dudarev, Botton, Savrasov, Humphreys, and
  Sutton]{dudarev1998electron}
Dudarev,~S.~L.; Botton,~G.~A.; Savrasov,~S.~Y.; Humphreys,~C.~J.; Sutton,~A.~P.
  Electron-energy-loss spectra and the structural stability of nickel oxide: An
  \mbox{LSDA+U} study. \emph{Phys. Rev. B} \textbf{1998}, \emph{57},
  1505--1509\relax
\mciteBstWouldAddEndPuncttrue
\mciteSetBstMidEndSepPunct{\mcitedefaultmidpunct}
{\mcitedefaultendpunct}{\mcitedefaultseppunct}\relax
\EndOfBibitem
\bibitem[Lutfalla \latin{et~al.}(2011)Lutfalla, Shapovalov, and
  Bell]{lutfalla2011calibration}
Lutfalla,~S.; Shapovalov,~V.; Bell,~A.~T. Calibration of the \mbox{DFT/GGA+U}
  Method for Determination of Reduction Energies for Transition and Rare Earth
  Metal Oxides of \mbox{Ti, V, Mo, and Ce}. \emph{J. Chem. Theory Comput.}
  \textbf{2011}, \emph{7}, 2218--2223\relax
\mciteBstWouldAddEndPuncttrue
\mciteSetBstMidEndSepPunct{\mcitedefaultmidpunct}
{\mcitedefaultendpunct}{\mcitedefaultseppunct}\relax
\EndOfBibitem
\bibitem[Weber \latin{et~al.}(2012)Weber, O'Regan, Hine, Payne, Kotliar, and
  Littlewood]{weber2012vanadium}
Weber,~C.; O'Regan,~D.~D.; Hine,~N. D.~M.; Payne,~M.~C.; Kotliar,~G.;
  Littlewood,~P.~B. Vanadium Dioxide: A \mbox{Peierls-Mott} Insulator Stable
  against Disorder. \emph{Phys. Rev. Lett.} \textbf{2012}, \emph{108},
  256402\relax
\mciteBstWouldAddEndPuncttrue
\mciteSetBstMidEndSepPunct{\mcitedefaultmidpunct}
{\mcitedefaultendpunct}{\mcitedefaultseppunct}\relax
\EndOfBibitem
\bibitem[Bensen \latin{et~al.}(2021)Bensen, Ciesielski, Gomes, Ortiz, Falke,
  Pavlosiuk, Weber, Braden, Steirer, Szyma{\'n}ski, \latin{et~al.}
  others]{bensen2021anomalous}
Bensen,~E.~A.; Ciesielski,~K.; Gomes,~L.~C.; Ortiz,~B.~R.; Falke,~J.;
  Pavlosiuk,~O.; Weber,~D.; Braden,~T.~L.; Steirer,~K.~X.; Szyma{\'n}ski,~D.,
  \latin{et~al.}  Anomalous Electronic Properties in Layered, Disordered
  \mbox{ZnVSb}. \emph{Phys. Rev. Mater.} \textbf{2021}, \emph{5}, 015002\relax
\mciteBstWouldAddEndPuncttrue
\mciteSetBstMidEndSepPunct{\mcitedefaultmidpunct}
{\mcitedefaultendpunct}{\mcitedefaultseppunct}\relax
\EndOfBibitem
\bibitem[Goh \latin{et~al.}(2017)Goh, Mah, and Yoon]{goh2017effects}
Goh,~E.; Mah,~J.; Yoon,~T. Effects of Hubbard Term Correction on the Structural
  Parameters and Electronic Properties of Wurtzite \mbox{ZnO}. \emph{Comput.
  Mater. Sci.} \textbf{2017}, \emph{138}, 111--116\relax
\mciteBstWouldAddEndPuncttrue
\mciteSetBstMidEndSepPunct{\mcitedefaultmidpunct}
{\mcitedefaultendpunct}{\mcitedefaultseppunct}\relax
\EndOfBibitem
\bibitem[Kanoun \latin{et~al.}(2012)Kanoun, Goumri-Said, Schwingenschl{\"o}gl,
  and Manchon]{kanoun2012magnetism}
Kanoun,~M.~B.; Goumri-Said,~S.; Schwingenschl{\"o}gl,~U.; Manchon,~A. Magnetism
  in \mbox{Sc}-doped \mbox{ZnO} with Zinc Vacancies: A Hybrid Density
  Functional and \mbox{GGA+U} Approaches. \emph{Chem. Phys. Lett.}
  \textbf{2012}, \emph{532}, 96--99\relax
\mciteBstWouldAddEndPuncttrue
\mciteSetBstMidEndSepPunct{\mcitedefaultmidpunct}
{\mcitedefaultendpunct}{\mcitedefaultseppunct}\relax
\EndOfBibitem
\bibitem[Zhang \latin{et~al.}(2004)Zhang, Yang, Ando, and
  Ohji]{zhang2004reactive}
Zhang,~G.-J.; Yang,~J.-F.; Ando,~M.; Ohji,~T. Reactive Hot Pressing of
  Alumina-Silicon Carbide Nanocomposites. \emph{J. Am. Ceram. Soc.}
  \textbf{2004}, \emph{87}, 299--301\relax
\mciteBstWouldAddEndPuncttrue
\mciteSetBstMidEndSepPunct{\mcitedefaultmidpunct}
{\mcitedefaultendpunct}{\mcitedefaultseppunct}\relax
\EndOfBibitem
\bibitem[Chamberlain \latin{et~al.}(2009)Chamberlain, Fahrenholtz, and
  Hilmas]{chamberlain2009reactive}
Chamberlain,~A.~L.; Fahrenholtz,~W.~G.; Hilmas,~G.~E. Reactive Hot Pressing of
  Zirconium Diboride. \emph{J. Eur. Ceram. Soc.} \textbf{2009}, \emph{29},
  3401--3408\relax
\mciteBstWouldAddEndPuncttrue
\mciteSetBstMidEndSepPunct{\mcitedefaultmidpunct}
{\mcitedefaultendpunct}{\mcitedefaultseppunct}\relax
\EndOfBibitem
\bibitem[May \latin{et~al.}(2008)May, Fleurial, and
  Snyder]{may2008thermoelectric}
May,~A.~F.; Fleurial,~J.-P.; Snyder,~G.~J. Thermoelectric Performance of
  Lanthanum Telluride Produced via Mechanical Alloying. \emph{Phys. Rev. B}
  \textbf{2008}, \emph{78}, 125205\relax
\mciteBstWouldAddEndPuncttrue
\mciteSetBstMidEndSepPunct{\mcitedefaultmidpunct}
{\mcitedefaultendpunct}{\mcitedefaultseppunct}\relax
\EndOfBibitem
\bibitem[Alinejad \latin{et~al.}(2020)Alinejad, Takagiwa, and
  Ikeda]{alinejad2020activated}
Alinejad,~B.; Takagiwa,~Y.; Ikeda,~T. Activated Reactive Consolidation Method
  as a New Approach to Enhanced Thermoelectric Properties of \textit{n}-type
  Nanostructured \mbox{Mg$_2$Si}. \emph{ACS Appl. Energy Mater.} \textbf{2020},
  \relax
\mciteBstWouldAddEndPunctfalse
\mciteSetBstMidEndSepPunct{\mcitedefaultmidpunct}
{}{\mcitedefaultseppunct}\relax
\EndOfBibitem
\bibitem[Dolyniuk \latin{et~al.}(2017)Dolyniuk, Whitfield, Lee, Lebedev, and
  Kovnir]{dolyniuk2017controlling}
Dolyniuk,~J.; Whitfield,~P.; Lee,~K.; Lebedev,~O.; Kovnir,~K. Controlling
  Superstructural Ordering in the Clathrate-\mbox{I Ba$_{8}M_{16}$P$_{30}$ ($M$
  = Cu, Zn}) Through the Formation of Metal--Metal Bonds. \emph{Chem. Sci.}
  \textbf{2017}, \emph{8}, 3650--3659\relax
\mciteBstWouldAddEndPuncttrue
\mciteSetBstMidEndSepPunct{\mcitedefaultmidpunct}
{\mcitedefaultendpunct}{\mcitedefaultseppunct}\relax
\EndOfBibitem
\bibitem[Gerke \latin{et~al.}(2013)Gerke, Hoffmann, and Poettgen]{gerke2013zn3}
Gerke,~B.; Hoffmann,~R.-D.; Poettgen,~R. \mbox{Zn$_3$} and \mbox{Ga$_3$}
  Triangles as Building Units in \mbox{Sr$_2$Au$_6$Zn$_3$} and
  \mbox{Sr$_2$Au$_6$Ga$_3$}. \emph{Z. Anorg. Allg. Chem.} \textbf{2013},
  \emph{639}, 2444--2449\relax
\mciteBstWouldAddEndPuncttrue
\mciteSetBstMidEndSepPunct{\mcitedefaultmidpunct}
{\mcitedefaultendpunct}{\mcitedefaultseppunct}\relax
\EndOfBibitem
\bibitem[Falmbigl \latin{et~al.}(2014)Falmbigl, Puchegger, and
  Rogl]{falmbigl2014mechanical}
Falmbigl,~M.; Puchegger,~S.; Rogl,~P. \emph{Mechanical Properties of
  Intermetallic Clathrates. In: Nolas G. (eds) The Physics and Chemistry of
  Inorganic Clathrates}; Springer, 2014; pp 277--326\relax
\mciteBstWouldAddEndPuncttrue
\mciteSetBstMidEndSepPunct{\mcitedefaultmidpunct}
{\mcitedefaultendpunct}{\mcitedefaultseppunct}\relax
\EndOfBibitem
\bibitem[Qiu \latin{et~al.}(2004)Qiu, Swainson, Nolas, and
  White]{qiu2004structure}
Qiu,~L.; Swainson,~I.~P.; Nolas,~G.~S.; White,~M.~A. Structure, Thermal, and
  Transport Properties of the Clathrates \mbox{Sr$_8$Zn$_8$Ge$_{38}$},
  \mbox{Sr$_8$Ga$_{16}$Ge$_{30}$}, and \mbox{Ba$_8$Ga$_{16}$Si$_{30}$}.
  \emph{Phys. Rev. B} \textbf{2004}, \emph{70}, 035208\relax
\mciteBstWouldAddEndPuncttrue
\mciteSetBstMidEndSepPunct{\mcitedefaultmidpunct}
{\mcitedefaultendpunct}{\mcitedefaultseppunct}\relax
\EndOfBibitem
\bibitem[Kim \latin{et~al.}(2015)Kim, Gibbs, Tang, Wang, and
  Snyder]{kim2015characterization}
Kim,~H.-S.; Gibbs,~Z.~M.; Tang,~Y.; Wang,~H.; Snyder,~G.~J. Characterization of
  \mbox{L}orenz Number With \mbox{S}eebeck Coefficient Measurement. \emph{APL
  Mater.} \textbf{2015}, \emph{3}, 041506\relax
\mciteBstWouldAddEndPuncttrue
\mciteSetBstMidEndSepPunct{\mcitedefaultmidpunct}
{\mcitedefaultendpunct}{\mcitedefaultseppunct}\relax
\EndOfBibitem
\bibitem[Qian \latin{et~al.}(2021)Qian, Zhou, and Chen]{qian2021phonon}
Qian,~X.; Zhou,~J.; Chen,~G. Phonon-Engineered Extreme Thermal Conductivity
  Materials. \emph{Nat. Mater.} \textbf{2021}, 1--15\relax
\mciteBstWouldAddEndPuncttrue
\mciteSetBstMidEndSepPunct{\mcitedefaultmidpunct}
{\mcitedefaultendpunct}{\mcitedefaultseppunct}\relax
\EndOfBibitem
\bibitem[Candolfi \latin{et~al.}(2012)Candolfi, Aydemir, Baitinger, Oeschler,
  Steglich, and Grin]{candolfi2012high}
Candolfi,~C.; Aydemir,~U.; Baitinger,~M.; Oeschler,~N.; Steglich,~F.; Grin,~Y.
  High Temperature Thermoelectric Properties of The Type-\mbox{I} Clathrate
  \mbox{Ba$_8$Au$_x$Si$_{46-x}$}. \emph{J. Appl. Phys.} \textbf{2012},
  \emph{111}, 043706\relax
\mciteBstWouldAddEndPuncttrue
\mciteSetBstMidEndSepPunct{\mcitedefaultmidpunct}
{\mcitedefaultendpunct}{\mcitedefaultseppunct}\relax
\EndOfBibitem
\bibitem[Lee \latin{et~al.}(2007)Lee, Yoshizawa, Avila, Hase, Kihou, and
  Takabatake]{lee2007neutron}
Lee,~C.; Yoshizawa,~H.; Avila,~M.; Hase,~I.; Kihou,~K.; Takabatake,~T. Neutron
  Scattering Study of Phonon Dynamics on Type-\mbox{I} Clathrate
  \mbox{Ba$_8$Ga$_{16}$Ge$_{30}$}. J. Phys.: Conf. Series. 2007; p 012169\relax
\mciteBstWouldAddEndPuncttrue
\mciteSetBstMidEndSepPunct{\mcitedefaultmidpunct}
{\mcitedefaultendpunct}{\mcitedefaultseppunct}\relax
\EndOfBibitem
\bibitem[Koza \latin{et~al.}(2010)Koza, Johnson, Mutka, Rotter, Nasir, Grytsiv,
  and Rogl]{koza2010vibrational}
Koza,~M.; Johnson,~M.; Mutka,~H.; Rotter,~M.; Nasir,~N.; Grytsiv,~A.; Rogl,~P.
  Vibrational Dynamics of the Type-\mbox{I} Clathrate
  \mbox{Ba$_{8}$Zn$_x$Ge$_{46-x-y}\square_y$}, ($x$ = 0, 2, 4, 6, 8).
  \emph{Phys. Rev. B} \textbf{2010}, \emph{82}, 214301\relax
\mciteBstWouldAddEndPuncttrue
\mciteSetBstMidEndSepPunct{\mcitedefaultmidpunct}
{\mcitedefaultendpunct}{\mcitedefaultseppunct}\relax
\EndOfBibitem
\bibitem[Stefanoski \latin{et~al.}(2014)Stefanoski, Beekman, and
  Nolas]{stefanoski2014inorganic}
Stefanoski,~S.; Beekman,~M.; Nolas,~G.~S. \emph{Inorganic Clathrates for
  Thermoelectric Applications. In: Nolas G. (eds) The Physics and Chemistry of
  Inorganic Clathrates}; Springer, 2014; pp 169--191\relax
\mciteBstWouldAddEndPuncttrue
\mciteSetBstMidEndSepPunct{\mcitedefaultmidpunct}
{\mcitedefaultendpunct}{\mcitedefaultseppunct}\relax
\EndOfBibitem
\bibitem[Wang \latin{et~al.}(2020)Wang, Dolyniuk, Krenkel, Niedziela, Tanatar,
  Timmons, Lanigan-Atkins, Zhou, Cheng, Ramirez-Cuesta, \latin{et~al.}
  others]{wang2020clathrate}
Wang,~J.; Dolyniuk,~J.-A.; Krenkel,~E.~H.; Niedziela,~J.~L.; Tanatar,~M.~A.;
  Timmons,~E.~I.; Lanigan-Atkins,~T.; Zhou,~H.; Cheng,~Y.;
  Ramirez-Cuesta,~A.~J., \latin{et~al.}  Clathrate \mbox{BaNi$_2$P$_4$}: an
  Interplay of Heat and Charge Transport Due to Strong Host--Guest
  Interactions. \emph{Chem. Mater.} \textbf{2020}, \emph{32}, 7932--7940\relax
\mciteBstWouldAddEndPuncttrue
\mciteSetBstMidEndSepPunct{\mcitedefaultmidpunct}
{\mcitedefaultendpunct}{\mcitedefaultseppunct}\relax
\EndOfBibitem
\bibitem[Myles \latin{et~al.}(2003)Myles, Dong, and Sankey]{myles2003rattling}
Myles,~C.~W.; Dong,~J.; Sankey,~O.~F. Rattling Guest Atoms in \mbox{Si, Ge, and
  S}n-based type-\mbox{II} Clathrate Materials. \emph{Phys. Status Solidi B}
  \textbf{2003}, \emph{239}, 26--34\relax
\mciteBstWouldAddEndPuncttrue
\mciteSetBstMidEndSepPunct{\mcitedefaultmidpunct}
{\mcitedefaultendpunct}{\mcitedefaultseppunct}\relax
\EndOfBibitem
\bibitem[Bedoya-Mart{\'\i}nez \latin{et~al.}(2016)Bedoya-Mart{\'\i}nez,
  Hashibon, and Els{\"a}sser]{bedoya2016influence}
Bedoya-Mart{\'\i}nez,~O.; Hashibon,~A.; Els{\"a}sser,~C. Influence of Point
  Defects on the Phonon Thermal Conductivity and Phonon Density of States of
  \mbox{Bi$_2$Te$_3$}. \emph{Phys. Status Solidi A} \textbf{2016}, \emph{213},
  684--693\relax
\mciteBstWouldAddEndPuncttrue
\mciteSetBstMidEndSepPunct{\mcitedefaultmidpunct}
{\mcitedefaultendpunct}{\mcitedefaultseppunct}\relax
\EndOfBibitem
\bibitem[Paudel and Lambrecht(2009)Paudel, and Lambrecht]{paudel2009calculated}
Paudel,~T.~R.; Lambrecht,~W.~R. Calculated Phonon Band Structure and Density of
  States and Interpretation of The \mbox{R}aman Spectrum in Rocksalt
  \mbox{ScN}. \emph{Phys. Rev. B} \textbf{2009}, \emph{79}, 085205\relax
\mciteBstWouldAddEndPuncttrue
\mciteSetBstMidEndSepPunct{\mcitedefaultmidpunct}
{\mcitedefaultendpunct}{\mcitedefaultseppunct}\relax
\EndOfBibitem
\bibitem[Erhart \latin{et~al.}(2015)Erhart, Hyldgaard, and
  Lindroth]{erhart2015microscopic}
Erhart,~P.; Hyldgaard,~P.; Lindroth,~D.~O. Microscopic Origin of Thermal
  Conductivity Reduction in Disordered van der Waals Solids. \emph{Chem.
  Mater.} \textbf{2015}, \emph{27}, 5511--5518\relax
\mciteBstWouldAddEndPuncttrue
\mciteSetBstMidEndSepPunct{\mcitedefaultmidpunct}
{\mcitedefaultendpunct}{\mcitedefaultseppunct}\relax
\EndOfBibitem
\bibitem[Callaway(1959)]{callaway1959model}
Callaway,~J. Model for Lattice Thermal Conductivity at Low Temperatures.
  \emph{Phys. Rev.} \textbf{1959}, \emph{113}, 1046\relax
\mciteBstWouldAddEndPuncttrue
\mciteSetBstMidEndSepPunct{\mcitedefaultmidpunct}
{\mcitedefaultendpunct}{\mcitedefaultseppunct}\relax
\EndOfBibitem
\bibitem[Ma \latin{et~al.}(2014)Ma, Li, and Luo]{ma2014examining}
Ma,~J.; Li,~W.; Luo,~X. Examining the \mbox{C}allaway model for Lattice Thermal
  Conductivity. \emph{Phys. Rev. B} \textbf{2014}, \emph{90}, 035203\relax
\mciteBstWouldAddEndPuncttrue
\mciteSetBstMidEndSepPunct{\mcitedefaultmidpunct}
{\mcitedefaultendpunct}{\mcitedefaultseppunct}\relax
\EndOfBibitem
\bibitem[Petersen \latin{et~al.}(2015)Petersen, Bhattacharya, Tritt, and
  Poon]{petersen2015critical}
Petersen,~A.; Bhattacharya,~S.; Tritt,~T.; Poon,~S. Critical Analysis of
  Lattice Thermal Conductivity of Half-\mbox{H}eusler Alloys Using Variations
  of \mbox{C}allaway Model. \emph{J. Appl. Phys.} \textbf{2015}, \emph{117},
  035706\relax
\mciteBstWouldAddEndPuncttrue
\mciteSetBstMidEndSepPunct{\mcitedefaultmidpunct}
{\mcitedefaultendpunct}{\mcitedefaultseppunct}\relax
\EndOfBibitem
\bibitem[Ciesielski \latin{et~al.}(2020)Ciesielski, Synoradzki, Veremchuk,
  Skokowski, Szyma{\'n}ski, Grin, and
  Kaczorowski]{ciesielski2020thermoelectric}
Ciesielski,~K.; Synoradzki,~K.; Veremchuk,~I.; Skokowski,~P.;
  Szyma{\'n}ski,~D.; Grin,~Y.; Kaczorowski,~D. Thermoelectric Performance of
  the Half-Heusler Phases \mbox{$R$NiSb ($R$ = Sc, Dy, Er, Tm, Lu):} High
  Mobility Ratio Between Majority and Minority Charge Carriers. \emph{Phys.
  Rev. Appl.} \textbf{2020}, \emph{14}, 054046\relax
\mciteBstWouldAddEndPuncttrue
\mciteSetBstMidEndSepPunct{\mcitedefaultmidpunct}
{\mcitedefaultendpunct}{\mcitedefaultseppunct}\relax
\EndOfBibitem
\bibitem[B{\"o}cher \latin{et~al.}(2017)B{\"o}cher, Culver, Peilst{\"o}cker,
  Weldert, and Zeier]{bocher2017vacancy}
B{\"o}cher,~F.; Culver,~S.~P.; Peilst{\"o}cker,~J.; Weldert,~K.~S.;
  Zeier,~W.~G. Vacancy and Anti-site Disorder Scattering in \mbox{AgBiSe$_2$}
  Thermoelectrics. \emph{Dalton Trans.} \textbf{2017}, \emph{46},
  3906--3914\relax
\mciteBstWouldAddEndPuncttrue
\mciteSetBstMidEndSepPunct{\mcitedefaultmidpunct}
{\mcitedefaultendpunct}{\mcitedefaultseppunct}\relax
\EndOfBibitem
\bibitem[Schrade and Finstad(2018)Schrade, and Finstad]{schrade2018using}
Schrade,~M.; Finstad,~T.~G. Using the \mbox{C}allaway Model to Deduce Relevant
  Phonon Scattering Processes: The Importance of Phonon Dispersion. \emph{Phys.
  Status Solidi B} \textbf{2018}, \emph{255}, 1800208\relax
\mciteBstWouldAddEndPuncttrue
\mciteSetBstMidEndSepPunct{\mcitedefaultmidpunct}
{\mcitedefaultendpunct}{\mcitedefaultseppunct}\relax
\EndOfBibitem
\bibitem[Pohl(1962)]{pohl1962thermal}
Pohl,~R. Thermal Conductivity and Phonon Resonance Scattering. \emph{Phys. Rev.
  Lett.} \textbf{1962}, \emph{8}, 481\relax
\mciteBstWouldAddEndPuncttrue
\mciteSetBstMidEndSepPunct{\mcitedefaultmidpunct}
{\mcitedefaultendpunct}{\mcitedefaultseppunct}\relax
\EndOfBibitem
\bibitem[Bagheri \latin{et~al.}(2020)Bagheri, Reddy, Kim, Rounds, Sochacki,
  Kirste, Bockowski, Collazo, and Sitar]{bagheri2020impact}
Bagheri,~P.; Reddy,~P.; Kim,~J.~H.; Rounds,~R.; Sochacki,~T.; Kirste,~R.;
  Bockowski,~M.; Collazo,~R.; Sitar,~Z. Impact of Impurity-Based Phonon
  Resonant Scattering on Thermal Conductivity of Single Crystalline \mbox{GaN}.
  \emph{Appl. Phys. Lett.} \textbf{2020}, \emph{117}, 082101\relax
\mciteBstWouldAddEndPuncttrue
\mciteSetBstMidEndSepPunct{\mcitedefaultmidpunct}
{\mcitedefaultendpunct}{\mcitedefaultseppunct}\relax
\EndOfBibitem
\bibitem[Pisoni \latin{et~al.}(2014)Pisoni, Jacimovic, Barisic, Spina,
  Ga{\'a}l, Forr{\'o}, and Horv{\'a}th]{pisoni2014ultra}
Pisoni,~A.; Jacimovic,~J.; Barisic,~O.~S.; Spina,~M.; Ga{\'a}l,~R.;
  Forr{\'o},~L.; Horv{\'a}th,~E. Ultra-Low Thermal Conductivity in
  Organic--Inorganic Hybrid Perovskite \mbox{CH$_3$NH$_3$PbI$_3$}. \emph{J
  Phys. Chem. Lett.} \textbf{2014}, \emph{5}, 2488--2492\relax
\mciteBstWouldAddEndPuncttrue
\mciteSetBstMidEndSepPunct{\mcitedefaultmidpunct}
{\mcitedefaultendpunct}{\mcitedefaultseppunct}\relax
\EndOfBibitem
\bibitem[Michalski and White(1995)Michalski, and White]{michalski1995thermal}
Michalski,~D.; White,~M.~A. Thermal Conductivity of a Clathrate With Restrained
  Guests: The \mbox{CCl$_4$} Clathrate of \mbox{D}ianin's Compound. \emph{J.
  Phys. Chem.} \textbf{1995}, \emph{99}, 3774--3780\relax
\mciteBstWouldAddEndPuncttrue
\mciteSetBstMidEndSepPunct{\mcitedefaultmidpunct}
{\mcitedefaultendpunct}{\mcitedefaultseppunct}\relax
\EndOfBibitem
\bibitem[Nolas \latin{et~al.}(1998)Nolas, Cohn, Slack, and
  Schujman]{nolas1998semiconducting}
Nolas,~G.; Cohn,~J.; Slack,~G.; Schujman,~S. Semiconducting \mbox{G}e
  clathrates: Promising Candidates for Thermoelectric Applications. \emph{Appl.
  Phys. Lett.} \textbf{1998}, \emph{73}, 178--180\relax
\mciteBstWouldAddEndPuncttrue
\mciteSetBstMidEndSepPunct{\mcitedefaultmidpunct}
{\mcitedefaultendpunct}{\mcitedefaultseppunct}\relax
\EndOfBibitem
\bibitem[Gurunathan \latin{et~al.}(2020)Gurunathan, Hanus, Dylla, Katre, and
  Snyder]{gurunathan2020analytical}
Gurunathan,~R.; Hanus,~R.; Dylla,~M.; Katre,~A.; Snyder,~G.~J. Analytical
  models of phonon--point-defect scattering. \emph{Phys. Rev. Appl.}
  \textbf{2020}, \emph{13}, 034011\relax
\mciteBstWouldAddEndPuncttrue
\mciteSetBstMidEndSepPunct{\mcitedefaultmidpunct}
{\mcitedefaultendpunct}{\mcitedefaultseppunct}\relax
\EndOfBibitem
\bibitem[H{\"a}rk{\"o}nen and Karttunen(2016)H{\"a}rk{\"o}nen, and
  Karttunen]{harkonen2016ab}
H{\"a}rk{\"o}nen,~V.~J.; Karttunen,~A.~J. Ab Initio Studies on the Lattice
  Thermal Conductivity of Silicon Clathrate Frameworks \mbox{II} and
  \mbox{VIII}. \emph{Phys. Rev. B} \textbf{2016}, \emph{93}, 024307\relax
\mciteBstWouldAddEndPuncttrue
\mciteSetBstMidEndSepPunct{\mcitedefaultmidpunct}
{\mcitedefaultendpunct}{\mcitedefaultseppunct}\relax
\EndOfBibitem
\bibitem[Lory \latin{et~al.}(2017)Lory, Pailh{\`e}s, Giordano, Euchner, Nguyen,
  Ramlau, Borrmann, Schmidt, Baitinger, Ikeda, \latin{et~al.}
  others]{lory2017direct}
Lory,~P.-F.; Pailh{\`e}s,~S.; Giordano,~V.~M.; Euchner,~H.; Nguyen,~H.~D.;
  Ramlau,~R.; Borrmann,~H.; Schmidt,~M.; Baitinger,~M.; Ikeda,~M.,
  \latin{et~al.}  Direct Measurement of Individual Phonon Lifetimes in the
  Clathrate Compound \mbox{Ba$_{7.81}$Ge$_{40.67}$Au$_{5.33}$}. \emph{Nat.
  Commun.} \textbf{2017}, \emph{8}, 1--10\relax
\mciteBstWouldAddEndPuncttrue
\mciteSetBstMidEndSepPunct{\mcitedefaultmidpunct}
{\mcitedefaultendpunct}{\mcitedefaultseppunct}\relax
\EndOfBibitem
\bibitem[Sofo and Mahan(1994)Sofo, and Mahan]{sofo1994optimum}
Sofo,~J.~O.; Mahan,~G. Optimum Band Gap of a Thermoelectric Material.
  \emph{Phys. Rev. B} \textbf{1994}, \emph{49}, 4565\relax
\mciteBstWouldAddEndPuncttrue
\mciteSetBstMidEndSepPunct{\mcitedefaultmidpunct}
{\mcitedefaultendpunct}{\mcitedefaultseppunct}\relax
\EndOfBibitem
\bibitem[Goldsmid and Sharp(1999)Goldsmid, and Sharp]{goldsmid1999estimation}
Goldsmid,~H.; Sharp,~J. Estimation of the Thermal Band Gap of a Semiconductor
  From \mbox{S}eebeck Measurements. \emph{J. Electron. Mater.} \textbf{1999},
  \emph{28}, 869--872\relax
\mciteBstWouldAddEndPuncttrue
\mciteSetBstMidEndSepPunct{\mcitedefaultmidpunct}
{\mcitedefaultendpunct}{\mcitedefaultseppunct}\relax
\EndOfBibitem
\bibitem[Snyder \latin{et~al.}(2020)Snyder, Snyder, Wood, Gurunathan, Snyder,
  and Niu]{snyder2020weighted}
Snyder,~G.~J.; Snyder,~A.~H.; Wood,~M.; Gurunathan,~R.; Snyder,~B.~H.; Niu,~C.
  Weighted Mobility. \emph{Adv. Mater.} \textbf{2020}, 2001537\relax
\mciteBstWouldAddEndPuncttrue
\mciteSetBstMidEndSepPunct{\mcitedefaultmidpunct}
{\mcitedefaultendpunct}{\mcitedefaultseppunct}\relax
\EndOfBibitem
\bibitem[Schmitt \latin{et~al.}(2015)Schmitt, Gibbs, Snyder, and
  Felser]{schmitt2015resolving}
Schmitt,~J.; Gibbs,~Z.~M.; Snyder,~G.~J.; Felser,~C. Resolving the True Band
  Gap of \mbox{ZrNiSn} half-Heusler Thermoelectric Materials. \emph{Mater.
  Horiz.} \textbf{2015}, \emph{2}, 68--75\relax
\mciteBstWouldAddEndPuncttrue
\mciteSetBstMidEndSepPunct{\mcitedefaultmidpunct}
{\mcitedefaultendpunct}{\mcitedefaultseppunct}\relax
\EndOfBibitem
\bibitem[Gibbs \latin{et~al.}(2015)Gibbs, Kim, Wang, and Snyder]{gibbs2015band}
Gibbs,~Z.~M.; Kim,~H.-S.; Wang,~H.; Snyder,~G.~J. Band Gap Estimation From
  Temperature Dependent Seebeck Measurement -- Deviations from the
  \mbox{$2e|S|_{max}T_{max}$} Relation. \emph{Appl. Phys. Lett.} \textbf{2015},
  \emph{106}, 022112\relax
\mciteBstWouldAddEndPuncttrue
\mciteSetBstMidEndSepPunct{\mcitedefaultmidpunct}
{\mcitedefaultendpunct}{\mcitedefaultseppunct}\relax
\EndOfBibitem
\bibitem[Rowe(2017)]{rowe2017materials}
Rowe,~D.~M. \emph{Materials, Preparation, and Characterization in
  Thermoelectrics}; CRC press, 2017\relax
\mciteBstWouldAddEndPuncttrue
\mciteSetBstMidEndSepPunct{\mcitedefaultmidpunct}
{\mcitedefaultendpunct}{\mcitedefaultseppunct}\relax
\EndOfBibitem
\bibitem[Xie \latin{et~al.}(2014)Xie, Wang, Fu, Liu, Snyder, Zhao, and
  Zhu]{xie2014intrinsic}
Xie,~H.; Wang,~H.; Fu,~C.; Liu,~Y.; Snyder,~G.~J.; Zhao,~X.; Zhu,~T. The
  Intrinsic Disorder Related Alloy Scattering in \mbox{ZrNiSn}
  Half-\mbox{H}eusler Thermoelectric Materials. \emph{Sci. Rep.} \textbf{2014},
  \emph{4}, 6888\relax
\mciteBstWouldAddEndPuncttrue
\mciteSetBstMidEndSepPunct{\mcitedefaultmidpunct}
{\mcitedefaultendpunct}{\mcitedefaultseppunct}\relax
\EndOfBibitem
\bibitem[Towns \latin{et~al.}(2014)Towns, Cockerill, Dahan, Foster, Gaither,
  Grimshaw, Hazlewood, Lathrop, Lifka, Peterson, \latin{et~al.}
  others]{towns2014xsede}
Towns,~J.; Cockerill,~T.; Dahan,~M.; Foster,~I.; Gaither,~K.; Grimshaw,~A.;
  Hazlewood,~V.; Lathrop,~S.; Lifka,~D.; Peterson,~G.~D., \latin{et~al.}
  XSEDE: accelerating scientific discovery. \emph{Comput. Sci. Eng}
  \textbf{2014}, \emph{16}, 62--74\relax
\mciteBstWouldAddEndPuncttrue
\mciteSetBstMidEndSepPunct{\mcitedefaultmidpunct}
{\mcitedefaultendpunct}{\mcitedefaultseppunct}\relax
\EndOfBibitem
\end{mcitethebibliography}

\newpage
\noindent \Large{\textbf{TOC Graphic}}

\begin{figure}
\includegraphics[height=4.45cm]{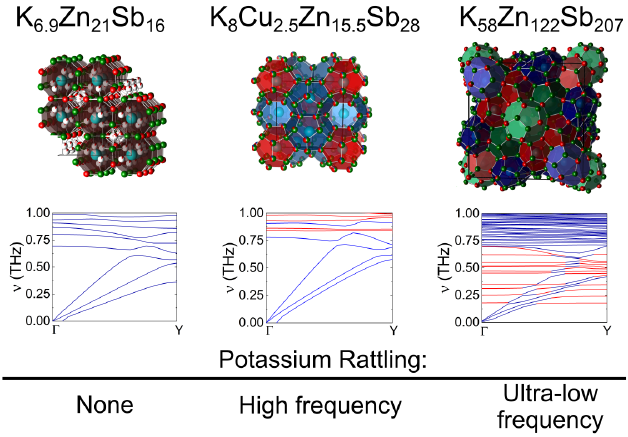}
\end{figure}

\end{document}